\documentclass[12pt]{iopart}
\usepackage[dvips]{graphicx}
\usepackage{url}

\newcommand{\gray}{$\gamma$-ray}
\newcommand{\grays}{$\gamma$-rays}
\newcommand{\de}{DE}

\newcommand{\pubjournal}[6] {#1 #5, #2, {\bf #3}, #4}

\newcommand{\icrc}{Int.\ Cosmic Ray Conf.}

\newcommand{\pubproc}[8]{#1 #7 {\it #2} ({\it #4}) #3 (#5) p~#6}

\newcommand{\apj}{ApJ}%
\newcommand{\apjl}{ApJ}%
\newcommand{\aap}{A\&A}%
\newcommand{\aaps}{A\&AS}%
\newcommand{\prl}{Phys.~Rev.~Lett.}%
\newcommand{\jgr}{J.~Geophys.~Res.}%
\usepackage{iopams}  

\begin{document}

\title[]{Pre-launch estimates for GLAST sensitivity to Dark Matter annihilation signals}
\author{E.~A.~Baltz$\ddag$, B.~Berenji$\ddag$, G.~Bertone$\S$, L.~Bergstr\"om $\P$, E.~Bloom$\ddag$, T.~Bringmann $\P$, J.~Chiang$\ddag$, J.~Cohen-Tanugi$\ddag$, J.~Conrad $\P$ $^{**}$, Y.~Edmonds$\ddag$, J. Edsj\"o $\P$, G.~Godfrey$\ddag$, R.~E.~Hughes$\|$, R.~P.~Johnson$^+$, A.~Lionetto$^*$, A.~A.~Moiseev$\sharp$, A.~Morselli$^*$, I.~V.~Moskalenko$\dagger\dagger$, E.~Nuss$\ddag\ddag$, J.~F.~Ormes$\S\S$, R.~Rando$\P\P\P$, 
A.~J.~Sander$\|$, 
A.~Sellerholm$\P$, P.~D.~Smith$\|$, A.~W.~Strong$\|\|$,
L.~Wai$\ddag$, 
P.~Wang$\ddag$, B.~L.~Winer $\|$}
\address{\ddag Stanford Linear Accelerator Center (SLAC), Kavli Institute for Particle Astrophysics \& Cosmology (KIPAC), Menlo Park, CA 94025,USA}
\address{\S Institut d'Astrophysique de Paris, 98bis Bd Arago, F-75014 Paris, France}
\address{\P Stockholm University, Physics Department, Albanova, SE-10691 Stockholm, Sweden}
\address{$\|$ The Ohio State University, Columbus, Ohio 43210, USA}
\address{$^+$ Santa Cruz Institute for Particle Physics, University of California, Santa Cruz, CA 95064, USA}
\address{$^*$ INFN Roma Tor Vergata, via della Ricerca Scientifica 1, Roma, Italy}
\address{$\sharp$ CRESST and AstroParticle Physics Laboratory, NASA/GSFC, Greenbelt, MD 20771, USA}
\address{$\dagger\dagger$ Hansen Experimental Physics Laboratory and Kavli Institute for Particle Astrophysics and Cosmology, Stanford University, Stanford, CA 94305, USA}
\address{$\ddagger\ddagger$LPTA, University of Montpellier,CNRS/IN2P3, Montpellier, France}
\address{$\S\S$ University of Denver, Denver, CO 80208, USA}
\address{$\|\|$ Max-Planck-Institut f\"ur extraterrestrische Physik, Postfach 1312, D-85741 Garching, Germany}
\address{$\P\P\P$ INFN, Sezione di Padova, Via Marzolo 8, Padova, I-35131, Italy}
\address{$^{**}$Corresponding author: conrad@physto.se}



\begin{abstract}
  We investigate the sensitivity of the Gamma-ray Large Area Space Telescope (GLAST)  to indirectly detect weakly interacting massive particles (WIMPs) through the $\gamma$-ray signal that their pair annihilation produces. WIMPs are among the favorite candidates to explain the compelling evidence that about 80\% of the mass in the Universe is non-baryonic dark matter (DM). They are serendipitously motivated by various extensions of the standard model of particle physics such as Supersymmetry and Universal Extra Dimensions (UED). With its unprecedented sensitivity and its very large energy range (20 MeV to more than 300 GeV) the main instrument on board the GLAST satellite, the Large Area Telescope (LAT), will open a new window of discovery. As our estimates show, the LAT will be able to detect an indirect DM signature for a large class of WIMP models given a cuspy profile for the DM distribution. Using the current state of the art Monte Carlo and event reconstruction software developed within the LAT collaboration, we present preliminary sensitivity studies for several possible sources inside and outside the Galaxy. We also discuss the potential of the LAT to detect UED via the electron/positron channel. Diffuse background modeling and other background issues that will be important in setting limits or seeing a signal are presented. 
\end{abstract}

\maketitle

\section{Introduction}
The Gamma-ray Large Area Space Telescope (GLAST) 
\cite{Atwood:1993zn,Michelson:1999,Michelson:2007zz,Meegan:2007zz} is a satellite-borne \gray \ detector launched on 11 June 2008. The Large Area Telescope
(LAT) is the main instrument on GLAST, which will also host the Gamma-ray
Burst Monitor (GBM) \footnote{For a summary
of overall design properties, general performance, planned operation
modes, data analysis tools and strategies of the LAT, 
as well as a full list of 
participating scientists, see the web site of the collaboration 
\protect\cite{glast_website}.}. In this paper, we investigate the potential of the LAT for confirming, or constraining, the most interesting models of 
particle dark matter  of the Universe, WIMP (weakly interacting massive particle) models. The model of the lightest supersymmetric particle is the most studied template, which we use for most of our analysis. For general reviews on supersymmetric and other models of particle candidates for dark matter, see \cite{reviews}. We also study some aspects 
of a completely different class of models, so-called Kaluza-Klein models
of Universal Extra Dimensions (see \cite{KK} and references therein).  
Although estimates of \gray \ signals exist in the literature
(see \cite{reviews}), the new feature in the present paper is that
 the theoretical predictions are fed through the experimental
response function of the LAT and the software analysis chain that will be
used for the actual data analysis after launch. Thus we give  good
estimates for the potential of GLAST for detecting or limiting 
dark matter models. We also discuss the most important astrophysical and instrumental backgrounds.\\

\noindent
The LAT will have more than an order of magnitude better sensitivity in the 20 MeV to 10 GeV region than
its predecessor, EGRET onboard the Compton Gamma-ray Observatory \cite{EGRET}, and furthermore will extend the high energy region to roughly 300 GeV. Therefore, the LAT emerges as an instrument that is well suited to search for signals from dark matter  annihilation, which in the case of WIMPs should be populate just this  energy range. As the Large Hadron Collider (LHC) at CERN will also start taking data by the end of 2008, there is a non negligible probability to detect a good WIMP candidate, and verify through the \gray \ signal that such particles constitute the dark matter halo of the Milky Way (or neighboring galaxies or sub-halos, which we also treat). For a thorough discussion
of the interplay between discovery at LHC and detection of dark matter 
through other methods, in particular \gray \ observations, see \cite{baltz}.\\

\noindent
It is now established beyond reasonable doubt from a combined study of the 
cosmic microwave background radiation \cite{Spergel:2006hy}, supernova
cosmology \cite{sn} and large Galaxy redshift surveys 
\cite{Tegmark:2006az,Sanchez:2005pi}
that non-baryonic dark matter
is needed, while other models such as modifications of the laws of gravity
have problems explaining the wealth of observations that now are in place.
For example, it seems that the combined X-ray and optical observations of 
the ``Bullet Cluster'' essentially exclude explanations not involving
dark matter \cite{Bradac:2006er}.\\

\noindent
Technically denoted CDM, Cold Dark Matter, the particles constituting the 
cosmologically required dark matter have to be moving non-relativistically 
at the epoch of structure formation to reproduce the observed structure of the Universe, especially at small scales. This property is always fulfilled by particles with masses in the GeV to TeV range that interact with the weak interaction strength, i.e. WIMPs. 
They will have
velocities which have redshifted since the time of thermal decoupling 
in the early Universe, and they will now move with typical Galactic
velocities $v/c\sim 10^{-3}$ in the Milky Way halo. 
This is in contrast with massive neutrinos, Hot Dark Matter, which
give an observationally disfavored top-down structure formation scenario
with relatively little structure on small scales. In fact, the agreement
of the cosmological power spectrum with that of CDM allows us to put stringent
bounds on the neutrino mass (see, e.g, \cite{Hannestad:2003xv}). 
On the other hand, the 
non-zero neutrino masses inferred from measured neutrino oscillations, although
not enough to explain more than a few percent of the dark matter, constitute 
a first demonstration that non-baryonic dark matter indeed is likely to exist.\\

\noindent
We thus have an excellent class of particle candidates, WIMPs,
which behaves as the cosmologically needed CDM, and which could have been
thermally produced in the early Universe to give the required relic density \cite{Spergel:2006hy} $\Omega_{CDM}h^2\sim 0.1$, where $h$ is the Hubble constant in units of $100$ kms$^{-1}$Mpc$^{-1}$. The ability of WIMPs to give the measured relic density from readily computed  thermal processes in the early Universe without much fine tuning is sometimes termed  the ``WIMP miracle''.\\

\noindent
Of course it has to be kept in mind that dark matter does not
necessarily have to be WIMPs in the mass range detectable by GLAST. 
The Warm Dark Matter model with thermal-relic particle masses above 2
keV may also explain the present observations in both sky surveys and
N-body simulations. In the case when, say, a gravitino is the LSP, no
signal would be observable with GLAST. Here we work, however, with the
assumption that the dark matter particle has detectable \gray \
couplings and present the discovery potential for GLAST.\\

\noindent
In Figure 1, we sketch how \gray s are produced
from DM annihilation, and also show a schematic of a typical simulation and 
analysis chain as used in this paper. Table \ref{tab:searches} shows the various approaches to the search for WIMP dark matter signals in \gray\ data explored in this paper. A ``smoking gun'' would be the detection of line emission in WIMP annihilation \cite{Bergstrom:1997fj}, through
the loop-induced annihilation into two photons, which for slow-moving 
dark matter particles would give rise to a striking, almost monoenergetic, 
photon signal.  However, the branching ratio for the annihilation into lines is typically about $10^{-3}$ or less 
in most models, as WIMPs turn out  to be electrically neutral, and thus do not 
couple directly to photons. There are, however, some exceptions to this estimate \cite{Gustafsson:2007pc}. 
In addition to considering signatures in \gray s we also illustrate the LAT capability to detect electron/positrons, which could provide signatures of Kaluza-Klein particle Dark Matter.\\

\begin{table}[t]
  \centering
  \caption{ \it The various venues GLAST will explore in its search for WIMPs, and the advantage/disadvantage of each method.}
  \vskip 0.1 in
  \begin{tabular}{|l|l|l|} \hline
   Search  &  Advantages & Challenges \\
           &             &            \\
    \hline
    \hline
    Galactic     & Good            & Source   confusions\\ 
    center      & statistics       & Uncertainty in  \\ 
                &                  & diffuse background prediction \\ \hline 
    Satellites & Low background,            & Astrophysical           \\
               & good source identification & Uncertainties  \\   \hline
    Galactic   & Very good           & Uncertainties            \\ 
    halo       & statistics      & in Galactic diffuse      \\ 
                &                 & background prediction    \\ \hline
    Extra      & very good          & Uncertainties in Galactic \\ 
    galactic   & statistics      & diffuse contribution  \\
               &                 & Astrophysical        \\ 
               &                 & uncertainties        \\ \hline
    Spectral   & No astrophysical & Potentially low         \\
    lines      & uncertainties    & statistics  \\    
               & ``Smoking gun'' signal  &       \\   \hline
  \end{tabular}                  
  \label{tab:searches}
\end{table}

\begin{figure}
\begin{center}
\includegraphics[height=8.5cm,width=7.5cm]{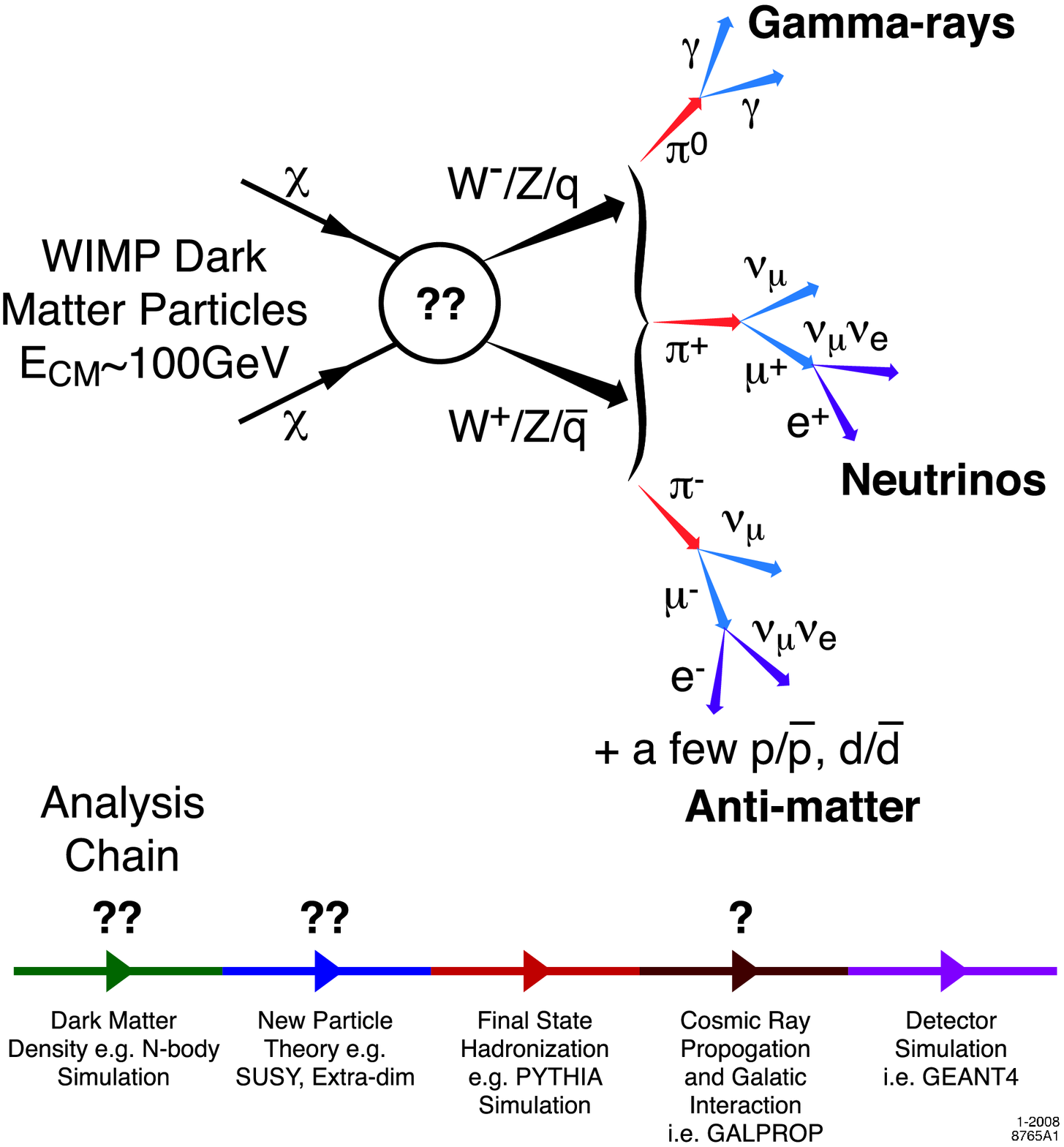}
\includegraphics[height=5.5cm,width=7.0cm]{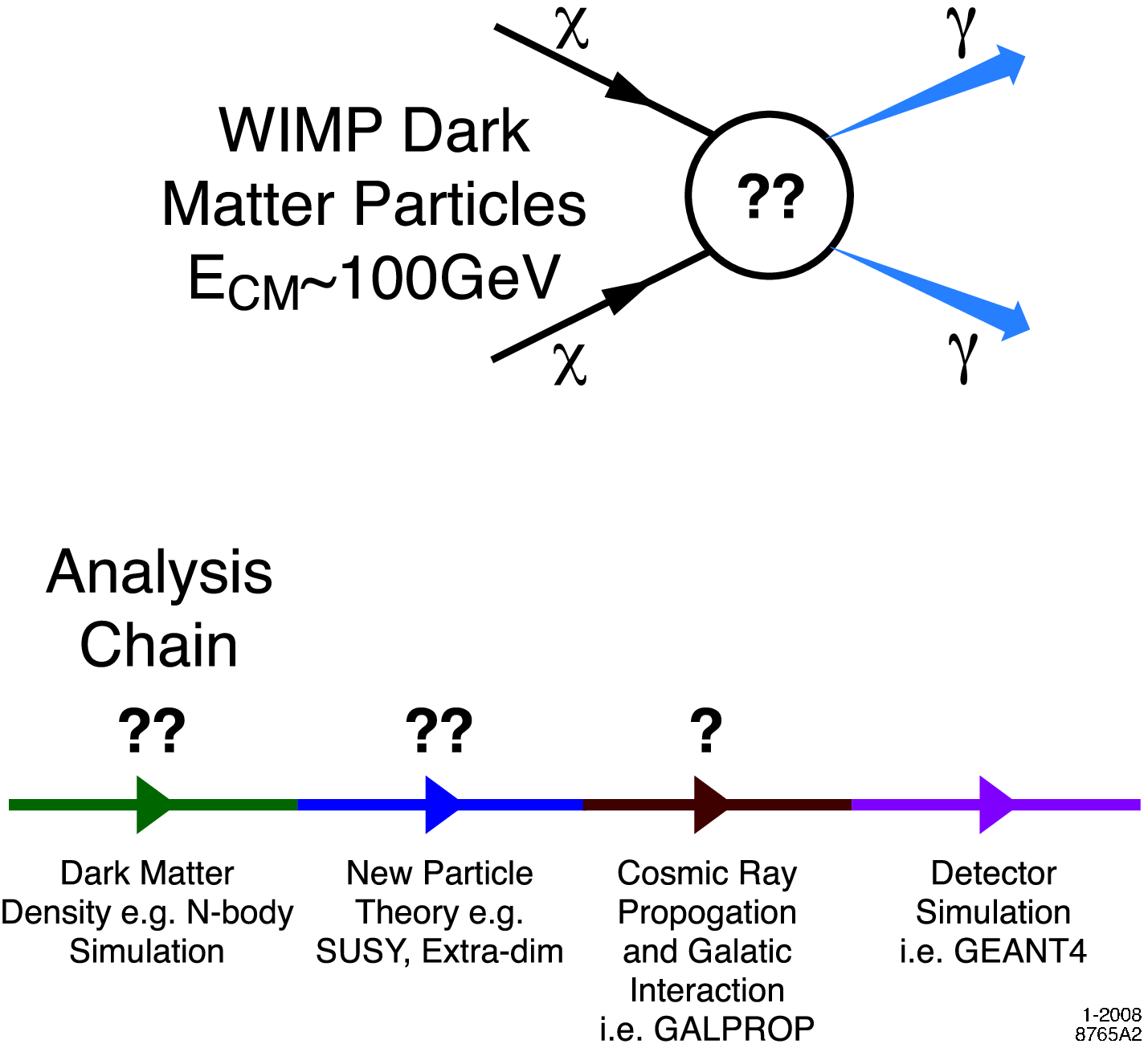}
\caption{A diagrammatic flow of how gamma rays are produced by annihilation of dark matter and elements of the
analysis chain used by the GLAST collaboration to detect them. The double question mark in the simulation chain indicates high uncertainty in the models of dark matter density and the new particle theories discussed in the
paper. The single question mark over the cosmic ray propagation and interaction models indicates lesser, although significant, uncertainty in those models that generate backgrounds to the potential dark matter gamma
ray signal. In this paper {\sffamily GALPROP} ({\sref{sec:bckgd}})  is used to estimate those backgrounds. In the next step, \gray \ detection is simulated using standard detector simulation packages (GEANT 4). Finally,these simulated LAT events are treated by various analysis software programs (event reconstruction and statistical analysis) to generate the results presented in this work. The same procedure is applied to the smoking gun signal of $\chi \chi \rightarrow \gamma \gamma$, except that in this case hadronization does not have to be taken into account.}
\label{schematics}
\end{center}
\end{figure}

\noindent
The  paper is organized as follows: In \sref{sec:LAT} we give a description of the LAT instrument and the software used for the analyses presented in this paper. In \sref{sec:calculation} the calculations of the WIMP signal flux are described and a discussion of the considered \gray \ background is given. In \sref{sec:sens} we summarize the sensitivities to generic WIMP annihilation signals achievable by the LAT for the search channels currently pursued by the LAT collaboration. Specific models are studied in \sref{sec:specific}.\\

\section{The LAT}
\label{sec:LAT}
The main instrument on board GLAST is the LAT, which is complemented by the GBM. The LAT is modular, consisting of a 4 $\times$ 4 array of identical towers. Each 40 $\times$ 40 cm$^2$ tower is composed of a tracker, a calorimeter, and a data acquisition module. The tracker array is covered by a segmented anti-coincidence shield (ACD). The tracking detector consists of 18 $xy$ layers of silicon strip detectors interleaved with 16 layers of tungsten foils.

Each calorimeter module has 96 CsI(Tl) crystals, arranged in an 8 layer hodoscopic configuration with a total depth of 8.6 radiation lengths, giving both longitudinal and transverse information about the energy deposition pattern \footnote{With the tracker the LAT presents 10 radiation lengths for normal incidence.}. 
The calorimeter's depth and segmentation enable the high-energy reach of the LAT and contribute significantly to background rejection. 
The ACD is the LAT's first line of defense against the charged cosmic ray background. It consists of 89 different size plastic scintillator tiles and 9 ribbons with wave-length shifting fiber readout. The segmentation is necessary to suppress self-veto effects caused by secondary particles emanating from the calorimeter showers of high energy \gray s \cite{Moiseev:2007hk}.

\subsection{LAT Exposure}\label{sec:exposure}
\begin{figure}
\begin{center}
\includegraphics[height=8cm,width=12cm]{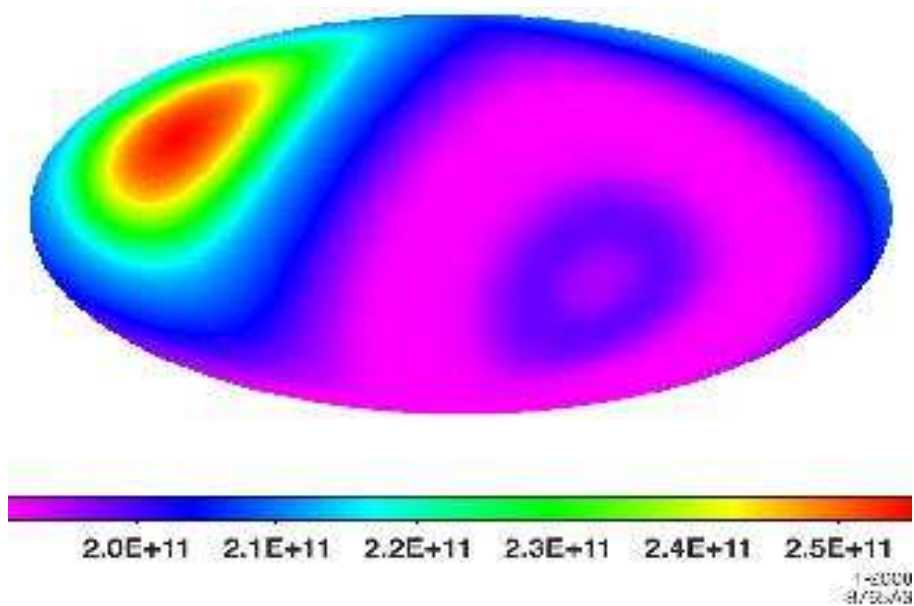}
\caption{The simulated LAT exposure {\bf for 5 years of all-sky scan}. The effect of turning off the LAT while in the SAA is included in the exposure. This exposure is calculated for a photon energy of 100 GeV. The plot is in Galactic coordinates with the values of exposure shown on the grey (coloured in coloured versions of the paper) bar in units of cm$^2\,$sec.}
\label{fig:exposure}
\end{center}
\end{figure}

For this paper, simulations of LAT all-sky ``exposures'' of  2 months, 1 year, 5 years and 10 years are used in the analyses. LAT exposure is defined as the amount of  cm$^2$ s the LAT effective area integrates over many orbits, which is a complex calculation. The GLAST orbit will be oriented at 25.4 deg. to the Earth's equator at an altitude of 565 km. Except where otherwise noted, all the analyses in this paper use codes that calculate the exposure from all-sky scanning consisting of alternating orbits with the LAT rocked by 35 deg toward the north celestial pole and then the south pole relative to local zenith. The GLAST orbit has about a 52.5-day precession period over which this all-sky scan achieves better than the required $\pm$ 20\% uniformity of exposure on the sky (not uniformity of sensitivity due to backgrounds that vary over the sky). Thus all of the analyses presented in this paper have the required uniformity of exposure since they all integrate longer than 52.5 days, but it should be noted that the only analyses making use of this fact are the line search (section \ref{sec:line}), the search for Galactic satellites (section \ref{sec:satellites}) and the sensitivity to mSUGRA models (section \ref{sec:specific}). The other analyses correct for the effects of exposure. Figure \ref{fig:exposure} shows the simulated LAT exposure in Galactic coordinates, for 5 years of all-sky scan, in units of cm$^2$s. The effect of turning off the LAT while in the South Atlantic Anomaly (SAA) is included in the exposure. This exposure is calculated for a photon energy of 100 GeV and is insensitive to the energy over the energy range of interest in this paper. The simulated exposure is uniform over the sky to $\pm$ 15 \%.

\subsection{LAT Analysis Software}
\label{sec:stools}
    The simulations and analyses presented in this paper make use of the
    LAT Monte Carlo simulation, GLEAM, and the ScienceTools software package.  GLEAM is a C++ package based on the GEANT 4 toolkit. GLEAM provides a detailed simulation of the physics processes and digitization within the LAT using a highly detailed geometrical description of the detector. It has been validated in extensive beam tests.
   The ScienceTools \cite{st} are being developed jointly by the
    LAT collaboration and the GLAST Science Support Center. The
    ScienceTools package comprises astronomical analysis software at a
    level similar to that available for other high energy satellite
    missions, such as Chandra or XMM, and these tools will be distributed
    in a similar manner for guest observers to perform analyses of LAT
    data.  The two main programs used in the current paper are the
    observation simulation tool {\tt gtobssim} and the source analysis
    tool {\tt gtlike}.
    The observation simulation tool allows users to define celestial
    \gray \ sources with any spectral properties.  These sources
    can be point-like or spatially extended.  For the latter, the angular
    distribution of the emission is given using FITS image files.  Given
    incident photons generated according to the input source models, {\tt
    gtobssim} uses the LAT instrument response functions (IRFs) to select
    photons as detected in the LAT and to assign {\em measured} energies
    and directions.  The IRFs, i.e., the effective area, energy
    dispersion, and point spread function, are ascertained from the
    detailed instrument Monte Carlo simulations. The source analysis tool, 
    {\tt gtlike}, uses a maximum likelihood method to fit source spectral parameters such as flux and power-law spectral index,
    though more complex spectral models are available.  Since the detected
    counts for sources near the detection limit will be fairly low, {\tt
    gtlike} calculates a likelihood function based on the Poisson probability 
    using the source model folded through the LAT IRFs  \cite{perf} 
 to provide the expected model count. 

    The process of the optimization of the event selection for high level analysis is ongoing.
    In the analyses presented in this
    paper, the event data are binned in direction and energy and {\tt
    gtlike} optimizes the binned log-likelihood.  A number of ancillary
    ScienceTools are also used that create the counts and exposure maps
    for the {\tt gtlike} program.

\section{Estimating the expected signal and background}
\label{sec:calculation}
In this section we give a general description of how we estimate the sensitivity of the LAT to \grays\ originating from WIMP annihilation. We describe the general way of estimating the signal flux and the backgrounds we consider in most of our searches. It should be noted, that in all our analyses we assume that the WIMP annihilation spectrum is the same no matter where observed (except for the redshift effect for the cosmological WIMP annihilation case), i.e. we assume the astrophysical environment does not affect the spectrum.\\

\noindent
After calculating WIMP annihilation and background fluxes, we use the fast GLAST detector simulation as described in the previous section for simulating LAT data. For the estimate of the sensitivity to specific particle physics models (see  \sref{sec:specific}), we assume a constant exposure to simplify computations. Finally, the sensitivity is calculated employing variants of the above mentioned maximum likelihood method unless otherwise stated.\\

\noindent
To begin with, we study the LAT sensitivity to generic WIMP annihilation models, i.e. we simply consider the annihilation of a  WIMP without any specific predictions on branching ratios and annihilation cross section, which could be provided by some underlying theory. This choice is motivated by lack of agreement between numerical computations. Various studies (see \cite{Allanach:2003jw,Belanger:2005jk} and  references therein) show, that in the framework of the widely popular and most restrictive supersymmetric extension of the Standard Model, the minimal supergravity framework (mSUGRA), there are important differences in the outputs of different spectrum calculators and  Renormalization Group Evolution codes. This is particularly true at large values of $tan(\beta)$, the ratio of vacuum expectation values of the two neutral component of the $SU(2)$ Higgs doublet (see  \sref{sec:specific}), or for large values of $m_0$, the common scalar masses. The practical consequence of these differences is uncertainty in the predictions of the various experimental observables like relic density or annihilation cross-sections.\\

\noindent
Nevertheless, especially in the context of comparison with accelerator experiments, it is interesting to consider specific particle physics models. Two examples are studied in  \sref{sec:specific}

\subsection{Calculation of the signal \gray\ flux}
\label{sec:calcflux}
The $\gamma$-ray continuum flux from WIMP annihilation at a given photon energy $E$ from a direction that forms an angle $\psi$ between the direction of  the Galactic center and that of observation is given by \cite{Gondolo:2004sc}
\begin{equation}
  \phi_{WIMP}(E,\psi)=\frac{1}{2}\frac{<\sigma v>}{4\pi} \sum_f \frac{dN_f}{dE} B_f
  \int_{\rm{l.o.s}} dl(\psi) \frac{\rho(l)^2}{m_{WIMP}^2} \ .
\label{eq:gammafluxcont}
\end{equation}
The particle physics model enters through the WIMP mass $m_{WIMP}$, the total mean annihilation cross-section $\sigma$ multiplied by the relative velocity of the particles (in the limit of $v \rightarrow 0 $), and the sum of all the photon yields
$dN_f/dE$ for each annihilation channel weighted by the corresponding branching ratio $B_f$. As pointed out in \cite{Cesarini:2003nr}, apart from the $\tau^+\tau^-$ channel, the photon yields are quite similar. \\  

\noindent
The integral in \Eref{eq:gammafluxcont} is the integral along the line of sight (l.o.s) of the assumed density squared, $\rho(l)^2$, of WIMPs. 
Among the kinematically allowed tree-level final states, the leading channels  are often $b\bar{b},t\bar{t},\tau^+\tau^-,W^+W^-,Z^0Z^0$. This is the case, e.g., for neutralinos and, more generically, for any Majorana fermion WIMP. For such particles the S-wave annihilation rate into the light fermion species is suppressed by the factor $m_{f}^2/m_{WIMP}^2$, where $m_{f}$ is the mass of the fermion in the final state. 
The fragmentation and/or the decay of the tree-level annihilation states gives rise to photons. The dominant intermediate step is the generation of neutral pions and their decay into $2\gamma$. 
The simulation of the photon yield is standard. We take advantage of a simulation performed with the Lund Monte Carlo program {\sffamily Pythia 6.202}~\cite{pythia} implemented in the {\sffamily DarkSUSY}\ package \cite{Gondolo:2004sc}.  
The density distribution in DM halos, from simulations, are well fitted by simple analytical forms, where the most common one is given by
\begin{equation}
\rho(r)=\frac{\rho_s}{\left(\frac{r}{r_s}\right)^\gamma\left(1+\left(\frac{r}{r_s}\right)^\alpha \right)^{(\beta-\gamma)/\alpha}}. \label{eq:profile} 
\end{equation}
This function behaves approximately as a broken power-law that scales as $r^{-\gamma}$ close to the center of the halo, $r^{-(\beta-\gamma)/\alpha}$ at an intermediate distance $r_s$, and $r^{-\beta}$ in the outskirts of the halo. In the remainder of the paper, we usually assume $(\alpha,\,\beta,\,\gamma, r_s) = (1,\,3,\,1,20\,\, \rm{kpc})$,  corresponding to the Navarro-Frenk-White (NFW) \cite{Navarro:1995iw} profile,  and $(\alpha,\,\beta,\,\gamma,r_s) = (1,\,3,\,1.5, 28\,\, \rm{kpc})$, corresponding to the Moore profile \cite{Moore:1999gc}. Unless otherwise stated,  we normalize $\rho_s$ to give 0.3 GeV cm$^{-3}$ at the Sun and our distance to the Galactic center is assumed to be 8 kpc \cite{Yao:2006px} . These kinds of profiles increase sharply towards the origin and are therefore called 'cuspy'.  Non-divergent profiles, referred to as cored, have also been proposed. We also give estimates for cored profiles, where applicable.

\noindent
For the case of the cosmological WIMP annihilation, the yield of \grays\ has to be redshifted and the line of sight integral has to be performed over halos of all redshifts, which implies modeling their structure as a function of redshift. For more details see  \sref{sec:cosmowimps}.

\subsection{Backgrounds}\label{sec:bckgd}

The sensitivities to a DM signal presented in this paper depend 
critically on accurate estimates of the following backgrounds:
diffuse Galactic \gray s, extragalactic diffuse \gray s, 
and charged particles in the instrument. In this section we discuss these different contributions and how we use them in the estimate of the sensitivity of GLAST to a DM signal. As we will see, the prediction of the conventional \gray \ diffuse emission is subject to considerable freedom. A quantitative treatment of the systematic uncertainties introduced in this prediction requires mapping a likelihood function in a multi-dimensional parameter space with non-linear dependences. We assume that the diffuse emission models that we use is the true background and the evaluation of the background systematics is beyond
the scope of this paper.\\ 

\noindent
We also do not take into account the contribution of unresolved point sources. The contribution of unresolved point sources is small compared with the true Galactic interstellar emission. The number and flux  distribution of EGRET point sources implies that the contribution of unresolved sources to the Galactic diffuse emission is about 10 \% or less.  The contribution will be evensmaller in a LAT measurement because fainter sources will be resolved. The contribution of point sources above the EGRET energy range is poorly constrained due to lack of observations, but is also likely to be small \cite{Strong:2006hf}\\

\noindent
For this reason (and considering the significant astrophysical uncertainties which enter the calculation of signal fluxes) we content ourselves with two illustrative examples, chosen to give a rough idea of what the systematic uncertainties might be: the ``conventional''  {\sffamily GALPROP} \cite{Strong:1998fr} and the ``optimized'' {\sffamily GALPROP} model \cite{Strong:2004de} , which are discussed in detail below.  For more details on the {\sffamily GALPROP} code, see \cite{Galprop}. For recent reviews on cosmic ray (CR) propagation and diffuse \gray\ emission, see \cite{Strong:2007nh,M04rev}.                                                                

\subsubsection{Diffuse Galactic gamma rays.}

The diffuse emission (\de) from the Milky Way dominates the \gray\ sky in the LAT energy range.
About 80\% of the high-energy luminosity of the Milky Way comes from
processes in the interstellar medium (ISM). 
The Galactic diffuse \gray\ emission is the product of CR particle 
interactions with gas in the ISM and the low energy photons in the 
interstellar radiation field
(ISRF). Therefore, its calculation
requires first the estimation of CR spectra throughout the
entire Galaxy \cite{Strong:2004de}. Major components of the \de\ are
$\pi^0$-decay, inverse Compton (IC), and bremsstrahlung.\\

\begin{figure}
\begin{center}
 \includegraphics[height=10cm,width=10cm]{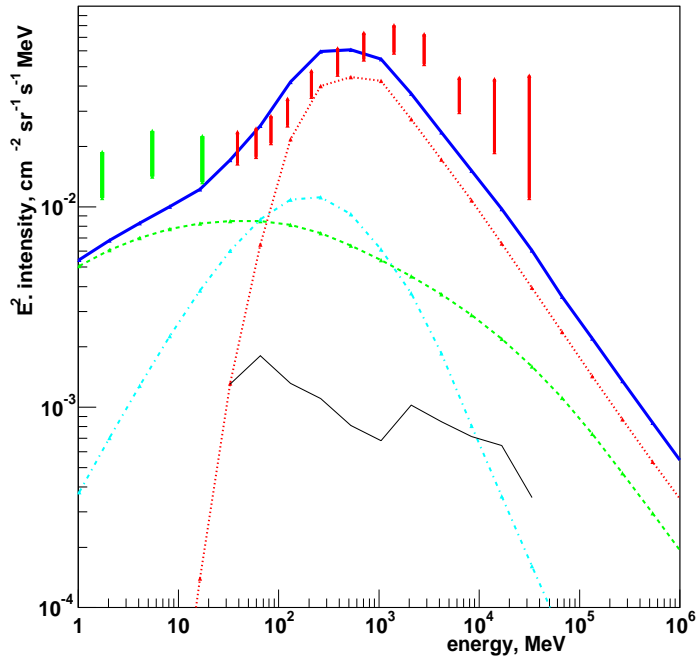}
 \caption{\gray \ spectrum of the inner Galaxy ($300^\circ < l < 30^\circ$,$|b| < 5$) derived from the ``conventional'' model (see text for more details). Dotted: contribution from $\pi^0$ decay, dashed: contribution from inverse Compton scattering, dash-dotted: contribution from bremsstrahlung, solid: extragalactic background, bold solid: total flux. Also shown are the data points from EGRET (dark bars, red in colored versions) and COMPTEL (light bars, green in colored versions). Figure taken from \cite{Strong:2007nh}.}
\label{fig:conventional}
\end{center}
\end{figure}

\noindent
The first detailed analysis of the diffuse emission from the Galactic plane $|b|\leq 10^\circ$ was made by Hunter \etal \cite{Hunter:1997we}. The spectrum of
\grays \ calculated under the assumption that the proton and electron
spectra in the Galaxy resemble those measured locally reveals an excess at
$>$ 1 GeV in the EGRET spectrum, the so-called "GeV excess".
An extensive study of the Galactic diffuse \gray \ emission in the context of cosmic ray propagation models has been carried out by Strong \etal \cite{Strong:2004de,Strong:1998fr}. This study confirmed that models based on locally measured electron and nucleon spectra and synchrotron constraints are consistent with \gray \  measurements in the 30 MeV - 500 MeV range, but outside this range excesses are apparent. In this paper the {\sffamily GALPROP} ``conventional'' model will be used as a representative of this class of models. The resulting \gray \ spectrum from this model is shown in \fref{fig:conventional}.\\

\noindent
The "GeV excess" is seen in all directions, not only in the Galactic 
plane. A simple re-scaling of the components ($\pi^0$-decay, IC) does
not improve the fit in any region, since the observed peak is at an energy higher than the $\pi^0$-peak. Although a possibility that the excess is an instrumental artifact due to the uncertainty in calibration can not be completely excluded \cite{Moskalenko:2006zy,Stecker:2007xp}, the detailed simulations by Baughman \etal \cite{Baughman:2007ck} indicate that the GeV excess is significantly larger when instrumental effects previously unaccounted for are considered.\\

\begin{figure}
\begin{center}
 \includegraphics[height=10cm,width=10cm]{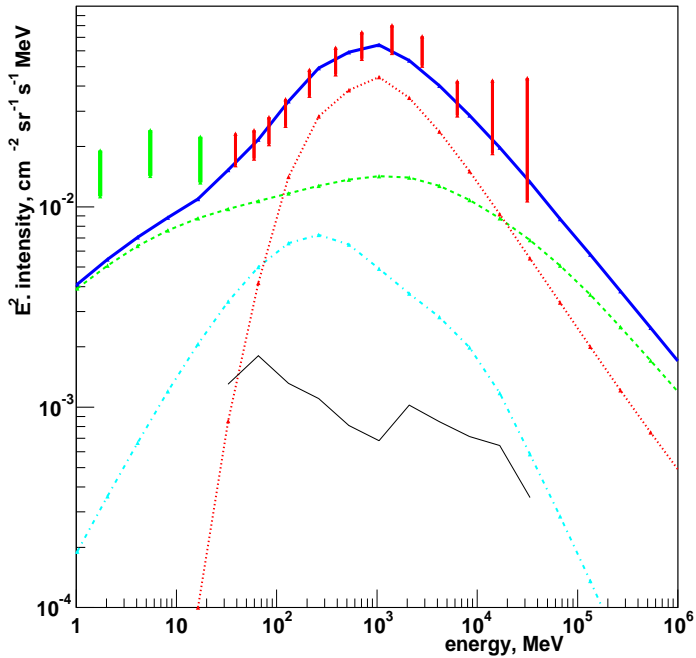}
 \caption{\gray \ spectrum of the inner Galaxy ($300^\circ < l < 30^\circ$,$|b| < 5$) derived from the ``optimized'' model (see text for more details). Dotted: contribution from $\pi^0$ decay, dashed: contribution from inverse Compton scattering, dash-dotted: contribution from bremsstrahlung, solid: extragalactic background, bold solid: total flux. Also shown are the data points from EGRET (dark bars, red in colored versions) and COMPTEL (light bars, green in colored versions). Figure taken from \cite{Strong:2007nh}.}
\label{fig:optimized}
\end{center}
\end{figure}

Alternatively, the assumption that locally measured electron and proton spectra
represent the Galactic average can be dropped. In that case, as we will discuss in detail in the next paragraph, the "GeV excess" can be reproduced. The resulting model, dubbed ``optimized {\sffamily GALPROP}'' model, is used as another benchmark model to represent possible diffuse \gray \ background. The resulting \gray \ spectrum from the ``optimized'' model is shown in \fref{fig:optimized}.

\paragraph{The ``optimized'' model.}\label{optimized}
There are a number of reasons why the CR intensity may fluctuate in
space and time. First are the stochastic spatial and temporal distributions of supernova (SN) events.
Dramatic increases in CR intensity, perhaps connected
with nearby SN explosions, are recorded in terrestrial concentrations
of  cosmogenic isotopes. Concentrations of $^{10}$Be in Antarctic and
Greenland ice core samples
\cite{yiou97} and $^{60}$Fe in a deep-sea
ferromanganese crust \cite{knie04} indicate highly significant
increases of CR intensity $\sim$40 kyr and 2.8 Myr ago. 
The SN rate is larger in the spiral arms
\cite{case96}. This may lead to lower CR intensity in the interarm
region where the solar system is located. In the case of anisotropic
diffusion or convection by the Galactic wind such fluctuations may be
even stronger (see \cite{M04rev} for a discussion).
The intensity variations and spectra of CR protons and  heavier CR nuclei may be uncorrelated \cite{Moskalenko:2005xu}.  The total inelastic cross
section for protons is $\sim$30 mb vs.\ $\sim$300 mb for Carbon,
so that their Galactic ``collecting areas'' 
differ by a factor of 10 or more for heavier nuclei. This means that carbon comes from Galactic regions that are about 10 times closer 
than the sources of protons, and thus implies that the directly measured CR protons and CR nuclei may come from different sources.\\

\noindent
Additionally, CR electrons and positrons suffer relatively large energy losses
\cite{Strong:1998pw} as they propagate away from their sources and thus their spectral and intensity fluctuations can be considerably larger than those of protons.\\ 

\noindent
Antiprotons in CR are presumably secondary and produced mostly by CR
protons in interactions with the interstellar gas.  Because of their
secondary origin in the ISM, which is far more uniform in space and time than the primary CR sources, 
their intensity fluctuates less than that of
protons.  The total cross-section of antiprotons is about the same as of 
protons, except at low energies due to annihilation, and thus they
trace the CR proton spectrum \emph{averaged} over a large region of the 
Galaxy.
Therefore, if the directly measured
local CR spectrum is not representative of the \emph{local} Galactic
average, then the \emph{antiproton measurements} can still be used instead
to derive this average intensity of CR \emph{protons}.\\

\noindent
In addition to spatial diffusion in the interstellar space,
the scattering of CR particles
on randomly moving MHD waves leads to stochastic acceleration
(second order Fermi acceleration or "reacceleration"),
which is described in the transport equation as diffusion in
momentum space. It has been shown \cite{SeoPtuskin1994} that the distributed acceleration may be strong enough to explain the peaks in the secondary-to-primary nuclei ratios
at approximately 1 GeV/nucleon. When normalized to the local CR proton spectrum, the reacceleration
model underproduces $\bar p$'s (BESS data \cite{Orito:1999re})  at $\sim$2 GeV and diffuse \grays\ above 1 GeV 
by the \emph{same} factor of $\sim$2 \cite{Strong:1998fr,Moskalenko:2001ya} 
while it works well for other CR nuclei. It is thus enough to renormalize\footnote{The BESS-Polar flight of 2004 \cite{Hams:2007} revealed that the CR antiproton flux is somewhat lower than previous measurements with lower
statistics. This reduces allowable variations in the CR nucleon
spectrum due to the uncertainties in the antiproton flux measurements.} the CR  proton spectrum up by
a factor of 1.8 to remove the excesses. The model then predicts  a
factor of 2 too many photons at $\sim$100 MeV. The 100 MeV photons are
produced mostly by $\sim$1 GeV protons, where many uncertainties
simultaneously come into play: poor knowledge of the
$\pi^0$-production cross-section at \emph{low} energies and/or
low-energy interstellar proton spectrum and/or solar modulation (see
\cite{Moskalenko:2001ya,Moskalenko:2002yx,Moskalenko:2003kq} for further discussion).  To get agreement with the EGRET photon
data effectively requires a corresponding adjustment of the
spectrum of CR protons at low energies. Since the IC and $\pi^0$-decay
photons have different distributions, the electron spectrum also needs
to be renormalized up by a factor of 4 in order to reproduce the EGRET diffuse \gray\  flux
itself. These adjustments in the injection spectra of protons and
electrons are enough to reproduce the spectrum of the EGRET diffuse
\grays\ in \emph{all directions} as well as the
latitude and longitude profiles for the whole EGRET energy range 30
MeV -- 50 GeV.

\subsubsection{Extragalactic diffuse \grays.}\label{sec:EGBR}

The extragalactic $\gamma$-ray background (EGRB) is a weak component
that is difficult to disentangle from the intense Galactic foreground.
Its spectrum depends on the model adopted for the Galactic diffuse
emission, which is itself not firmly determined.  The isotropic,
presumably extragalactic, component of the diffuse \gray \ flux was
first discovered by the SAS-2 satellite \cite{Thompson92} and
confirmed by EGRET \cite{Sreekumar:1997un}. Even at the Galactic
poles, the EGRB does not dominate, and its flux is comparable to the
Galactic contribution from IC scattering of the Galactic plane photons
(interstellar radiation field) and cosmic microwave background photons by CR electrons \cite{Strong:2004ry,Moskalenko:1998gw}.
Additionally, Compton-scattered solar
photons by CR electrons make broadly distributed emission
with maximum brightness in the direction of the Sun 
\cite{Moskalenko:2006ta,Orlando:2006zs,Orlando:2007qx}.
This all-sky average $\sim 1\cdot 10^{-6}$ photons
cm$^{-2}$ s$^{-1}$ sr$^{-1}$ above 100 MeV is about 10\% of the 
extragalactic emission (as inferred from EGRET data). Finally, the \gray \
albedo of small solar system bodies (asteroids) may contribute
at high Galactic latitudes as an additional component of the
\gray \ foreground distributed around the ecliptic \cite{Moskalenko:2007tk}.
 The determination of the EGRB is thus model-dependent an
influenced by the assumed size of the Galactic halo, the electron
spectrum there, and the spectrum of low-energy background photons,
each of which must be derived independently. 
It can also be impacted by instrumental backgrounds, which are at a comparable level, in case they are  not accurately determined and subtracted.\\

\noindent
More recently, Strong \etal \cite{Strong:2004ry} reanalyzed the EGRET data using the 
``optimized'' model discussed in \sref{optimized}.  
Since this ``optimized'' model is not exact, the same method  as in \cite{Sreekumar:1997un} 
was used. Reasonable agreement was obtained,  considering the EGRET systematic uncertainty of $\sim$15\%.
A larger deviation, was found only in one region, 2--4 GeV.
This new estimate of the EGRB spectrum
is lower and steeper than found by \cite{Sreekumar:1997un} in most energy 
ranges.
It is not consistent with a power-law and shows some positive curvature, 
as expected in blazar population studies and
various scenarios of cosmological neutralino annihilation
\cite{Stecker:1996ma,Elsaesser:2004ap}.
The integral flux above 100 MeV is $(1.11\pm0.01)\cdot10^{-5}$ photons
cm$^{-2}$ s$^{-1}$ sr$^{-1}$ \cite{Moskalenko:2003kq}. \\

\noindent
The extragalactic background models, that are used as benchmarks in the present work are those presented by \cite{Sreekumar:1997un} and \cite{Strong:2004ry}, which are consistent with EGRET data. The EBGR is most important in the analysis of cosmological WIMP annihilation (see section \ref{sec:cosmowimps}). For this analysis we also consider a model which takes into account GLAST's ability to resolve many of the sources that contributed to the EGRET EGRB measurement

\subsubsection{Particle backgrounds.}
Charged particles, mainly protons, electrons and positrons, as well as a smaller number of neutrons and Earth albedo photons, present a major instrumental background to potential DM  signals.  These background particles greatly dominate the flux of cosmic photons incident on the  LAT, but the background rejection capability of the instrument is such  that less than 1 in $10^5$ end up in the final sample of \gray \ candidates. The necessary cuts are included in the simulated instrument performance. \\

\noindent
We take into account the remaining charged particle background in the analyses where it can have a significant effect (i.e. for cosmological WIMP annihilation and the search for DM satellites).

\section{Sensitivity to generic WIMP annihilation} 
\label{sec:sens}

\subsection{The Galactic Center}
\label{sec:gc}
The Galactic center (GC) is expected to be the strongest source of \grays\ from DM annihilation, due to its coincidence with the cusped part of the DM halo density profile, see \Eref{eq:profile}. 
As a result, the GC is frequently proposed for DM searches \cite{Bergstrom:1997fj} and
literature devoted to possible signatures of a DM \gray\ signal at the GC is extensive.
The ``GeV excess'' discussed in the previous section has been interpreted as the possible signature of WIMP annihilation \cite{Cesarini:2003nr, deBoer:2004es}, although such an interpretation relies heavily on the underlying model for the Galactic diffuse background (see discussion in \sref{sec:bckgd}).\\ 

\noindent
The neighborhood of the GC harbors 
numerous objects capable of accelerating CR to very high energies and thus producing \grays\ by 
inverse Compton scattering on electrons or pion decays following proton-proton or proton-$\gamma$ 
interactions.  Most notably, the bright, very high energy \gray\ point source observed 
by the H.E.S.S. \cite{HessGC}, MAGIC \cite{Albert06}, VERITAS \cite{VeritasGC} and CANGAROO 
\cite{CangarooGC} collaborations represents a formidable background for DM studies
in this region of the sky. Initially proposed as the DM signature of a very heavy ($\gtrsim$ 6 TeV) WIMP  (\cite{Profumo:2005xd} and references therein), such a signal would rule out a Moore profile for neutralino masses above $\sim$400 GeV, for   $<\sigma v>\ \sim 10^{-24}\ \rm{cm}^3\, \rm{s}^{-1}$ or the expected cross-section for Kaluza-Klein particles  \cite{Bergstrom:2004cy}. However, this source is now widely considered to be a standard astrophysical source, either associated with the  $\sim 2.6\cdot 10^{6} M_\odot$ black hole at the kinematic centre of our Galaxy, commonly identified with the bright compact radio source Sgr A${}^*$, or with the candidate pulsar wind nebula G359.95-0.04 \cite{vanEldik:2007yi} recently discovered in a deep Chandra survey and only at $8.7''$ from Sgr A${}^*$. One should also take into account the presence of  a possible EGRET source at approximately $0.2^\circ$ from the GC \cite{Hooper:2002ru}.\\ 

\noindent
At this point, we do not know the behavior of these high energy sources
in the GLAST energy range, which makes extrapolation for the sake of
including the source flux as an additional background doubtful. The  
following analysis is based on the hypothesis that the main task will be to 
distinguish the DM signal from a Galactic
diffuse background after the astrophysical sources are disentangled and subtracted using the information provided by spectral and angular analysis and multiwavelength observations. This
subtraction will introduce a systematic uncertainty in the assessment of the diffuse flux 
which we neglect here. For a recent attempt to include HESS and EGRET sources into a sensitivity calculation, see \cite{Dodelson:2007gd}. \\

\begin{figure}
\begin{center}
 \includegraphics[height=10cm,width=16cm]{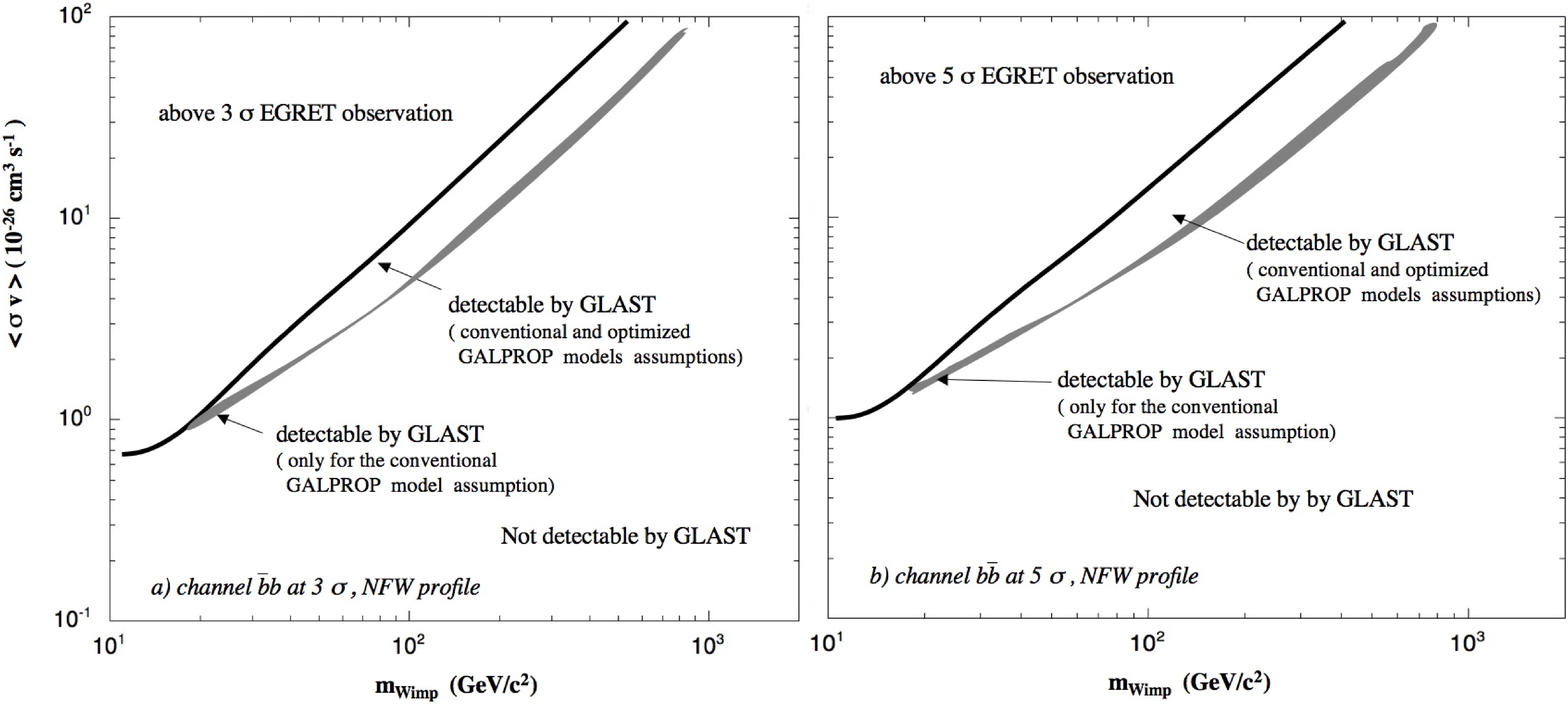}
 \caption{
Cross-sections $<\sigma v>$ ($v \rightarrow 0)$ versus the WIMP
mass $m_{WIMP}$ for the { \sl $b \bar{b}$ } annihilation channel.
Left panel shows the result for $3 \sigma$ significance, right panel shows the result for $5 \sigma$ significance for {\bf 5 years of GLAST operation,} The upper part of the plots corresponds to
regions which are already excluded by the EGRET data around the GC
and the lower part corresponds to regions not detectable by GLAST.
The "detectable by GLAST region"
corresponds to models detectable by GLAST for both ``conventional'' and
``optimized'' astrophysical background. The shaded region represents models which can be detected only under the assumption of ``conventional'' Galactic diffuse background. See text for more details.}
\label{bbar2}
\end{center}
\end{figure}

\begin{figure}
\begin{center}

\includegraphics[height=10cm,width=16cm]{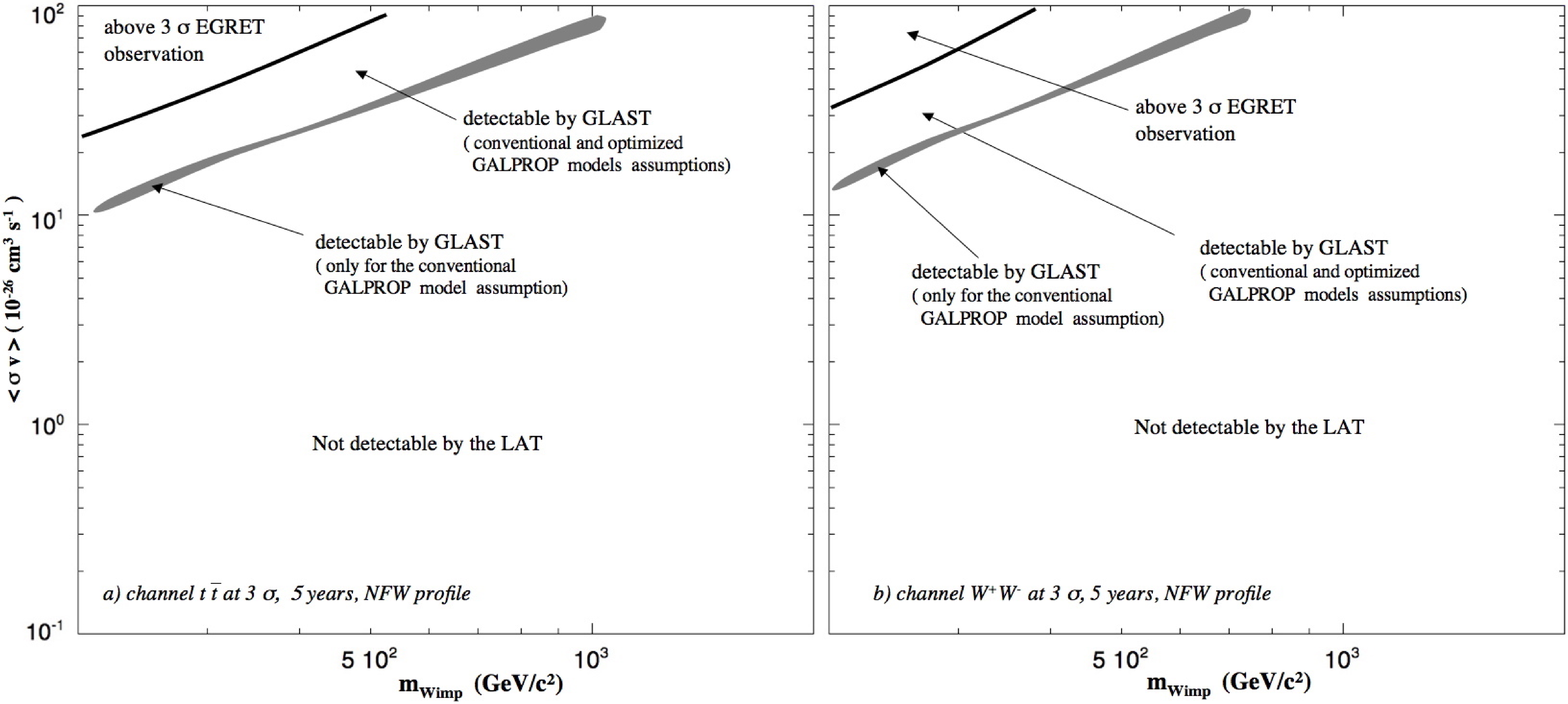}
\caption{3 $\sigma$ sensitivity regions for the $t\bar{t}$ (left panel) and the $ W^+ ~ W^- $annihilation channel (right panel). Definitions of regions are the same as in  \fref{bbar2}. Note the difference in $x$-axis scale as compared to \fref{bbar2}. }
\label{top}
\end{center}
\end{figure}

\begin{figure}
\begin{center}
\includegraphics[height=10cm,width=12cm]{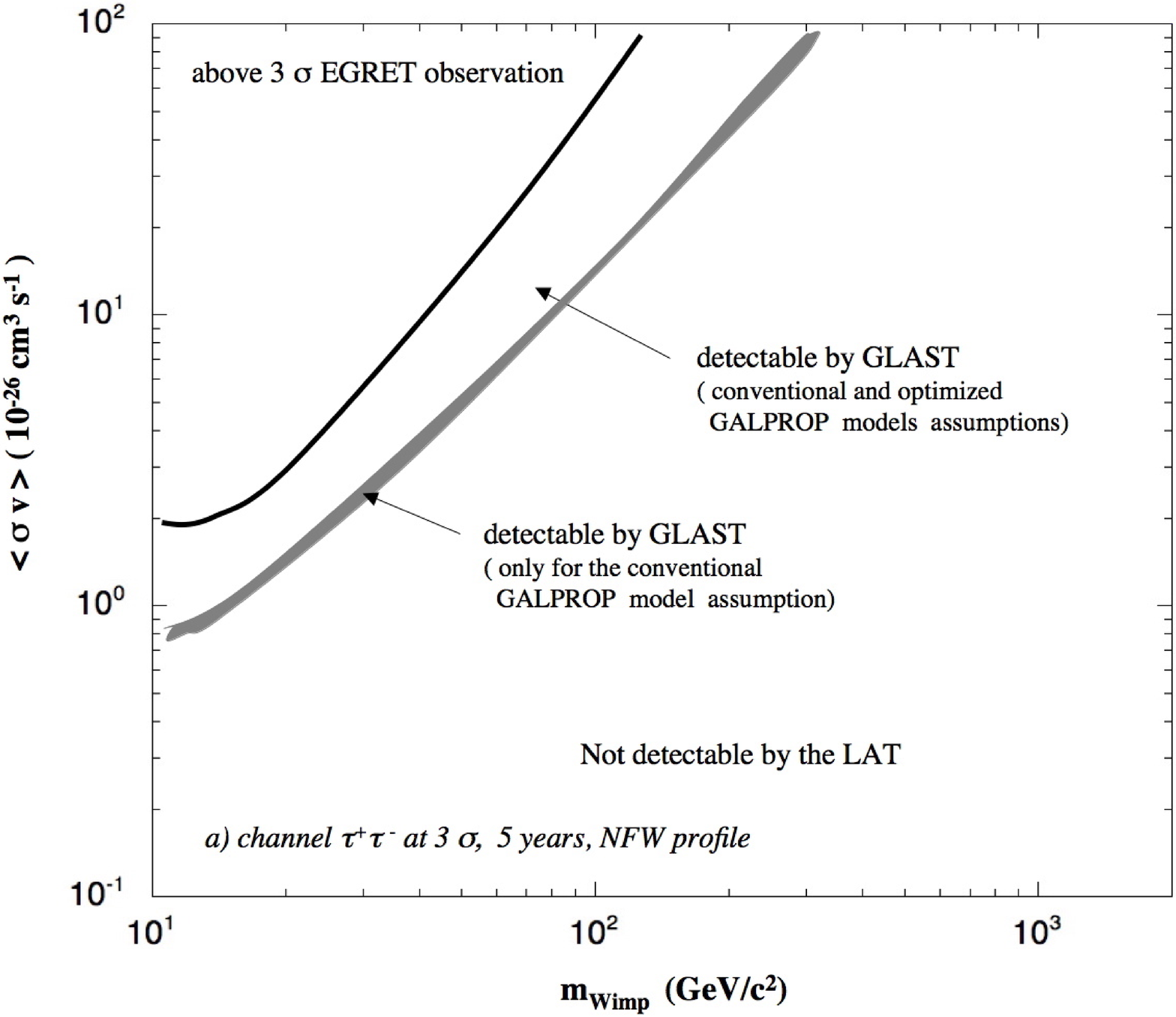}
\caption{3 $\sigma$ sensitivity regions for the $\tau^+ ~  \tau^-$ annihilation channel. Definitions of regions are the same as in  \fref{bbar2}.}
\label{tauW}
\end{center}
\end{figure}

\noindent
As described in  \sref{sec:calculation},
the expression of the $\gamma$-ray continuum flux for a generic WIMP at a given energy $E$ is given by  \Eref{eq:gammafluxcont}. A truncated NFW profile as defined in \cite{Gondolo:2004sc} is assumed for the WIMP distribution, and only one dominant annihilation channel ($W^+W^-$, $b\bar{b}$, $t\bar{t}$, $\tau^+\tau^-$) is considered at a time. Care has also been taken in order not to violate the EGRET flux constraint around the GC \cite{Mayer}. WIMP annihilation differential fluxes above 1 GeV and in a region of 0.5 degs radius around the GC (corresponding to the angle for $68\%$ containment at this energy threshold) have been generated using {\sffamily DarkSusy v. 4.15} \cite{DarkSusy}. The expected DM from this region, incident on the LAT for 5 years of all sky scanning operation is simulated using {\sffamily gtobssim} (see \sref{sec:stools}). This flux was simulated for 10000 grid points ($m_{{\rm WIMP}}$, $<\sigma v>$) on a  $100\times100$ logarithmic grid. The range of chosen is $<\sigma v>\in[10^{-28},10^{-24}]$ 
and $m_{\rm{WIMP}}\in[10,3000]$ GeV for the $b\bar{b}$ and $\tau^+\tau^-$ channels, and $m_{\rm{WIMP}}\in[200,3000]$  for the  $W^+W^-$ and $t \bar{t}$ channels.  The EGRET flux modeled in \cite{Mayer} is also simulated at the GC and
a standard $\chi^2$ statistical analysis is performed to check if a given WIMP model conflicts with EGRET data at the 5$\sigma$ level. For models compatible with EGRET data, a second $\chi^2$ test is performed to check if GLAST is able to disentangle the WIMP contribution from either the ``conventional'' or the ``optimized'' Galactic diffuse background. For the $\chi^2$ calculation, 20 logarithmic bins in the energy range [1 GeV, 100 GeV] are used for $b\bar{b}$ and $\tau^+\tau^-$, [1 GeV, 300 GeV] for $W^+W^-$ and $t \bar{t}$. Figures \ref{bbar2} to \ref{tauW} show the results of the scans for different annihilation channels. These results are given at a $3\sigma$ confidence level. We also show the  $5\sigma$ significance in \fref{bbar2} for the $b \bar{b}$ final state. As expected, sensitivity regions for the $b\bar{b}$, $t \bar{t}$ and $W^+W^-$ annihilation channels are quite similar, whereas differences are apparent for the $\tau^+\tau^-$ annihilation channel\footnote{For the other analyses presented in this paper, we usually assume $b\bar{b}$ as the dominant annihilation channel. Strictly, those regions might not be useful for models with dominant annihilation to $\tau\tau^{-}$.}. Furthermore, it can be seen that regions of $\mathcal{O}(10^{-26})$cm$^3$s$^{-1}$ are within the reach of GLAST.\\

\noindent
In the GC study presented here, a NFW profile is assumed. Recently, Hooper \etal \cite{Hooper:2007gi} have studied the observational consequences  for GLAST of an excess in the WMAP foreground coined the ``WMAP haze'' \cite{Finkbeiner:2003im}, when interpreted as the synchrotron emission from highly relativistic electron-positron pairs produced by WIMP annihilation. 
 The author's conclusions are encouraging. If a slightly steeper inner slope of the NFW profile is considered, $\rho(r) \propto r^{-1.2}$ rather than $r^{-1}$, as suggested in \cite{Hooper:2002ru}, the LAT sensitivity should be improved by a factor of $\sim 10$. Assuming a Moore profile, the line of sight integral increases by a factor of $~$200 (with corresponding improvement in sensitivity), under the assumption of cored profile the sensitivity decreases by a factor of $\sim$100.
Compared to this, a simplified introduction of a 20 \% uncertainty in the Galactic diffuse background (by adding a term to the $\chi^2$), only leads to a decrease in sensitivity by about 5 \%.  It is noteworthy, that observations of the GC at x-ray and radio wavelengths might lead to strong constraints on the DM signal \cite{Regis:2008ij}.  

\subsection{The Galactic Halo}
\label{sec:halo}
The detection of the continuum WIMP signal from the Galactic halo is complicated by difficult systematic effects due to the presence of comparably uncertain astrophysical backgrounds (see \sref{sec:bckgd}). However, analyzing the halo is complementary to analyses of the immediate region of the GC, where the DM rates are significantly greater in the standard halo models, but where there are also issues of source crowding and large backgrounds. A potential continuum WIMP signal has the advantage that its spectral shape would be the same no matter where observed in the Galaxy assuming effects of the astrophysical environment can be neglected.  In addition, in most scenarios one expects a sharp cutoff in this signal at the mass of the WIMP that is very difficult to achieve with expected astrophysical backgrounds.\\ 

\noindent
This section presents an analysis method for the WIMP continuum contribution to the Galactic diffuse emission in the halo. For this analysis we include the full Galactic Halo except for a region excluded because of large diffuse backgrounds and
systematic errors associated with this background.  We consider two possible  exclusions: the first is the region within 10$^\circ$ of the GC and the second is the region within 10$^\circ$ of the Galactic plane. Both of these choices will avoid the GC where the proposed dark matter density profiles diverge from one another.\\

\noindent
The statistical sensitivity to a WIMP signal is quantified as a function of the WIMP mass and other parameters such as the annihilation cross-section and DM density profile.

\begin{figure}
\begin{center}  
\includegraphics[height=.3\textheight]{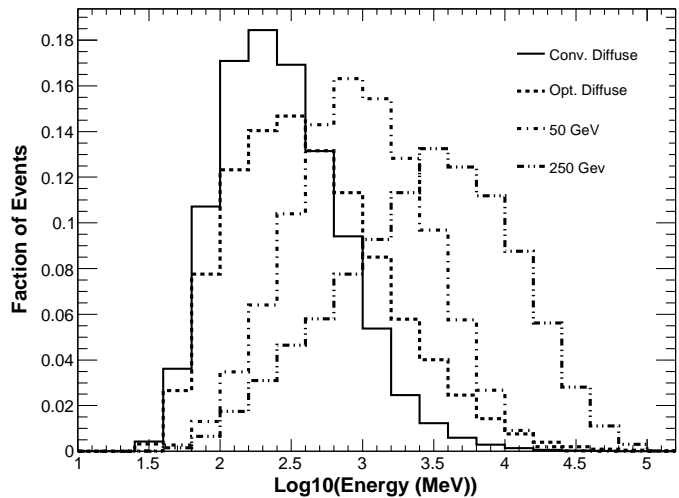}  
\caption{
The energy distributions for the diffuse ``conventional''  and ``optimized'' diffuse-emission models and two DM models for WIMP masses at 50 and 250 GeV/$c^2$. 
Each distribution is individually normalized to unit area to highlight the shape differences. 
}
\label{fig:counts}
\end{center}
\end{figure}

Since the diffuse \gray \ emission is the largest background for a halo DM search, two models (``optimized'' and ``conventional'') of the Galactic diffuse background intensity maps (see section 3.2) are used in the analysis. We generate DM models using a NFW density profile. With the intensity maps calculated from the diffuse model plus the DM model computer simulations of the detector are performed to obtain GLAST events with the expected distributions in space, time and energy. \Fref{fig:counts} shows the shapes of the energy distributions observed in the LAT  for various DM masses versus the two models of the diffuse background, ``conventional'' and ``optimized''.\\

\noindent
The Galactic diffuse background peaks at about 150 MeV, while the
DM distribution peaks at about 1 GeV for a WIMP mass of 50 GeV$/c^2$,
and increases with increasing WIMP mass. Background and signal spectra are thus distinguishable. In addition, the DM has a different spatial distribution. We can take advantage of these differences
by performing a  simultaneous fit over angles (i.e. Galactic latitude and longitude) and in energy. 
 We perform a log-likelihood fit of the dark matter and diffuse model to numerous pseudoexperiments generated 
 from the same models binned $3 \times 3 \times 30$ in $l$, $b$, and $\log E$). In the fit there are no constraints to the normalization of either signal or background.\\

\begin{figure}
\begin{center}  
\includegraphics[height=.25\textheight,width=7.5cm]{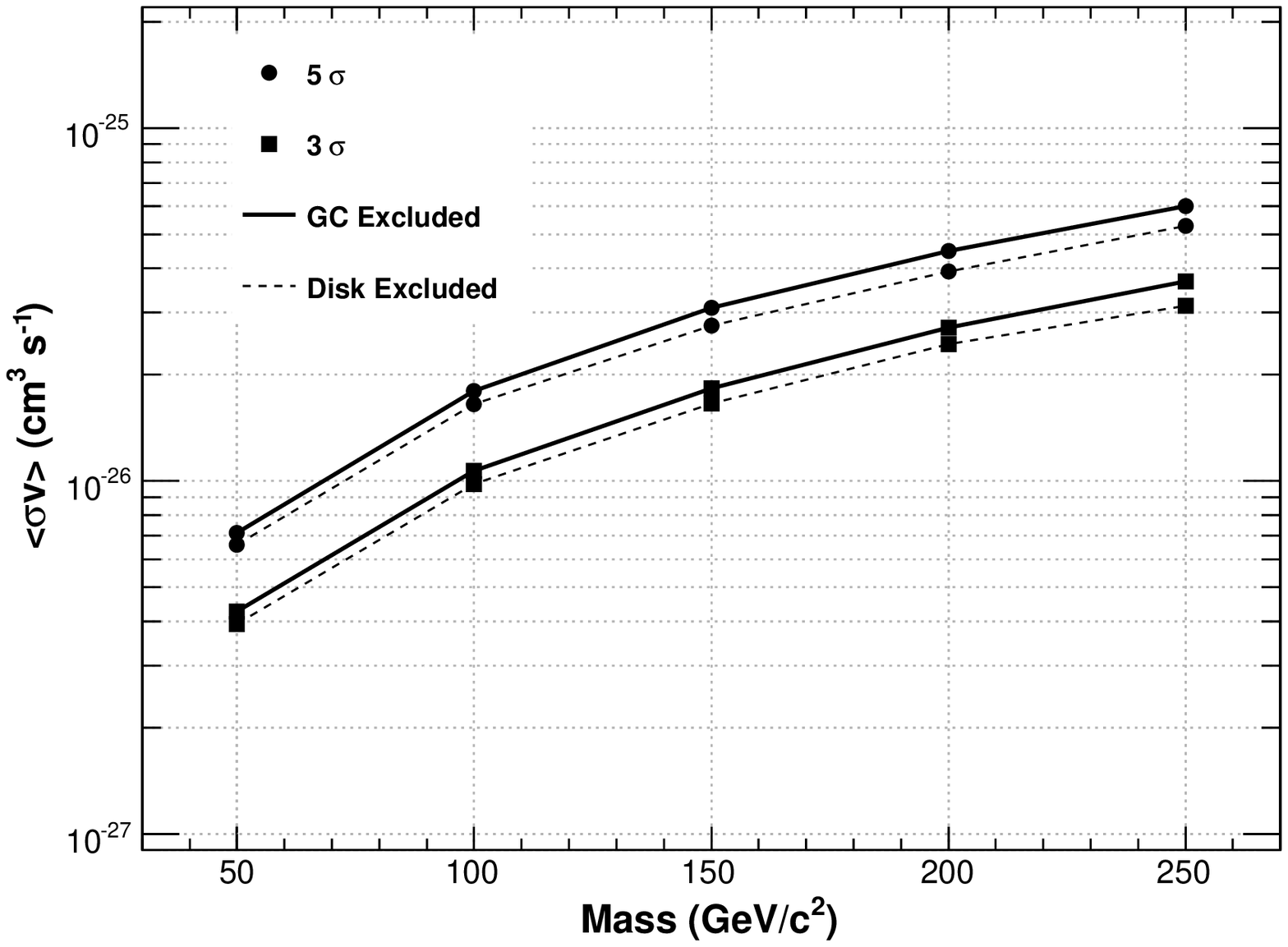}  
\includegraphics[height=.25\textheight,width=7.5cm]{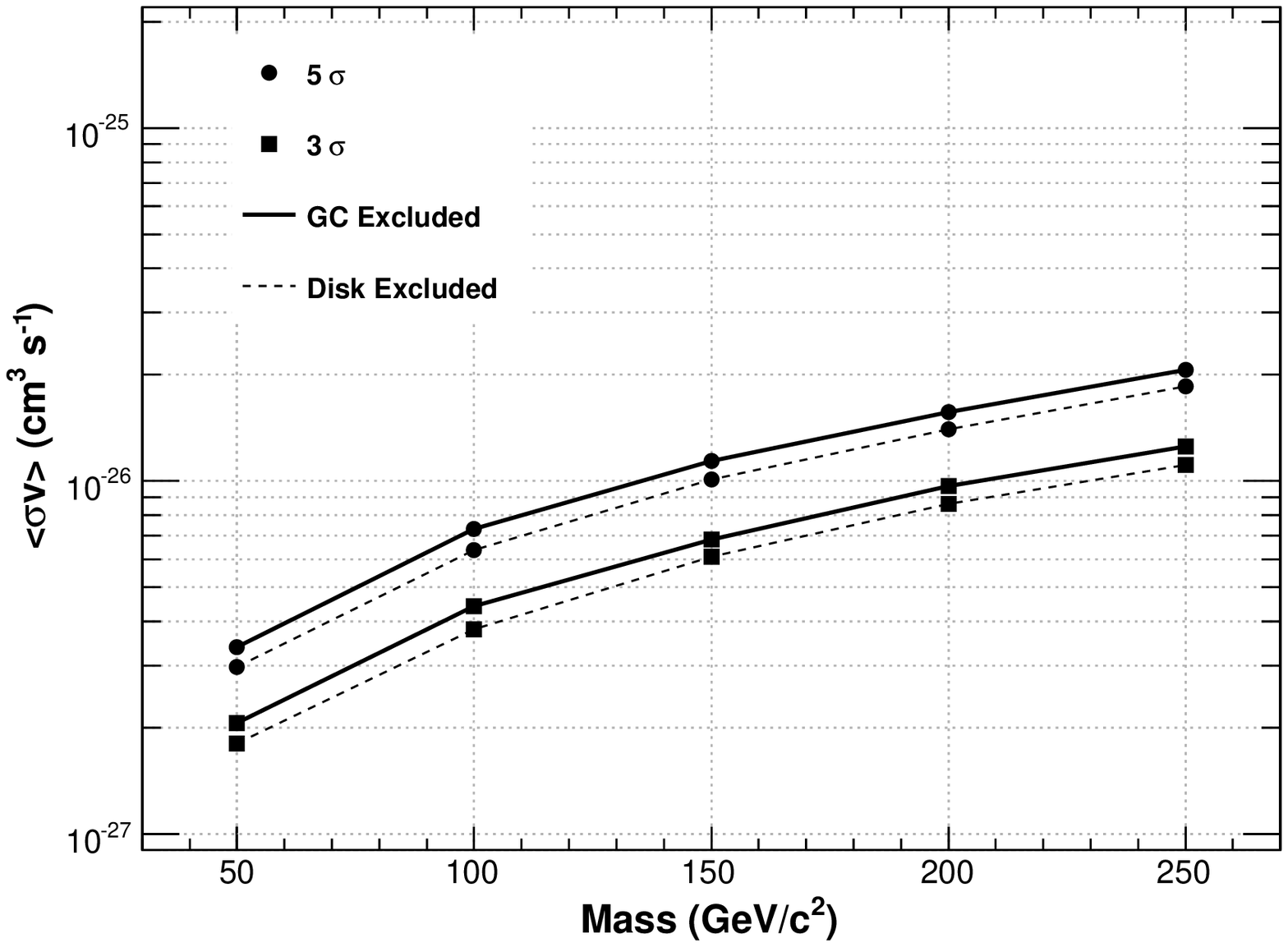}
\caption{The $<\sigma v>$ required to obtain an observation of WIMP annihilation at either 3$\sigma$ (square) and 5$\sigma$ (circle) significance  for {\bf one year of GLAST data}, as a function of WIMP mass. Left panel: considering the ``optimized'' diffuse model as background, right panel: considering the ``conventional'' diffuse model. The dashed line corresponds to the 10$^\circ$ cut above and below the Galactic plane; the solid line corresponds to a 10$^\circ$ radial cut around the Galactic center.}
  \label{fig:signif}
  \end{center}
 \end{figure}

\noindent
 We estimate the LAT sensitivity for an observation of DM from WIMP
  annihilation by running sample experiments
 (or pseudo-experiments), assuming data sample sizes consistent with one
 year of running. Each pseudo-experiment includes photons from WIMP
 annihilation with fixed mass and an assumed $<\sigma v>$,   along with photons
 expected from diffuse gamma ray background.   Other backgrounds are smaller and are presently ignored.  
 We then fit the resulting pseudo-experiment to the
 background and signal shape in latitude, longitude, and energy. 
 We construct ensembles of $\sim 1000$ such pseudo-experiments, with each ensemble
 corresponding to a fixed WIMP mass from 50 to 250 GeV$/c^2$.  For each
 mass, we vary $<\sigma v>$ until 50\% of the experiments have observations at
 either 3$\sigma$ or 5$\sigma$ significance.   
 \Fref{fig:signif} shows the results for both the ``conventional'' and the ``optimized'' diffuse backgrounds.\\

\noindent
In our fiducial region, a Moore profile predicts a factor 1.3 times the flux of NFW, which would enter linearly into our sensitivity. While the uncertainty in the signal flux (due to lack of knowledge on DM density distribution) is the dominant systematic uncertainty entering the sensitivity calculation, the question if a halo DM signal is distinguishable from a the cosmic-ray induced \gray \ background will crucially depend on the background uncertainty, for which careful studies with data will have to be performed and which requires sophisticated statistical tools due to non-linear dependences in the parameters entering the background model.\\ 

\noindent
Nevertheless, in order to estimate the impact of a systematic uncertainty on the shape of the diffuse background in a rough fashion, we introduce a  nuisance parameter into the likelihood function.  The parameter allows for a continuous change in the background distribution from the ``conventional'' model to the ``optimized''  model. The likelihood is minimized for the nuisance parameter and the signal and background components. A flat prior is assumed for the nuisance parameter. When using the ``conventional'' background model and allowing for a systematic variation in shape, primarily in the energy spectra, the required $<\sigma v>$ for a $3\sigma$ ($5\sigma)$ observation is increased by $\sim$70\% ($\sim$45\%) at low mass and $\sim$40\% ($\sim$25\%) at high mass. When using the ``optimized'' background model the change due to the systematic variation of shape is less, about a $\sim$30\% change at low mass and a $\sim$20\% at high mass for both the $3\sigma$ and $5\sigma$ observations.\\

\noindent
It can be concluded, that (with the caveat of only preliminary treatment of systematic uncertainties) the LAT would probe a large region of the MSSM and mSUGRA parameter space, as shown in \fref{fig:scan} in \sref{sec:specific}.\\

\begin{figure}[tb]
\begin{center}
\begin{minipage}[t]{7cm}
\vspace*{0.0in}
\includegraphics[width=8cm]{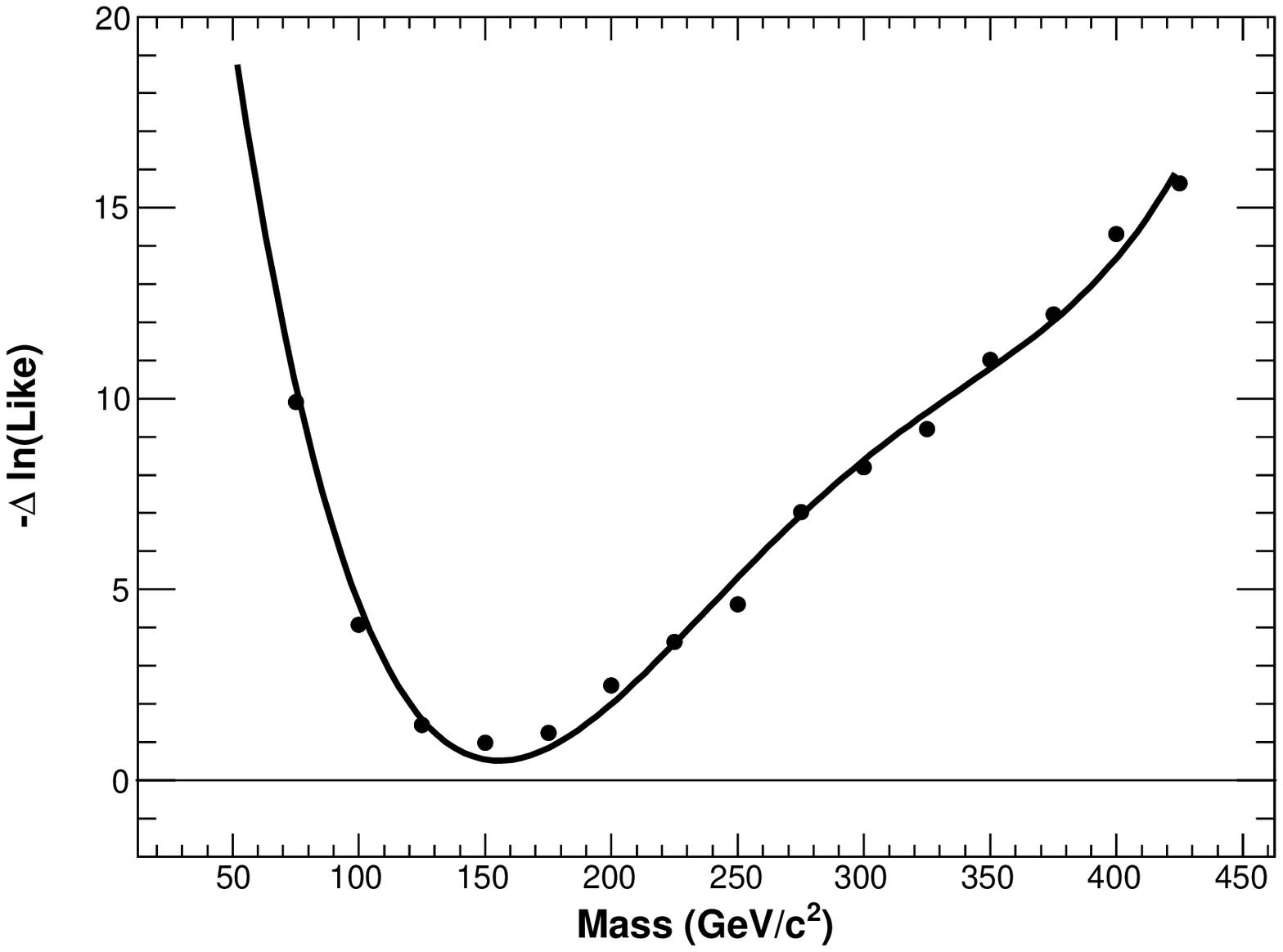}
\end{minipage}
\begin{minipage}[t]{7cm}
\vspace*{0.0in}
\includegraphics[width=8cm]{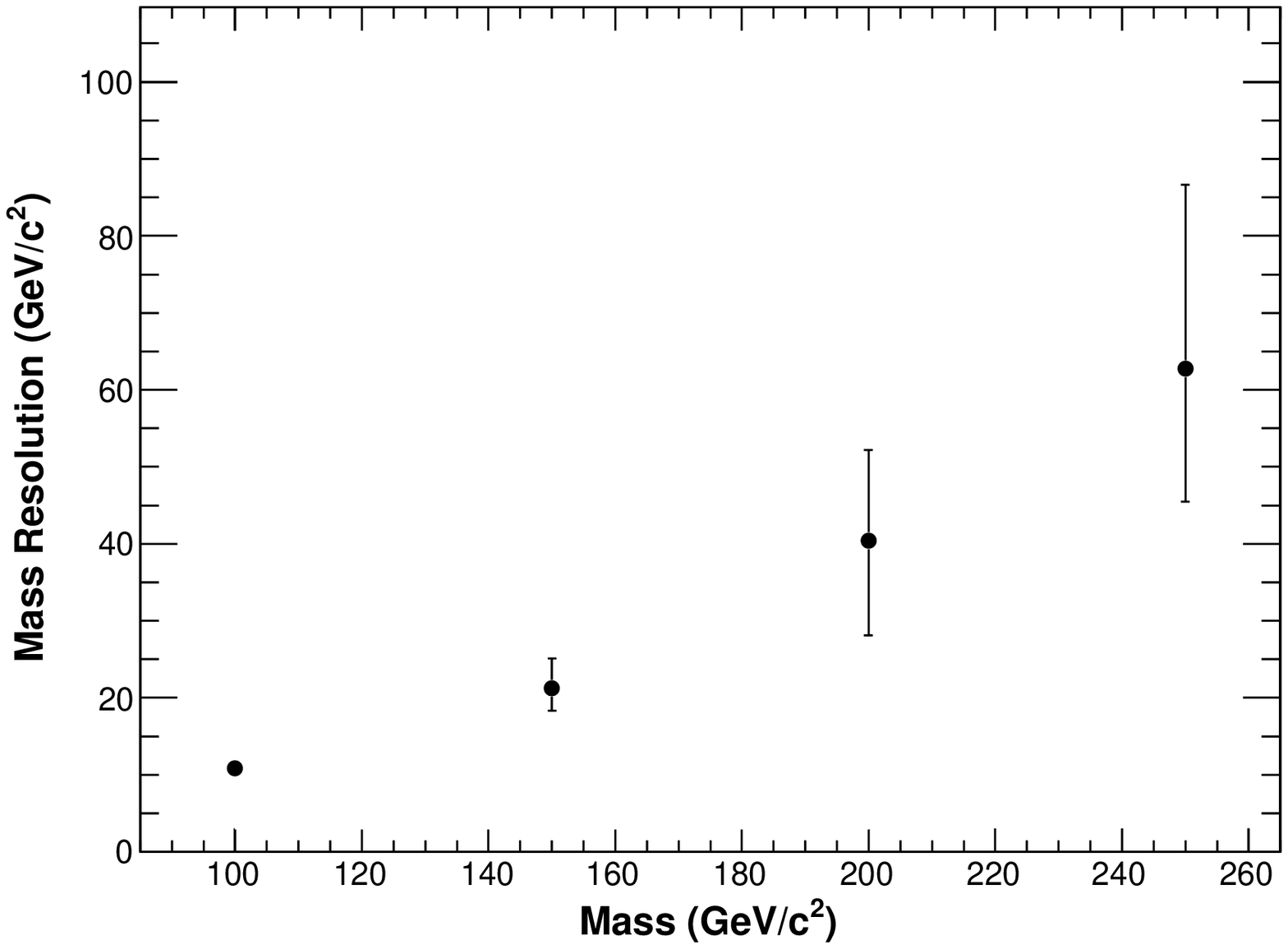}
\end{minipage}
\caption{\label{fig:pseudo}
The left plot shows the fitted $-\Delta$ log-likelihood versus mass for a single pseudo-experiment assuming the true mass of the WIMP to be 150 $ GeV/c^2$. The resulting fitted mass for this 
pseudo-experiment is 
$155 \pm 23$ GeV/$c^2$. 
The right plot shows the mean fitted error returned by the fit,
versus the true WIMP mass. The error bars show the 68\% containment interval for the pseudoexperiments, which demonstrates the range of possible outcomes from a single experiment. The parameters describing the pseudoexperiments are given in the text.
}
\end{center}
\end{figure}

\noindent
Using a similar technique we obtain the expected LAT mass resolution provided that $<\sigma v>$  would be at a high enough level. Again, we construct  ensembles of 1000 pseudo-experiments, this time with  fixed WIMP masses  and $<\sigma v> =2.3\cdot10^{-26}cm^{3}s^{-1}$.  For an  observation time of 1 year, this results in approximately 3.3 $ \cdot 10^3$ photons
from a 250 GeV WIMP signal,  and $10^7$ photons from diffuse background  (``conventional'' model).  A likelihood fit of each pseudo-experiment is made to a combination of diffuse background and signal, varying the assumed  WIMP mass.  An
example of such a fit  for a single pseudo-experiment is  shown in the left side of
\fref{fig:pseudo}.   The resulting fitted mass for this pseudo-experiment is
$155 \pm 23$ GeV/$c^2$. 
We then repeat this procedure for additional WIMP masses
of 100, 200  and 250 GeV$/c^2$.  Assuming the same  $\sigma
v=2.3\cdot10^{-26}cm^{3}s^{-1}$ and 1 year of observation 
 results in signal counts which vary from $1.5 \cdot 10^{4}$ (at
100 GeV$/c^2$) down to $3 \cdot 10^{3}$ (at 250 GeV$/c^2$). The right side of \fref{fig:pseudo}
shows the average returned uncertainty on the fitted mass  versus WIMP
mass.  From this figure, we see that our expected mass  resolution varies from about
11 GeV$/c^2$ (at 100 GeV$/c^2$) to 63 GeV$/c^2$ (at 250 GeV$/c^2$). The bars
in   \fref{fig:pseudo}  show the 68\% containment interval for possible outcomes
of a single experiment. 
The mass resolution shows that from the consideration of the statistical errors it would be possible to resolve the mass of the WIMP, which could be used to guide further $\gamma$-line and accelerator searches. Note that the mass resolution obtained in this analysis is not dominated by the energy resolution of the detector (in contrast to the case of detection of a line signal), but rather intrinsic to the signal.

\subsection{Galactic Satellites}
\label{sec:satellites}
\subsubsection{Galactic sub-halos.}

In the CDM paradigm \cite{Blumenthal:1984bp}\cite{Peebles:1984ge}, structure forms hierarchically, so the dark halos of galaxies such as the Milky Way should contain large numbers of subhalos.  For WIMPs the subhalo mass spectrum is expected to extend as low as $10^{-4}$ to $10^{-12}$ solar masses depending on the free-streaming scale of the considered DM particle, see \cite{Profumo:2006bv}. It is improbable that GLAST will  detect halos of this size as they would need to be very close to the Earth.\\

\noindent 
The substructure is expected to be nearly isotropic. Thus, annihilation in the subhalos can be significant away from the Galactic plane, where astrophysical sources are concentrated. These substructures are commonly called DM satellites.\\

\begin{figure}[here]
\begin{center}
\includegraphics[height=.3\textheight]{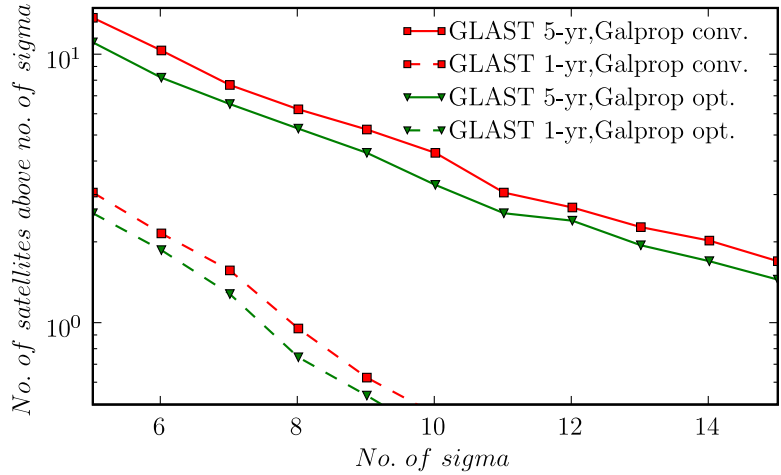}
\caption{Estimated number of observable DM satellites for the LAT in the Milky Way for {\bf 1 and 5 year of GLAST operation}. The background consists of the isotropic extragalactic diffuse \cite{Sreekumar:1997un} and {\sffamily GALPROP} ``conventional'' / ``optimized'' Galactic diffuse model. The significance is estimated as $N_S/\sqrt{N_S+N_B}$  within the satellite tidal radius (or the PSF 68\% containment radius if larger ) at $E_\gamma > $ 1 GeV. }
\label{fig:sigma}
\end{center}
\end{figure}

\noindent
To take into account the tidal stripping of DM satellites, we use a truncated NFW profile \cite{Hayashi:2002qv}, which approximates a structure of a stripped halo by a simple modification of the NFW profile:
\begin{equation}
 \rho(r) = \frac{f_t}{1+(r/r_t)^3}\ \rho_{NFW}(r)\ ,
\end{equation}
where $r$ is the distance from the center of the structure, $f_t$ is a dimensionless measure of the reduction in central density, and $r_t$ is an ``effective'' tidal radius. 
A Milky Way-sized dark halo is simulated based on the above satellite profile and a satellite distribution generated by Taylor \& Babul \cite{Taylor:2004gq}, which was chosen as representative among many possible theoretical predictions (see for example \cite{Eke:2000av},\cite{Diemand:2005vz}).  In their simulations, about 30 \% of the mass is in satellites. Then, a generic WIMP model, in which the WIMP mass is 100 GeV, the annihilation cross-section is $2.3\cdot10^{-26}\rm{cm}^3\rm{s}^{-1}$ (consistent with the WMAP DM density estimate) \cite{Jungman:1995df}, and the annihilation channel is $b\bar{b}$, is used to estimate the number of Milky Way DM satellites with $>10^6$ solar masses observable by the LAT after  1 and 5 year exposures in scanning mode, described in section 2.1. The DM satellite distribution is roughly spherically symmetric about the GC and extends well beyond the solar orbit. Thus, the observable satellites are located mostly at high Galactic latitudes. The background is estimated using the isotropic extragalactic diffuse background described in Sreekumar \etal \cite{Sreekumar:1997un} plus the {\sffamily GALPROP} ``conventional'' / ``optimized'' model as discussed in \sref{sec:bckgd}.  The significance of the DM signal is then estimated to be the number of signal events within the satellite tidal radius (or the PSF 68\% containment radius, whichever is larger) divided by the square root of the number of total events, including both background and signal, within the same radius, at energy larger than 1 GeV. \Fref{fig:sigma} shows the cumulative number of DM satellites with significance of at least 5 $\sigma$. For this generic WIMP model, the LAT is expected to observe $\sim$12 such DM satellites within 5 years. (assuming uniform exposure). For discussion on observability for other models see for example \cite{Pieri:2007ir},\cite{Diemand:2006ik}.
Adding the charged particle background to this analysis, reduces the number of detectable satellites by about 10 \%. Uncertainties in the charged particle background prediction have negligible effect.\\ 

\begin{figure}[here]
\begin{center}
\includegraphics[height=.3\textheight]{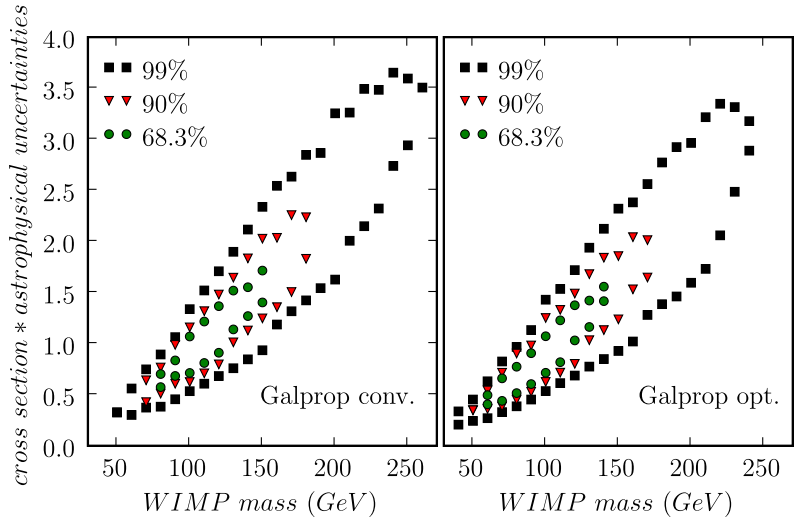}
\caption{LAT error ellipses for a simulated ``5 sigma'' DM satellite with the generic WIMP model. The vertical axis is given by \Eref{eq:sigmav}.}
\label{fig:contour}
\end{center}
\end{figure}

\noindent
We also calculate the LAT error ellipses for WIMP mass vs. the quantity given in \Eref{eq:sigmav} for this generic WIMP model.  We choose one ``5 $\sigma$'' satellite for 1 year of GLAST data (this significance is estimated using the back of the envelope method mentioned above), which is at high latitude (b = -39 deg), 8.9 kpc away from the Earth, with the mass $2.7\cdot10^7 M_\odot$, and with the tidal radius 0.2 kpc. We use the likelihood ratio test statistic and profile likelihood to extract the 99\%, 90\% and 68 \% error contours on WIMP mass and annihilation cross-section jointly, as shown in \fref{fig:contour}, for this generic WIMP DM satellite, where the unit of the vertical axis is given by 
\begin{equation}\label{eq:sigmav}
\frac{<\sigma v>}{2.3 \times 10^{-26}\,\rm{cm}^3\rm{s}^{-1}} \times \left(\frac{8.9\, \rm{kpc}}{d}\right)^2 \times \left(\frac{M_{\rm{satellite}}}{2.7 \times 10^7 M_\odot}\right)^2 \times \left(\frac{0.2\, \rm{kpc}}{r_{\rm{t}}} \right)^3
\end{equation}
which can be used to rescale the error ellipse for different satellite masses, distances and tidal radii. The above scaling factors are calculated from the gamma ray flux from WIMP annihilation, for point-like sources, for which
\begin{equation}
 \phi_{WIMP}  \propto \frac{<\sigma\,v>}{m_\chi^2}\frac{1}{d^2} \bar{\rho}^2Vf(V)
\label{eq:scaling}
\end{equation}
 where
\begin{equation}
f(V) = \frac{1}{\bar{\rho}^2\,V}\int \rho d\,V
\end{equation}
is a dimensionless flux multiplier \cite{Taylor:2002zd} and $\bar{\rho}$ is the average density within volume $V$. In Taylor and Babul's simulations, the observable satellites for the LAT have $\frac{r_t}{r_s} \sim 0.1$, therefore it is accurate enough to consider f(V) over the volume within $r_t$. For the NFW profile, for which $\rho \propto r^{-1}$ in that region, f(V) is a constant. Consequently:
\begin{equation}
 \phi_{WIMP} \propto \frac{<\sigma\,v>}{m_{\chi}^2}\frac{1}{d^2} \bar{\rho}^2\,V = \frac{<\sigma\,v>}{m_{\chi}^2}\frac{1}{d^2} \frac{M_{\rm{satellite}}}{V}
\end{equation}
\noindent
However, the Moore profile \cite{Moore:1999gc} with the same scale radius and scale density, exhibits $\rho \propto r^{-1.5}$ within $r_{t}$. In this case, $f(V) \propto \ln{r_t/r_{min}}$. Integrating \Eref{eq:scaling} from $r_{min}$ to $r_{t}$, taking into account the change in $\bar{\rho}$, yields that the example satellite will emit $\sim$50 times more photons under the assumption of a Moore profile than under the assumption of a NFW profile (for a cutoff radius of $r_{min} = 10^{-5}$ kpc). It should be noted that the Moore profile is rather optimistic.\\

\begin{figure}[here]
\begin{center}
\includegraphics[height=.3\textheight]{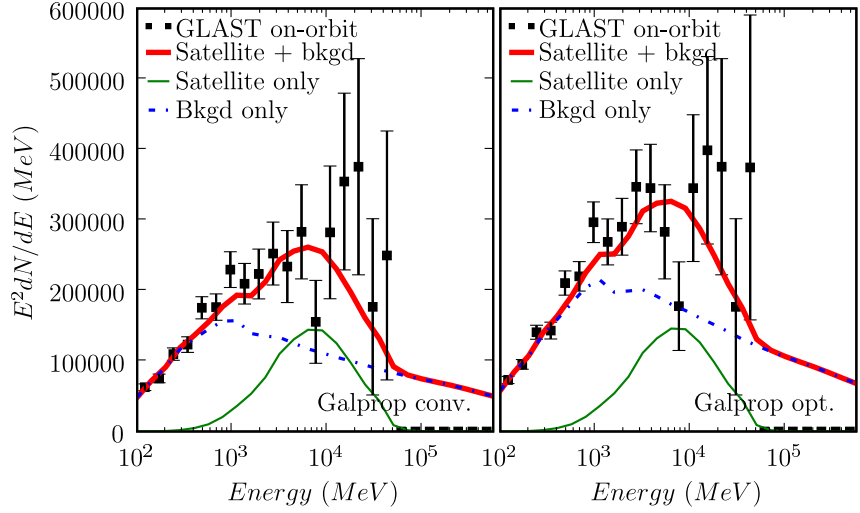}
\caption{Counts spectra for the same generic WIMP satellite within its tidal radius. The squares with 1 sigma error bars show the spectrum of simulated {\bf 1 year GLAST data}. The thick lines show the background plus the satellite signal. The thin lines show the satellite predictions only. The dash-dot lines show the background predictions only.}
\label{fig:spec}
\end{center}
\end{figure}

\begin{figure}[here]
\begin{center}
\includegraphics[height=.3\textheight]{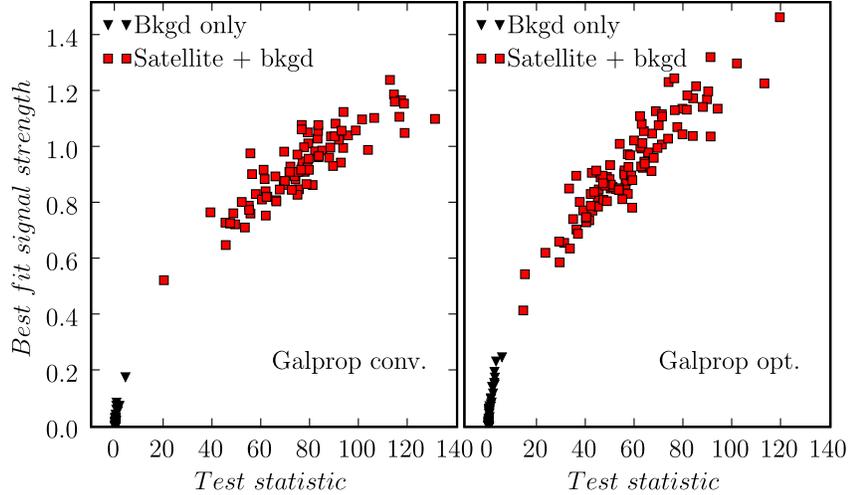}
\caption{Null hypothesis test to distinguish the same generic WIMP satellite from a background fluctuation. The test statistics for background only and for background plus satellite signal could be separable. }
\label{fig:likelihood}
\end{center}
\end{figure}

\noindent
In order to identify DM satellites, we need firstly to find possible candidates for DM satellites; secondly to make sure these candidates are not fluctuations of the background emission; and finally to distinguish these real candidates from other typical astrophysical sources, such as pulsars or molecular clouds. The typical extent of the observable DM satellites is on the order of 1 deg.  For bright enough sources, the LAT should be able to resolve them using  WIMP annihilation photons above 1 GeV, for which the PSF is $< 0.5$ deg. {\tt  SExtractor} \cite{bertin96}, which can build a catalog of objects from an astronomical image, is used to search for possible candidates for DM satellites. We then use the binned likelihood analysis (see \sref{sec:stools}) to reject false sources and other typical classes of sources.  \Fref{fig:spec} shows the counts spectra for the same generic WIMP satellite, plus diffuse background. The diffuse background model consists of the {\sffamily GALPROP} ``conventional'' / ``optimized'' model and the isotropic extragalactic diffuse background \cite{Sreekumar:1997un}.  In future data analysis we will use a hypothesis test using a likelihood ratio test statistic to distinguish the satellite from background. As a sanity check of the methodology developed by the GLAST collaboration, we generate 120 simulated GLAST experiments for the background only and also for the example satellite. Then, we compute our test statistic (TS) as the maximum log-likelihood difference between background only and background plus signal, where the free fitting  parameters are the signal and background normalizations. \Fref{fig:likelihood} shows a two dimensional plot of the TS for background only (filled triangles) and for background plus signal (filled squares) versus best-fit signal strength, which is the ratio of the fitted flux to the actual simulated flux of the satellite. Indeed the TS seems to behave as expected. The signal satellite can be detected with an efficiency close to 100 \%. Only one (two) out of the 120 realizations gives a TS value which falls below the threshold of TS = 25 for the ``conventional'' (``optimized'') GALPROP background. Using the unique WIMP pair annihilation spectrum, which is an extremely hard non-power-law and has an end point at the mass of the WIMP, we will use the same TS to distinguish DM satellites from other astrophysical sources. Also, note that real DM satellite candidates are expected to have no counterparts in radio, optical, X-rays and TeV energies. For more details on separating DM satellite signals from possible backgrounds, see \cite{Baltz:2006sv}.


\subsubsection{Dwarf Galaxies.}

\begin{figure}[ht]
\begin{center}
\includegraphics[height=10cm, width=14cm]{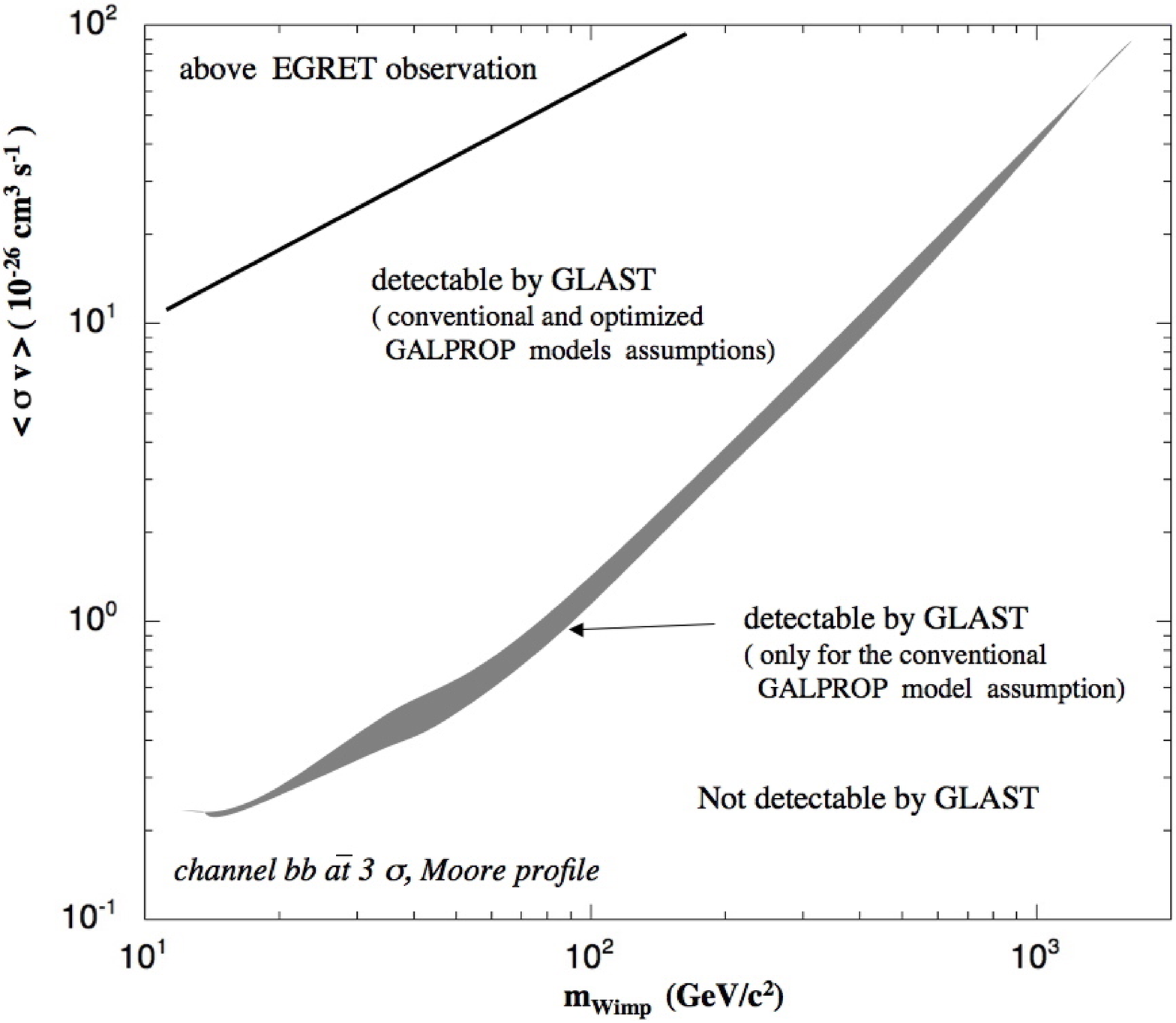}
\caption{Sensitivity to a Sagittarius Dwarf DM signal for {\bf 5 years of GLAST operation} assuming Moore profile 
as described in \cite{Evans:2003sc}. The region labeled ``above EGRET observation'' is calculated with respect to the upper limit map provided in  \cite{Hartman:1999fc}.}
\label{SagPlot}
\end{center}
\end{figure}

As described in the previous section, N-body simulations predict a large amount of substructure which represent good targets 
for DM annihilation $\gamma$-ray studies with the LAT.
Dwarf spheroidal galaxies (dSph) may be manifestations of the largest clumps predicted by the CDM  scenario.
They are potential candidates for indirect searches of DM because
they are apparently among the most extremely DM dominated environments.
For example, the mass-to-light ratio of Draco is $\sim 250$ in solar units 
\cite{Kleyna}, while it is  $\sim 100$ \cite{Ibata} for the Sagittarius Dwarf. \\

\noindent
As an illustration of the LAT capability to detect DM $\gamma$-ray 
signals from dSph galaxies,
we compute the 3$\sigma$ sensitivity for 5 years operation
for Sagittarius, which is the closest dSph to the Sun.
Discovered in 1994, Sagittarius 
is one of the nine dwarf spheroidal galaxies orbiting
our own Galaxy.
The core of the Sagittarius dwarf is located at l=$5.6^o$ and b=$-14^o$ in Galactic coordinates  at a distance of about 24 kpc from the Sun. As most dSphs do not contain significant amounts of gas, the structure of the dark halos 
must be inferred from stellar motion. For this study we use the Moore cusped halo profile derived for Sagittarius in \cite{Evans:2003sc}.   For compatibility with EGRET observations, we require here that the total integrated flux given in \Eref{eq:gammafluxcont} does not exceed the value(0.4 $\cdot$ 10$^{-7}$ cm$^{-1}$ s$^{-1}$) extracted from the upper limit maps presented in \cite{Hartman:1999fc}.\\

\noindent
The result reported in \fref{SagPlot} shows
the LAT capability to detect a DM signal from
the Sagittarius dSph in the case of the
favorable cusped Moore profile proposed in \cite{Evans:2003sc}.\\

\noindent
Using a Moore profile for Sagittarius is a very optimistic assumption, though it has been argued that the number of passes through the Galactic disk is low enough, so that it is not completely disrupted \cite{Evans:2003sc}. More conservatively assuming a NFW profile will decrease the sensitivity by about a factor 10.\\
\noindent
For example, the assumption of a Moore profile is less optimistic for the Draco dwarf. However, Draco is about a factor 3-4 more distant. Correctly taking into account the difference in Galactic diffuse background, leads to an estimated decrease in sensitivity by about a factor 10, when comparing Draco with Sagittarius assuming a Moore profile. Since Draco on the other hand might have a more favourable densitiy profile, it is one of the prime candidates for DM searches.\\

\noindent
In conclusion, nearby dwarf spheroidals are attractive targets for DM searches with the LAT observatory, if cuspy profiles can be considered.

\subsection{Point Sources}
In this section we study the possibility for GLAST to detect point sources of Dark Matter. These sources can be motivated by scenarios invoking intermediate size black holes \cite{IMBH2}, i.e. wandering black holes with masses $10^2 \lesssim M/M_\odot \lesssim 10^6$, which would adiabatically grow ``mini-spikes'' of DM. \\ 

\begin{figure}[ht]
   \centering
     \includegraphics[height=9cm]{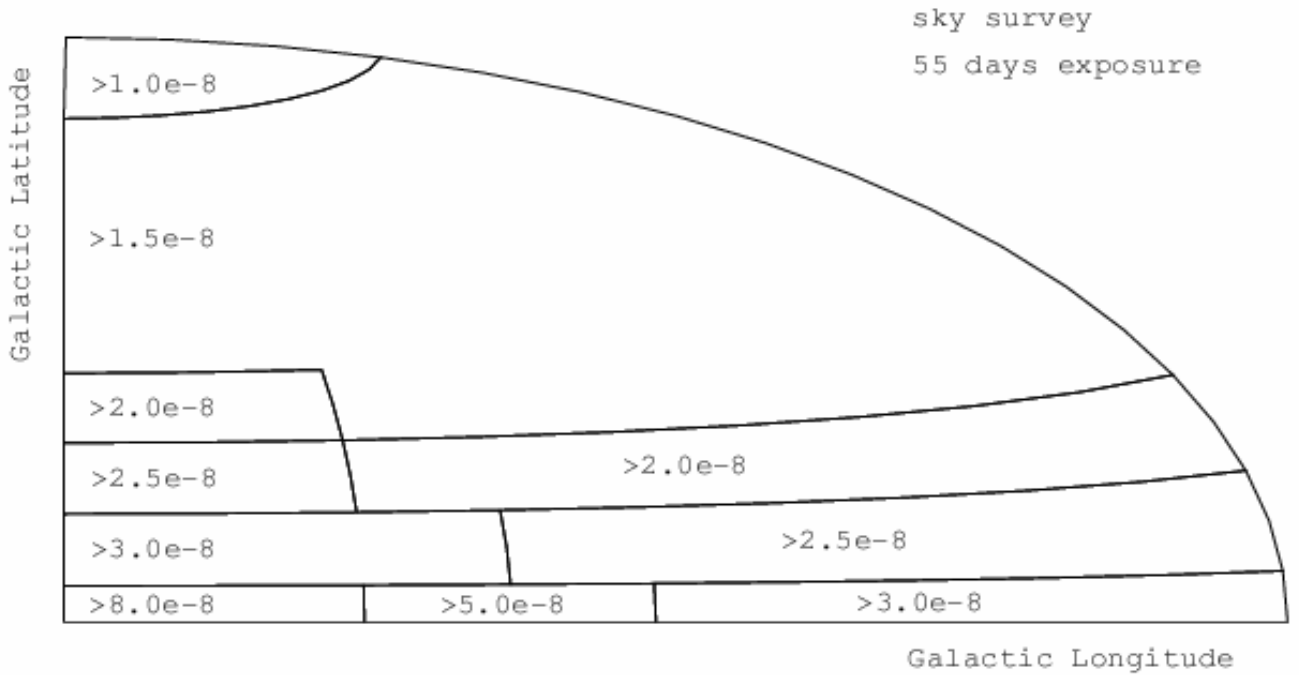}
     \caption{Flux needed for separating a point sources of DM annihilation from the background, i.e. full-sky map in Galactic coordinates of the minimum flux above 100 MeV, in units of [ph cm$^{-2}$s$^{-1}$], that is required for a $5\sigma$ detection of an annihilation spectrum, assuming a DM particle with mass $m_{WIMP}=150~$GeV annihilating into $b \bar b$ (note, however, that the map does not depend very sensitively on DM properties). The map is relative to {\bf a 2 month operation period}; for longer operation times, fluxes scale approximately as $t_{obs}^{-1/2}$.}
   \label{sensmap}
\end{figure}

\noindent
In order to study the LAT sensitivity to such objects, we perform a two- month scanning mode simulation of the \gray \ sky. For the backgrounds we employed the ``optimized'' {\sffamily GALPROP} model for Galactic diffuse emission and the EGRET measurement \cite{Sreekumar:1997un} for the extragalactic contribution. We divide the sky into 23 regions of about 10 degs in radius, and in each region we placed one DM point source. Then, we consider each source separately and let the flux intensity above 100~MeV vary from $10^{-8}$ to $10^{-7}$ ph cm$^{-2}$s$^{-1}$. These flux values are typical for scenarios with mini-spikes as for example discussed in \cite{IMBH2}. 
For each intensity, we calculate the significance of the observed signal, given the local background counts, with a maximum likelihood analysis assuming Poisson statistics as described in \sref{sec:stools}.
By estimating the minimum flux required to discriminate the DM source from the background at a $5\sigma$ level on a grid of points uniformly distributed over the sky, we obtain the sensitivity map shown in \fref{sensmap} (where we adopt a DM particle mass $m_{WIMP} = 150$~GeV and assume annihilation mainly into $b \bar{b}$).\\

\begin{figure}[ht]
   \centering 
   \includegraphics[width=0.4\textwidth]{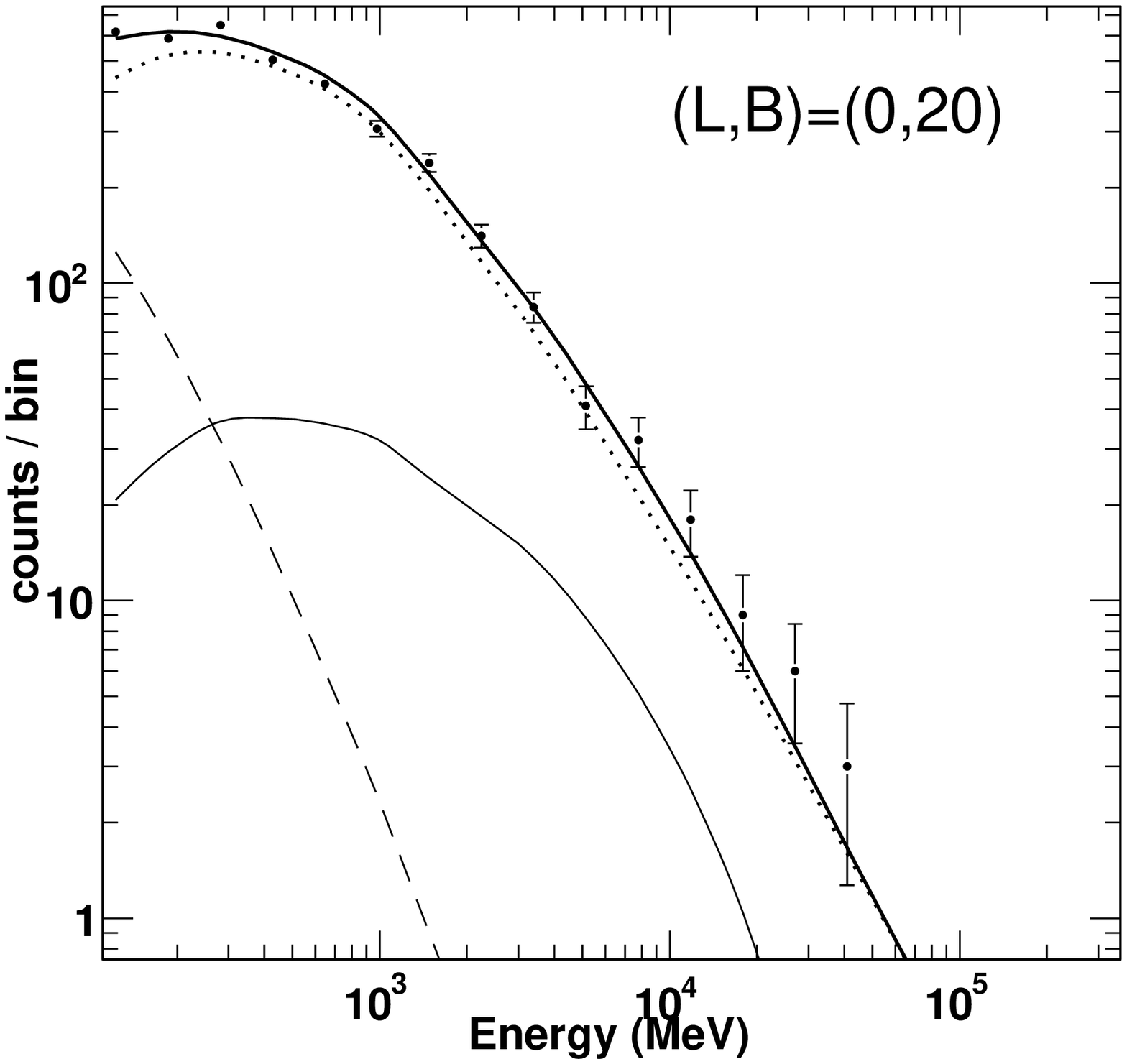}
   \includegraphics[width=0.4\textwidth]{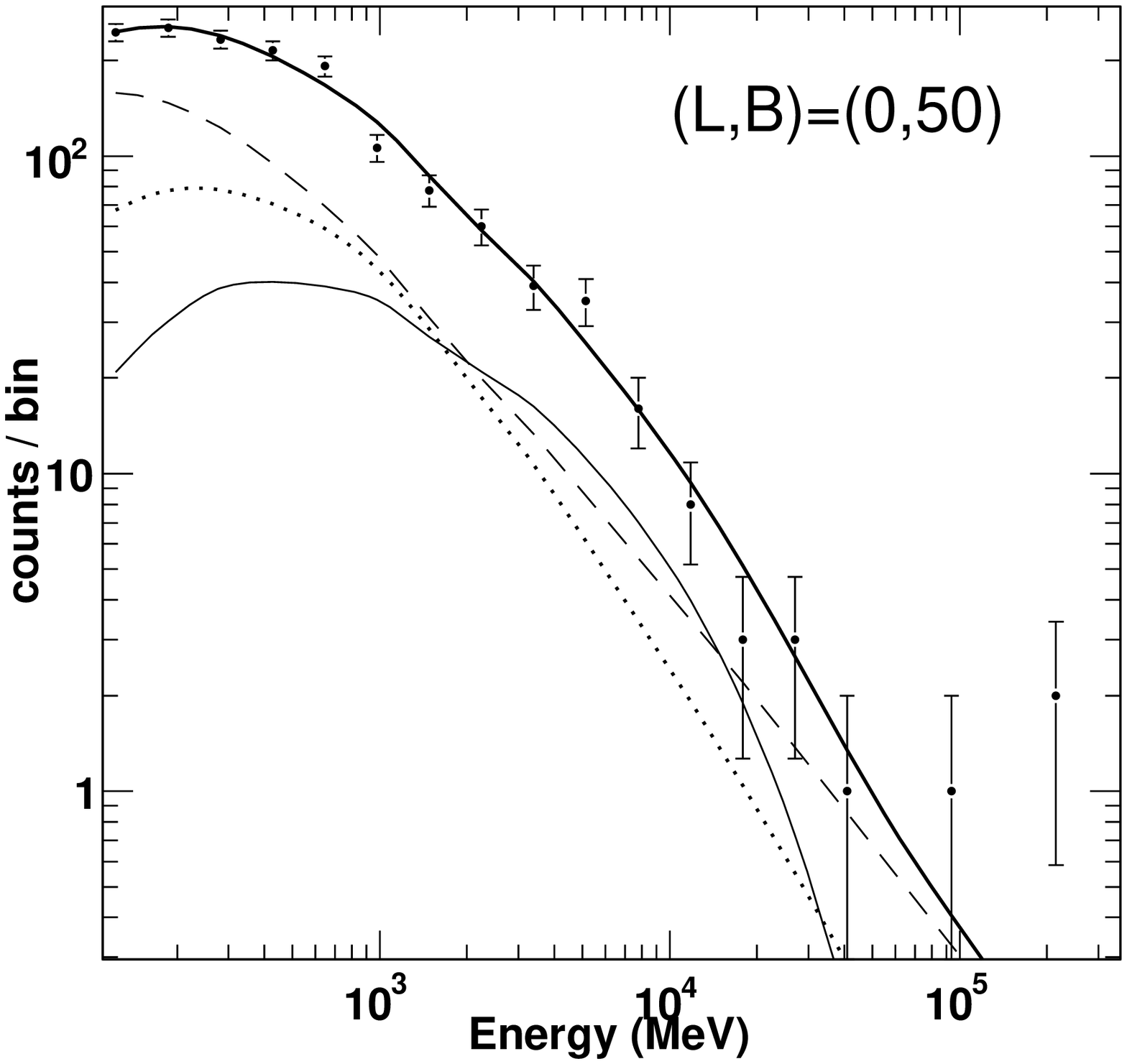}
   \includegraphics[width=0.4\textwidth]{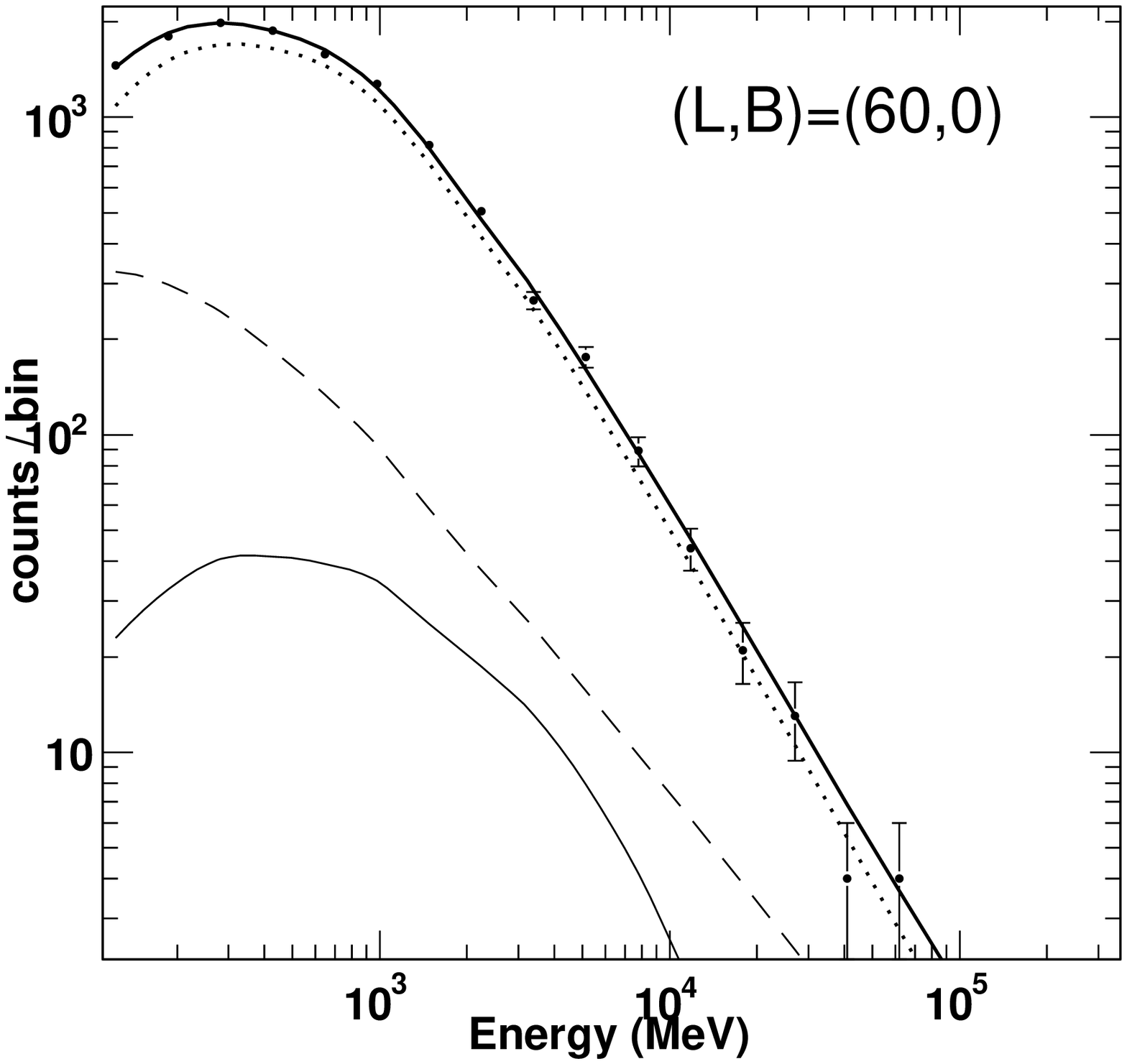}
   \caption{ Examples of spectral fits of simulated  DM point sources of intensity $\phi = 8 \cdot 10^{-8}$ photons cm$^{-2}$s$^{.1}$ above 100 MeV, m$_{WIMP}$ = 150 GeV $b\bar{b}$ annihilation channel. Upper left for (l,b) = (0,20); upper right for (l,b)= (0,50) and on the bottom (l,b)=(60,0). Thin solid lines: DM signal, dashed: Galactic diffuse contribution, dotted: extragalactic contribution (from \cite{Sreekumar:1997un}), points with error bars are photon counts from the simulated observation.}
   \label{spectra}
\end{figure}

\noindent
The sensitivity appears to depend significantly on the Galactic longitude only along the Galactic disk, as expected. At high Galactic latitudes a source as faint as $1\cdot 10^{-8}$ ph cm$^{-2}$s$^{-1}$ above 100 MeV is resolved, while close to the GC a minimum flux of $8\cdot 10^{-8}$ ph cm$^{-2}$s$^{-1}$ is required. \Fref{spectra} shows some illustrative examples of simulated sources. All sources detected with a significance above 5 $\sigma$ show a clear indication of a spectral cutoff at high energy, i.e. the fit with a simple power line spectrum is disfavored. Mini-spikes scenarios such as the one  discussed in \cite{IMBH2} predict  a population of $\sim 100$ DM mini-spikes. Using the likelihood approach to discriminate against usual astrophysical background we expect to be able to unambiguously detect about 25 \% of the sources (depending on mini-spike distribution and assumptions on background).



\subsection{GLAST sensitivity to a line signal from DM annihilation}
\label{sec:line}

\begin{figure}[ht]
\begin{center}
\includegraphics{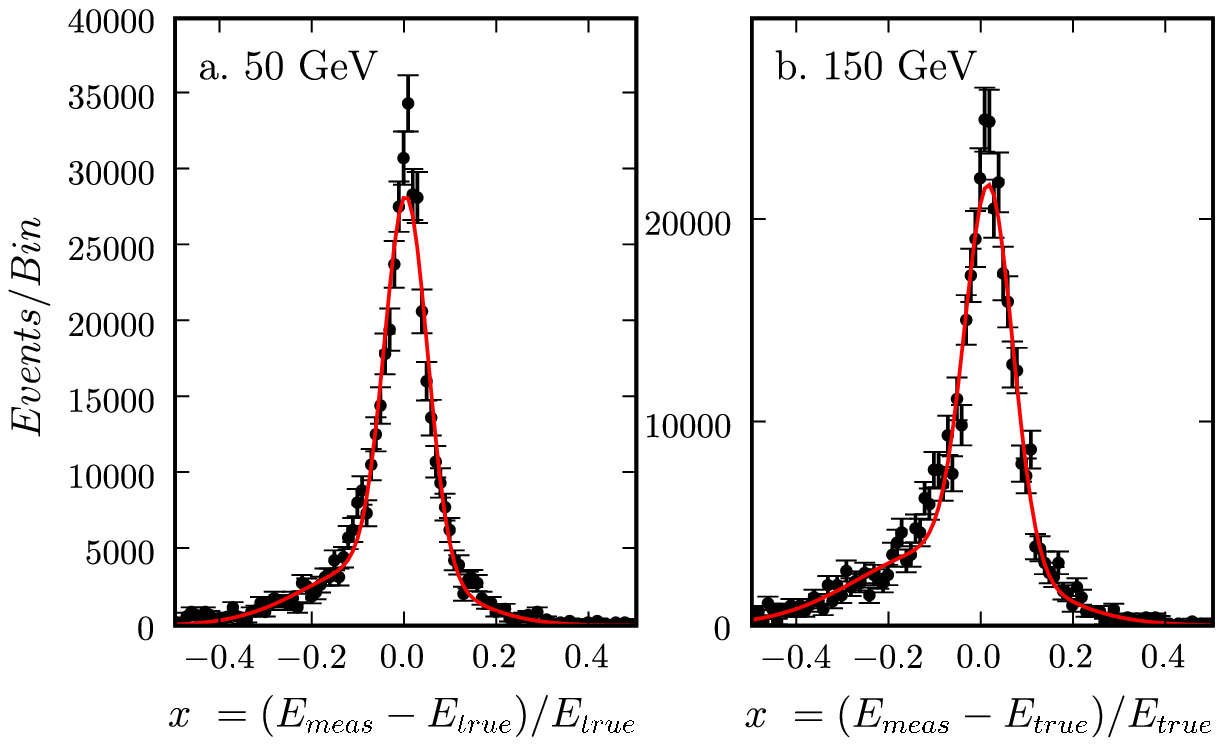}
\caption{Double Gaussian fits to the simulated LAT energy dispersion for \gray\ energies of a) 50 and b) 150 GeV.
\label{fig:line1}
}
\end{center}
\end{figure}
\begin{figure}[ht]\begin{center}
\includegraphics{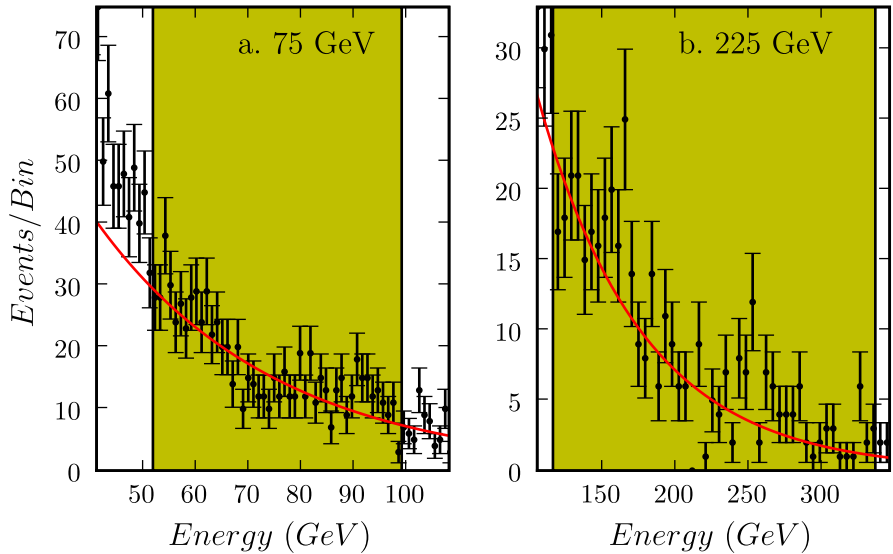}
\caption{``Conventional'' Galactic plus extragalactic background fit to a powerlaw over the range $[E_0-6\sigma_E, E_0+6\sigma_E]$, for a) $E_0$ = 75 GeV and b) $E_0$ = 225 GeV.  The fit range is shown as a shaded area.
\label{fig:line2}
}
\end{center}
\end{figure}
\begin{figure}[ht]\begin{center}
\includegraphics{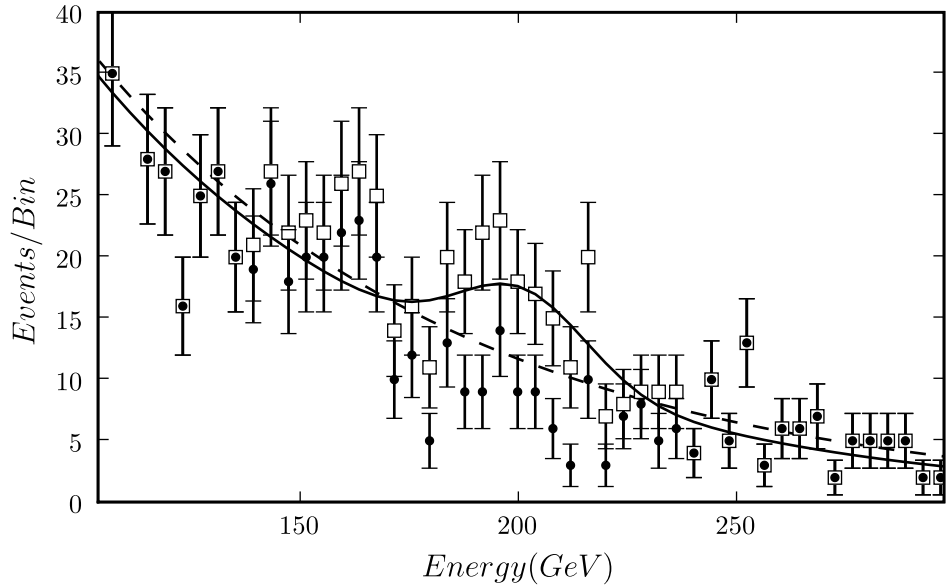}
\caption{``Optimized'' diffuse background and a 5$\sigma$ signal at 200 GeV.  The black dots and open squares correspond to the diffuse background and the diffuse background plus MC signal, respectively. Full and dotted lines correspond to the signal plus background fit to $\phi2$ and $\phi1$+$\phi2$, respectively.  $<\Delta \chi^2>$=25.0 for this run.
\label{fig:line3}
}
\end{center}
\end{figure}   
\begin{figure}[ht]\begin{center}
\includegraphics{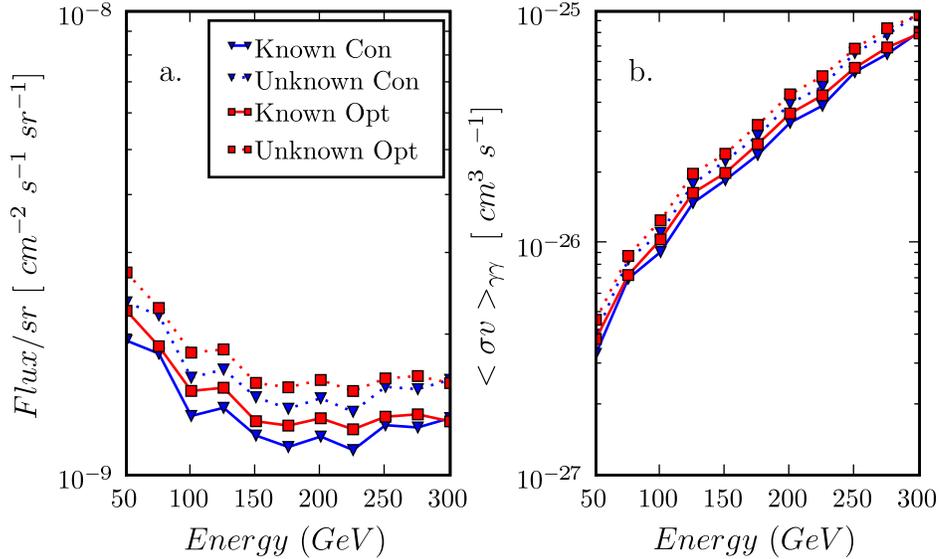}
\caption{ 5$\sigma$ sensitivity contours {\bf (5 years of GLAST operation)}  in a) flux and b) velocity-averaged effective cross-section. Triangles and squares correspond to the ``conventional'' and ``optimized'' Galactic background model, respectively. Full and dotted lines correspond to the case of known and unknown WIMP energy, respectively.
\label{fig:line4}
}
\end{center}
\end{figure}
\begin{figure}[ht]\begin{center}
\includegraphics{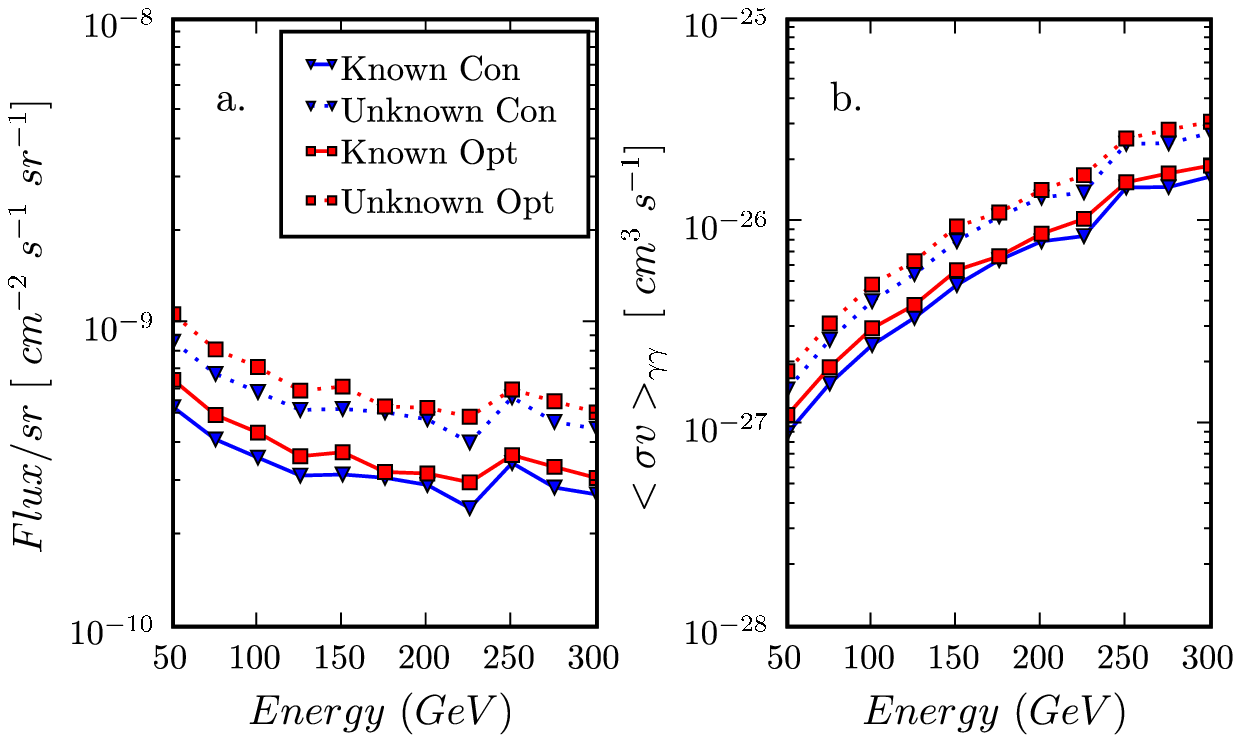}
\caption{95\% confidence level upper limit contours {\bf (5 years of GLAST operation)} in a) flux and b) velocity-averaged effective cross-section, as a function of the WIMP energy. The legend is identical to \fref{fig:line4}.
\label{fig:line5}
}
\end{center}
\end{figure}
In this section an approach for the indirect astrophysical detection of DM annihilation lines is presented. A line at the WIMP mass, due to the 2$\gamma$ production channel, could be observed as a feature in the astrophysical source spectrum \cite{Bergstrom:1997fj}. Such an  observation is  a ``smoking gun'' for WIMP DM as it is difficult to explain by a process other than WIMP annihilation or decay\footnote{The presence of a feature due to annihilation into $\gamma Z$ in addition would be even more convincing.}.  In addition, it should be free of astrophysical uncertainties, since the background can be determined from the data itself. Remember however, that in the most popular models branching ratio for the annihilation into lines is typically about $10^{-3}$ or less (see \cite{Gustafsson:2007pc}, however).\\ 

\noindent
We calculate sensitivities here in units of  photons/cm$^2$/s/sr which thus are independent of the signal model. However, they do depend on the diffuse background model. The Galactic diffuse models used here are the {\sffamily GALPROP} ``optimized'' and ``conventional'' models, discussed in \sref{sec:bckgd}.  An isotropic power law is used for the extragalactic diffuse component \cite{Sreekumar:1997un}.  Theoretical predictions can be compared with these sensitivities assuming a WIMP model (e.g., WIMP mass and many other theoretical parameters), a DM halo model (e.g. the NFW profile) and other astrophysical backgrounds. To estimate the photon background, this analysis uses diffuse-model simulated data from the Galactic centered annulus(r $\in$ [20$^\circ$, 35$^\circ$]), excluding the region within 15$^{\circ}$ of the Galactic plane. In the context of an NFW DM profile \cite{Navarro:1995iw}, including data in such an annulus minimizes the contribution of the Galactic diffuse emission and could give a signal-to-noise ratio as much as 12 times greater than at the GC \cite{Stoehr:2003hf}. The region we use in this analysis is not necessarily optimal for real data, but is shown as an example of how one might optimize signal to background. LAT line energy sensitivities are calculated for 5$\sigma$ detection and for the 95\% confidence level upper limit (CLUL). We address cases for which the line energy is known (e.g. supplied by a discovery at the Large Hadron Collider [LHC]), as well as for which the line energy is unknown, requiring a search over the energy range of interest. The latter gives somewhat poorer limits as the number of energy bins introduces a trial factor.  The LAT resolved signal is simulated using approximate delta functions as input for the full detector Monte Carlo (MC) simulation and reconstruction framework, GLEAM and applying event selections corresponding to the ones developed by the LAT collaboration.  To calculate known line energy sensitivities, LAT resolved lines are generated from  50 GeV to 300 GeV at 25 GeV intervals.  The results for uniform exposure over the LAT are reasonably well fit by a double Gaussian distribution $\phi_1$:
\begin{equation}
\phi_1 (E;E_{1,2},\sigma_{1,2},N_T,r) = \frac{N_T}{\sqrt{2\pi}} \left[ \frac{1-r}{\sigma_1} e^{-(E-E_1)^2/2\sigma_1^2}  + \frac{r}{\sigma_2} e^{-(E-E_2)^2/2\sigma_2^2} \right]
\end{equation}
where $N_T = N_1 + N_2$, and $r = N_2/N_T$. We use a double Gaussian instead of the official parameterization of the energy dispersion, since it is simpler to handle and since it gives a good fit over the entire energy range of interest. The center of the signal $E_0$ is defined by the peak position.  \Fref{fig:line1} shows some examples of fits. The full width at half maximum (FWHM) is used to define an equivalent single Gaussian $\sigma_E$ for each line. We use {\sffamily gtobssim} to generate LAT resolved 5 year all-sky diffuse background spectra in the region of interest, the Galactic annulus excluding the Galactic plane as defined above. \Fref{fig:line2} shows, for $E_0$ = 75 and 225 GeV, photon count spectra for this background, together with an exponential fit ($\phi_2(E;a,b) = a \cdot e^{E/b}$) over the range $[E_0-6\sigma_E, E_0+6\sigma_E]$. The spectra are well fit over this energy range.\\

\noindent
Next, the 5 year 5$\sigma$ signal sensitivity is estimated.  For each line energy the input background is resampled (bootstrapped) 1000 times with a $\phi_1$ MC signal and fit successively to ($\phi_1+\phi_2$) and $\phi_2$ with free parameters $a, b$ and $N_T$, for the range $[E_0-6\sigma_E,E_0+6\sigma_E$ (see \fref{fig:line2}). Each bootstrap is randomly sampled from the original background resulting in mildly correlated background realizations.  This series of 1000 bootstraps is rerun varying the number of events thrown into the  $\phi_1$ signal MC until $<\Delta \chi^2> = <\chi^2_{\phi_2+\phi_1} - \chi^2_{\phi_2} > \approx 25$ (5 $\sigma$). An example of a line signal just fulfilling this condition is shown in \fref{fig:line3}.\\

\noindent
The average number of signal counts needed at each energy is then converted to the known line energy sensitivity using average exposures over the annulus (see \fref{fig:line4}).  The LAT detection sensitivity for a line of unknown energy in the range [40,350] GeV for a 5$\sigma$ above background signal corresponds to a confidence level of $99.99997\%$.  To calculate the number of counts needed when the line energy is unknown, the probability of no signal detection in a single bin is used:
\begin{equation}
(1-P)^{\frac{1}{n_{bins}}} = \frac{1}{\sqrt{2\pi}} \int_{-\infty}^{n_\sigma} e^{\frac{-x^2}{2}}\,dx\ ,
\end{equation}
where P = $3\cdot10^{-7}$ is the probability of detecting a signal greater than the ``number of $\sigma$'', $n_\sigma$, in one or more of $n_{bins}$ energy bins.  A bin width of the LAT FWHM/E  $=$ 0.19, ($\sigma$/E = 0.08), was used, giving 15 bins over the energy range [40, 350] GeV.  The value of the number of $\sigma$s needed for a 5$\sigma$ detection over 15 bins is found to be 5.5.  The unknown line energy sensitivity is estimated by scaling the known line energy sensitivity by $(5.5/5)^2$ (see \fref{fig:line4}). \\

\noindent
To obtain the effective $<\sigma v>$ in the WIMP annihilation, we assume an NFW profile and calculate the line of sight integral (see \Eref{eq:gammafluxcont}) within the annulus region with the mass of the WIMP equal to the line energy in each bin that we considered. The right panel in \fref{fig:line4} shows the 5$\sigma$ cross-sections.\\

\noindent
The 5 year 95\% confidence level upper limit sensitivity for known line energies is obtained similarly to the 5$\sigma$ case by bootstrapping the diffuse background (with no MC generated signal), and by fitting each bootstrap sample to $\phi_1 + \phi_2$ with free parameters $a$, $b$ and $N_T$, for $E \in [E_0-6\sigma_E,E_0+6\sigma_E]$.  $N_T \pm \sigma_{N_T}$ 
is retrieved from the fit,  and the 95\% CLUL is obtained as 1.64$\sigma_{NT}$.  The case of unknown line energy ($E \in [40,350]$ GeV) 95\% CLUL is calculated as in the 5$\sigma$ case with scale factor 2.71/1.64. Over one thousand bootstraps were required to produce 1000 convergent fits for $E_0$ = 50 and 75 GeV. Results are shown in   \fref{fig:line5}.\\  

\noindent
These results are preliminary as they lack the WIMP continuum contribution to $\phi_2$, the event selections and corresponding parameterizations of the response are being updated, the optimum $(l,b)$ range of integration is still being considered including the GC, and refined statistical methods utilizing Poisson statistics are being explored. 
Note that with the annulus and the NFW profile used in this paper, boosts of 500 or more are needed to set limits or see signals in currently interesting MSSM cross-section ranges. With the (optimistic) Moore profile there is not much change in the result for the region of the sky integrated over. This is because the
lower boundary of this region is 15 $^\circ$ from the GC, and the line of
sight integral does not probe the cuspy region of Moore vs. NFW. Indeed,
an isothermal-cored profile also give about the same result in this case. However, as already mentioned one should note that models with enhanced branching fraction into lines have recently been suggested \cite{Gustafsson:2007pc}.

\subsection{Cosmological WIMP annihilation}
\label{sec:cosmowimps}

WIMP pair annihilation, taking place in DM Halos at all 
redshifts might contribute to the extragalactic \gray \ background  
(EGRB, see \sref{sec:EGBR}). The 2\gray\ channel could then  
result in a distinctive feature, a line which is distorted by the  
integration over all redshifts. The signal of cosmological WIMP annihilation is subject to significant astrophysical uncertainties but less sensitive to the exact choice of halo profile due to the integration over large volumes.\\ 

\noindent
There are several ingredients necessary to calculate the flux of  
\gray s from cosmological WIMP annihilation. In addition to the \gray \ yield per annihilation (see \Eref{eq:gammafluxcont}),  
assumptions need to be made on the distribution and structure of DM halos on  
cosmological scales as well as on the model for the evolution of the  
Universe. One also needs to take into account absorption of the high-energy \gray s on the extragalactic background light. Following \cite{Ullio:2002pj}, the flux of photons can be calculated as
\begin{equation}
\frac{d\phi_\gamma}{dE_0} = \frac{<\sigma v>}{8 \pi} \frac{c}{H_0} \frac 
{\bar{\rho}_0^2}{m_{WIMP}^2}\int{dz (1+z)^3 \frac{\Delta^2(z)}{h(z)} 
\frac{dN_\gamma(E_0(1+z))}{dE} e^{-\tau(z,E_0)}}
\label{eq::1}
\end{equation}
where $c$ is the speed of light, $H_0$ the Hubble constant, $<\sigma v>$ the annihilation cross-section, $m_{WIMP}$ is the WIMP mass, $\bar{\rho}_0$ the average DM density, $h(z)=\sqrt{\Omega_M(1+z)^3 + \Omega_K(1+z)^2 + \Omega_\Lambda}$ parameterizes the energy content of the Universe, $\Delta^2(z)$ describes the averaged squared over-density in halos as a function of redshift and $\tau(z,E_0)$ is the optical depth. The \gray \ yield is the same as discussed in section \ref{sec:calcflux}, where we assume dominant annihilation into $b\bar{b}$ and a branching ratio of $5 \cdot 10^{-4}$ into lines. \\

\begin{figure}[h]
\begin{center}
\includegraphics[width=12cm, height=10cm]{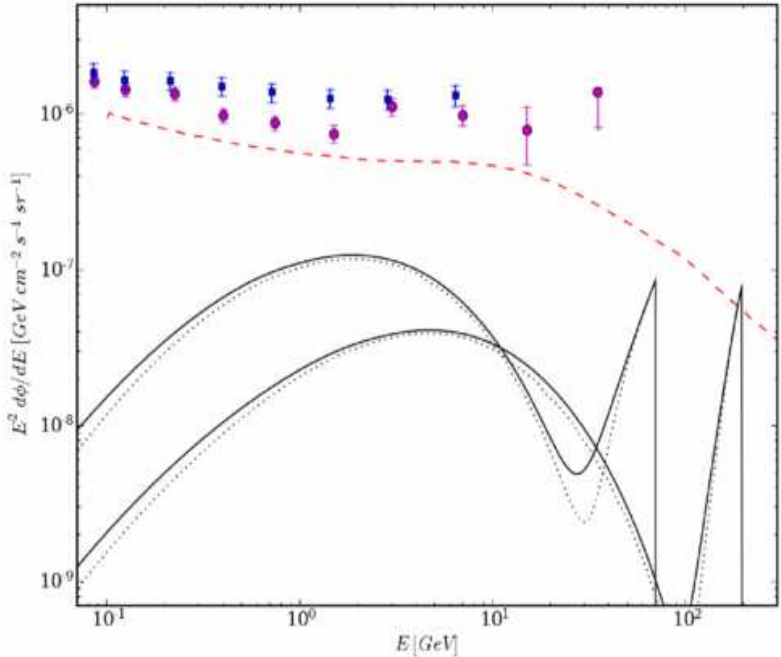}
\caption{Spectral shapes of the EGRET measurements of the EGRB, with \cite{Sreekumar:1997un} (dots) and \cite{Strong:2004ry} (squares) used for our conservative background, and \cite{Ullio:2002pj} (dashed) for the the unresolved blazar model, and two examples of 
cosmological WIMPs, \Eref{eq::1} (including the effect of substructures, see text), with masses of 70 and 200 GeV.
The dotted WIMP spectra are calculated with the absorption from \cite{Stecker:2007nv}.}
\label{fig:ExPlot_spec}
\end{center}
\end{figure}

\noindent
The extragalactic \gray \ signal is strongly  affected by  
absorption in the inter-Galactic medium, especially at high energies.  
The dominant contribution to the absorption in the tens of GeV-TeV  
range is pair production on the extragalactic background light  
emitted in the optical and infrared range. For the optical depth as a 
function of both redshift and observed energy we use the results of  
\cite{Primack:2000xp}. Newer results imply a slightly lower optical depth at low  
redshifts and a slightly higher depth at high redshifts, which in turn  
slightly enhance and weaken our signal at high and low energies  
respectively \cite{Stecker:2007nv}. The cosmic evolution of the energy content of the Universe, $h(z)$, is estimated from the WMAP three-year data \cite{Spergel:2006hy}. Example spectra of a cosmological WIMP signal together with the backgrounds considered in the present note are shown in \fref{fig:ExPlot_spec}. \\

\noindent
The structure of the DM density is encoded in the quantity  $\Delta^2(z) 
$ in \Eref{eq::1}. To derive this quantity one has  
to integrate the contribution from halos of all masses, weighted by  
the halo mass function, and for each mass take into account the  
spread of halo shapes as a function of redshift. Doing this  
it is convenient to parametrize a halo by its virial mass, $M$, and  
concentration parameter, $c$, instead of the parameters $\rho_s$ and  
$r_{s}$ given in \Eref{eq:profile}.
The concentration parameter is defined as $c = R/r_{-2}$, where $r_ 
{-2}$ is the distance where the profile falls as $r^{-2}$ ($r_{-2}=r_s 
$ for the NFW profile) and $R$ is the virial radius within which the  
halo has the mass $M$ and a mean density of $\delta(z)\bar\rho(z)$, where $ 
\bar\rho(z)$ again is the mean background matter density. 
The concentration parameter is dependent both on mass and redshift  
and we model this dependence according to \cite{Bullock:1999he}. Here  
the concentration parameter is also treated as a stochastic variable  
with a log-normal distribution for a fixed mass. Simplified, we can  
write the quantity $\Delta^2(z)$ as:
\begin{equation}
\Delta^2(z)=\int\,dM\,\frac{dn}{dM}\int\,dc\,P(c)\frac{<\rho^2(M,\,c) 
 >}{<\rho(M,\,c)>^2}
\end{equation}
where $dn/dM$ is the halo mass function and $P(c)$ the log-normal  
distribution. Clumping the DM into halos typically yields a boost of $10^4<\Delta^2 
(z=0)<10^6$, depending on the choice of halo profile and the model of  
halo concentration parameters. Both choices contribute  
about a factor of ten each to the uncertainty in the normalization of 
the cosmological WIMP signal.
The concentration parameter model, together with the optical depth,  
changes not only the normalization of the cosmological signal but  
also its shape. This can be compared to point-sources where the  
spectrum can be divided into a product of two factors: one  
determining the shape, resulting only from particle physics, and one  
determining the normalization, originating from astrophysics, i.e.  
the shape of the density profile. The largest contribution to $\Delta^2(z)$ comes from small halos  
formed in an earlier, denser Universe. However, our understanding of  
halos at the low mass end is limited due to finite resolution of the  
N-body simulation. Therefore we have to use a cut-off mass, below  
which we do not trust our toy models for the halo concentration  
parameters. We put this cut-off at $10^5 M_\odot$. Lowering the cut-off might boost the signal 
even further but will also introduce  
further uncertainties. Obviously, the observation of a significant signal could yield 
important information about the DM structure at all redshifts.\\

\noindent
As mentioned in \sref{sec:satellites}, there should exist smaller, bound halos that have survived tidal  stripping within larger halos \cite{Hayashi:2002qv}. Although not as massive as the primary  halos, the substructure halos arise in higher density environments, which makes them denser than  
their parent halos. Also, the subhalos are tidally stripped from the  
outside inward, which further increases their concentration. The phenomenon  
of halos within halos seems to be a generic feature, since detailed  
simulations reveal substructures even within sub-halos \cite{Diemand:2007qr}.
We model the subhalo structure by imposing that a certain fraction of  
the parent halo mass is in substructures, and we associate higher concentration parameters with them. Also here we only consider halos down to masses of approximately $10^5 M_\odot$.\\

\begin{figure}[h]
\begin{center}
\includegraphics[width=12cm,height=6cm]{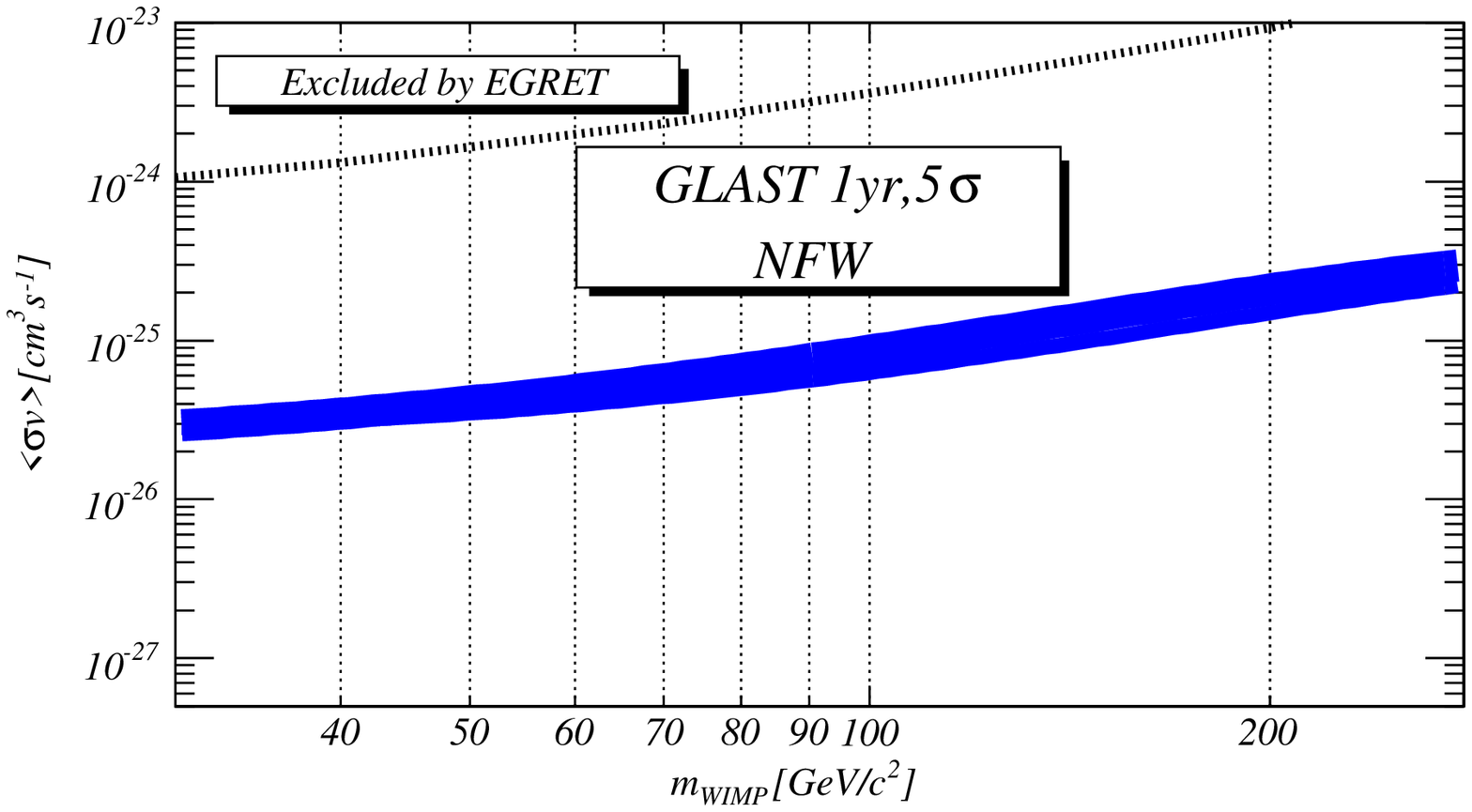}
\includegraphics[width=12cm,height=6cm]{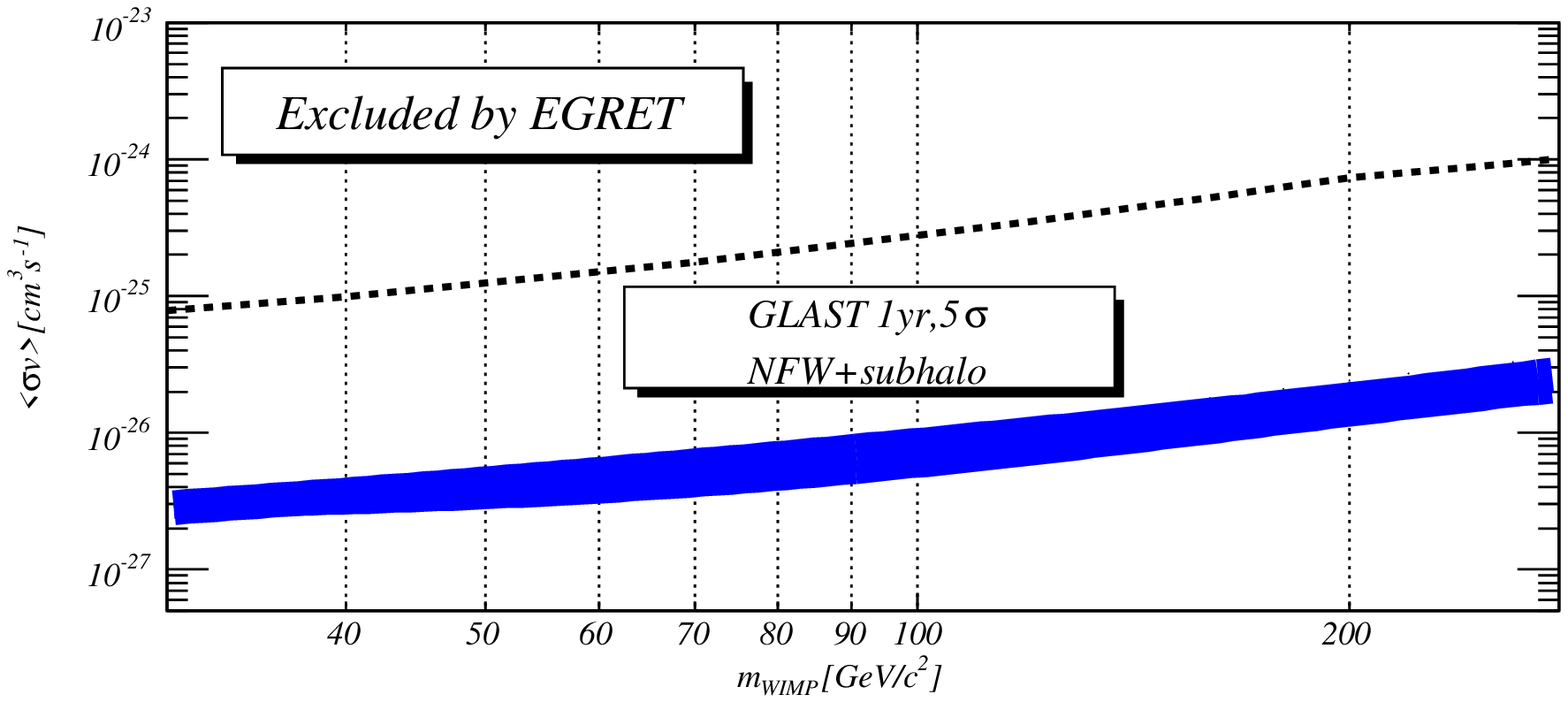}
   \caption{5$\sigma$ exclusion curves for one year of GLAST  
simulated data. The lower edge of the shaded band corresponds to a background as predicted by the Blazar model presented by \cite{Ullio:2002pj}, and the upper bound corresponds to the conservative case, where the background flux is given by the analysis of EGRET data \cite{Sreekumar:1997un}. NFW denotes computation of  the WIMP signal using a NFW profile \cite{Navarro:1995iw} and NFW+subhalo includes  the effect of having 5 \% of the host halo mass in substructures with four times higher concentration parameter than the parent halo. Note that a Moore profile would lead to an improvement in sensitivity by about a factor of ten.}
\label{fig:sensitvities}
\end{center}
\end{figure}

\noindent
Fast detector simulations are performed (using the tools described in \sref{sec:stools}) for a generic model of WIMPs with masses ranging from 50 GeV to 250 GeV. A $\chi^2$ analysis was performed to obtain a sensitivity plot in $<\sigma v>$ {\em vs} $m_{WIMP}$.  
The WIMP signal is computed using the NFW profile and with  
and without the effect of substructures. For the substructures, we assume that they  constitute $5\%$ of the mass and have four times the concentration parameter of their parent halo. As background to our signal we use  
both an optimistic spectrum of unresolved blazars \cite{Ullio:2002pj} and  
a conservative assumption that the background is the EGRB as  
measured by \cite{Sreekumar:1997un}. We  include the residual  
charged particle background in the detector at a level of 10 \% (30 \%) of  
the respective extragalactic diffuse gamma ray background above (below) 10 GeV.\\

\noindent
The result (see \fref{fig:sensitvities})
shows that GLAST is sensitive to total annihilation cross- 
sections of the order $10^{-26}-10^{-25}$ cm$^3$ s$^{-1}$, depending  
on the presence or absence of sub-halos.It should be noted that if the dominant fraction of DM  
were indeed  thermal WIMPs annihilating according to our simplified  
model, cosmology would, to first order, constrain the cross-section  
to be $<\sigma v> \approx 3\cdot10^{-26}$ cm$^3$ s$^{-1}$ in order to  
get  the correct relic density, which would be within the reach of GLAST.\\

\noindent
It should be noted that the  predictions for background spectra from astrophysical sources are very uncertain,  
especially at high energies. Our sensitivity estimate also  
optimistically assumes that we have a perfect analysis in our  
determination of the EGRB.  For the estimate of the signal expectation (as mentioned earlier), the exact shape of the halo is here less important than in other analyses: when assuming a Moore profile the sensitivity improves by a factor of roughly ten.

\section{GLAST sensitivity to specific particle physics models}
\label{sec:specific}
\subsection{GLAST and mSUGRA}

\begin{figure}[ht]
\begin{center}
\includegraphics[width=10cm,height=10cm]{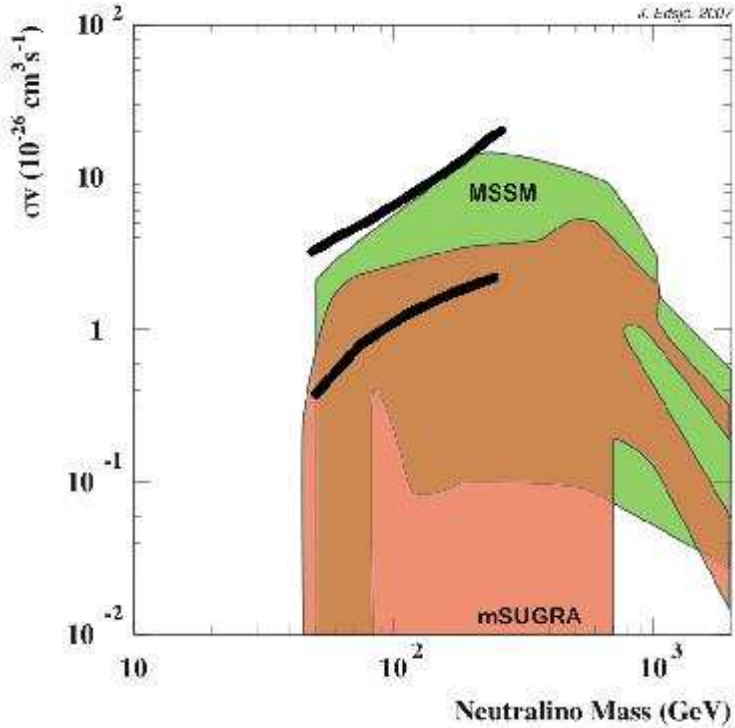}
\caption{MSSM and mSUGRA models in the $<\sigma v>, m_{WIMP}$ plane. The models included in these regions are consistent with accelerator constrains and WMAP data. The lines represent the 5 $\sigma$ sensitivity from the GC (upper) and the 5 $\sigma$ sensitivity from a Galactic halo analysis (lower) corresponding to the best and worst sensitivities estimated in this paper for a NFW profile.}
\label{fig:scan}
\end{center}
\end{figure}

In the preceding sections we have considered a generic model for DM that 
causes the \gray \ signal. We only assumed a WIMP that gives a mono-energetic quark (or fermion) spectrum or line features. In this context the cross-section and mass of the WIMP are free parameters.
However, there are several extensions to the Standard Model of Particle Physics (SM) which predict a particle state which could constitute the WIMP. 
The most studied class of such models is supersymmetry, in particular minimal supersymmetric extensions to the SM (MSSM) and its constrained version mSUGRA, in which the soft supersymmetry breaking terms derive from a high-energy supergravity theory with a common scalar
mass $m_0$ and a common gaugino mass $m_{1/2}$ at the GUT scale (for
details see~\cite{Hall:1983iz,Ohta:1982wn}). In \fref{fig:scan} we show the set of  mSUGRA and MSSM models which pass all accelerator constraints and are consistent with WMAP data in the parameters in which we calculated the LAT sensitivity, i.e. the ($<\sigma v>,m_{WIMP}$) plane.\\

\begin{figure}[ht]
\begin{center}
\includegraphics[scale=1]{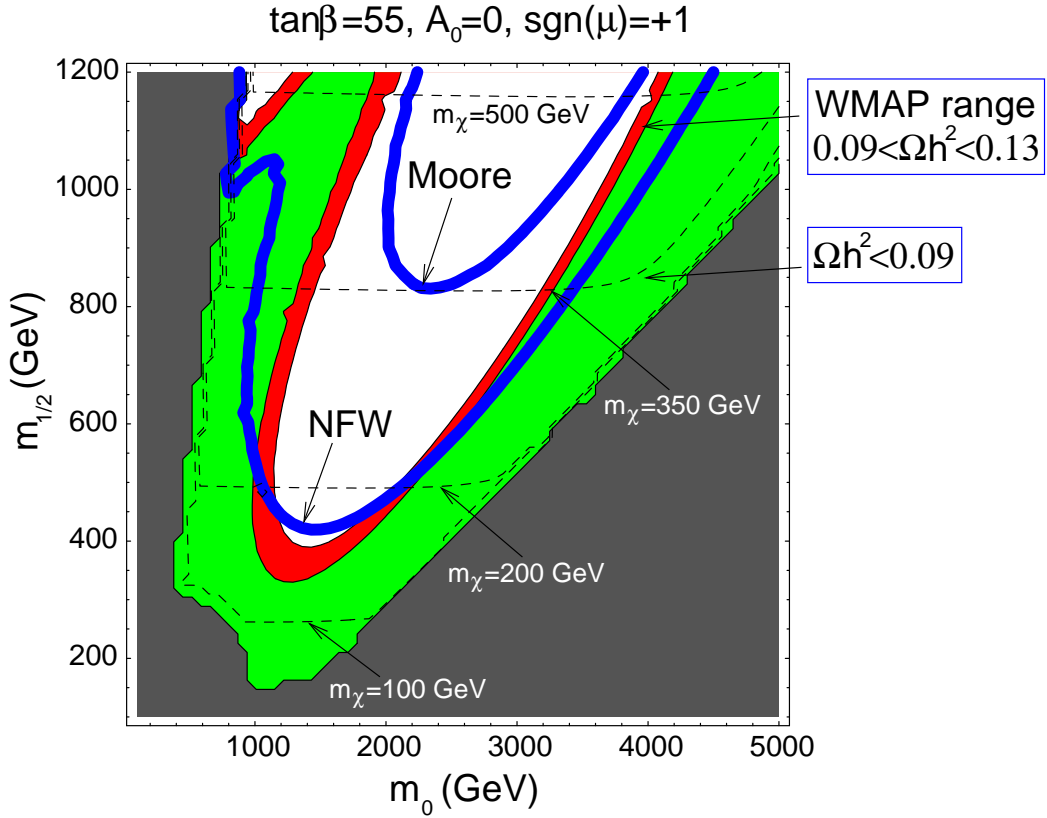}
\caption{GLAST reach (solid lines) in the mSUGRA parameter space for $\tan\beta=55$,
$A_0=0$ and $sgn(\mu)=+1$. The region below the lines can be excluded at 3$\sigma$ level after {\bf 5 years of data taking}.
Solid lines represent the GLAST reach for NFW (lower line) and Moore (upper line) profiles, while dashed lines represent the neutralino isomass contours (expressed in GeV). Dark 
shaded regions are parameter space regions excluded by either theoretical or experimental 
(accelerator) constraints. The upper left region is excluded because in that region the lightest 
stau (not the neutralino) is the LSP,  the lower left region is excluded due the accelerator bounds on the Higgs boson masses, $b\to s \gamma$, slepton and squark masses, etc., while the right lower region is excluded because for these parameters there would be no electroweak symmetry breaking.}
\label{tanbe55cmssm}
\end{center}
\end{figure}

\noindent
For comparison with other experiments (for example accelerators) mSUGRA sensitivities  are often presented in the plane ($m_0$,$m_{1/2}$), where the other free parameters ($\tan\beta$, $A_0$ and $sgn(\mu)$) are fixed. Here $\tan\beta$ denotes the ratio of the vacuum expectation values of the two neutral components of the SU(2) Higgs doublet, $A_0$ is the proportionality factor between the supersymmetry breaking trilinear couplings and the Yukawa couplings, while $\mu$ is determined (up to a sign) by imposing the Electroweak Symmetry Breaking (EWSB) conditions at the weak scale. To give an example we consider the LAT sensitivity to a signal from the GC. The analysis is similar to the one presented in \sref{sec:gc}, except that for ease of computations we use an approximation of the effective area instead of using the detector simulations. In particular, we assume a total exposure of $3.7\cdot 10^{10}\, {\rm cm}^2\, s$, an angular resolution (at $10$ GeV) of $\sim 3\cdot 10^{-5}$ sr, and 5 years of data taking. The error in the sensitive area occupied in the model space introduced by this approximation is estimated to be about 30 \%. The result, at a $3\sigma$ confidence level, for $\tan\beta=55$, $A_0=0$ and $sgn(\mu)=+1$ is shown in  ~\fref{tanbe55cmssm}. It can be seen that for this particular class of models, which is neither particularly optimistic nor pessimistic, GLAST is sensitive to a significant fraction of the cosmologically relevant parameter spaces.\\

\subsection{GLAST and UED}
Universal extra dimension (UED) theories \cite{Appelquist:2000nn} provide another natural candidate for DM. In UED theories all Standard Model fields are allowed to propagate in a higher-dimensional bulk. After compactification of the internal space, these additional degrees of freedom appear as towers of new, heavy states in the effective four-dimensional theory. The lightest of these Kaluza-Klein particles (LKP) is usually stable due to KK parity, an internal symmetry analogous to R parity in the supersymmetry. Since for UED the chiral suppression of annihilation to light fermions is not present, significant yields  ($\sim 20 \%$) of electrons can be expected. We therefore study the KKDM case in the context of the LAT's capability to detect electrons (positrons). Based on the simulations that Baltz and Hooper \cite{Baltz:2004ie} made using a NFW DM distribution with boost factor of 5 and $\rho_{local}$=0.4 GeV $/cm^{3}$ we can estimate the magnitude of the electron flux (after propagation) generated by the LKP annihilation (with the mass of $m_{LKP}$) as
\begin{equation}
\left( \frac{dN_{e}}{dE_{e}}\right)
\approx \frac{9.5\cdot10^{8}}{m_{LKP}^{6}\left[ {\rm GeV}\right] }{\rm m}^{-2}{\rm s}^{-1}{\rm sr}^{-1}{\rm GeV}^{-1}.
\end{equation}

\begin{figure}[ht]
\begin{center}
\includegraphics[scale=0.6]{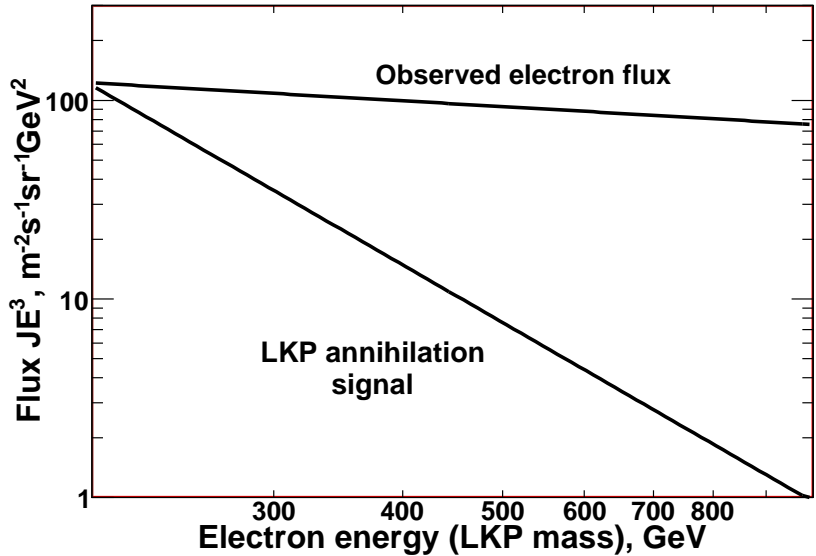}
 \caption {Expected electron flux from LKP annihilation, along with the observed electron flux (see \cite{Torii:2001aw} and references therein).}
\label{fig:moiseev-fig1}
\end{center}
\end{figure}

\noindent
The flux produced by LKP annihilation would have a sharp cutoff in the electron spectrum at the energy corresponding to the LKP mass. The observational goal would be to see a signal with a sharp cutoff riding atop the ``conventional'' electron spectrum. With the large LAT effective area, the number of electrons detected could be sufficient to provide a statistically detection of such a spectral feature. \Fref{fig:moiseev-fig1} shows the comparison of that ``conventional'' electron flux \cite{Torii:2001aw},\cite{Moiseev:2007js} with the flux from LKP annihilation (peak value).\\

\begin{figure}[ht]
\begin{center}
\includegraphics[scale=0.6]{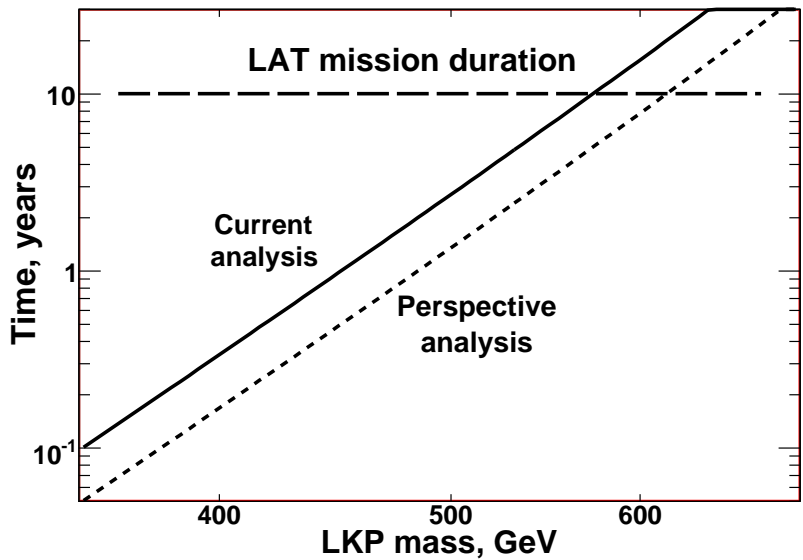}
\caption{LKP detection in the LAT electron spectrum. The time needed to detect an LKP feature with $5 \sigma$ significance.}
\label{moiseev-fig2}
\end{center}
\end{figure}

\noindent
Using a simple sequential cut analysis using topological variables provided by the tracker and the calorimeter, it is possible to detect cosmic ray electrons efficiently \cite{Moiseev:2007js}. In the energy range from 20 GeV to $\sim 1$ TeV the LAT effective geometric factor for electrons is $0.2 - 2 m^{2}sr$, and the energy resolution is 5-20\%, depending on the energy. The residual hadron contamination in this analysis is $\approx 3\%$ of the remaining electron flux. The expected number of electrons is $\sim 10^{7}$ electrons per year in 20 GeV - 1 TeV energy range. The observation time needed to detect an LKP feature within the assumed model with $5\sigma$ significance is shown in \fref{moiseev-fig2}. An LKP with a mass of $\sim$ 600 GeV is probably the heaviest one which can be observed within the assumed constraints and the GLAST mission duration. Taking into account that for thermal freeze-out \cite{Baltz:2004ie} the preferred LKP mass is in the range 600-700 GeV, the window for detecting a LKP mass in the LAT search is not large. However, similar analyses of the LAT capability to detect DM annihilation can be applied to any model in which electrons are produced.\\

\noindent
As an illustration,  we simulate the electron spectrum to be detected by the LAT in 1 year of observation ($\approx 10^{7} $ electrons). For this illustration we consider a signal from the nearest DM clump at a distance of 100 pc, taking into account diffusive propagation of electrons {\cite{moiseev-frascati} and masses of the LKP of 300 and 600 GeV. The signal can be identified with high statistical significance on top of the ``conventional'' electron flux \cite{Torii:2001aw},\cite{Moiseev:2007js} (as can be seen in figure \fref{fig:moiseev-fig3}). This is a very favorable model, but recent results of the balloon-borne instruments ATIC \cite{ATIC} and PPB-BETS \cite{BETS} indicate the presence of such a feature in the electron spectrum  at 300-500 GeV.

\begin{figure}[ht]
\begin{center}
\includegraphics[scale=0.7]{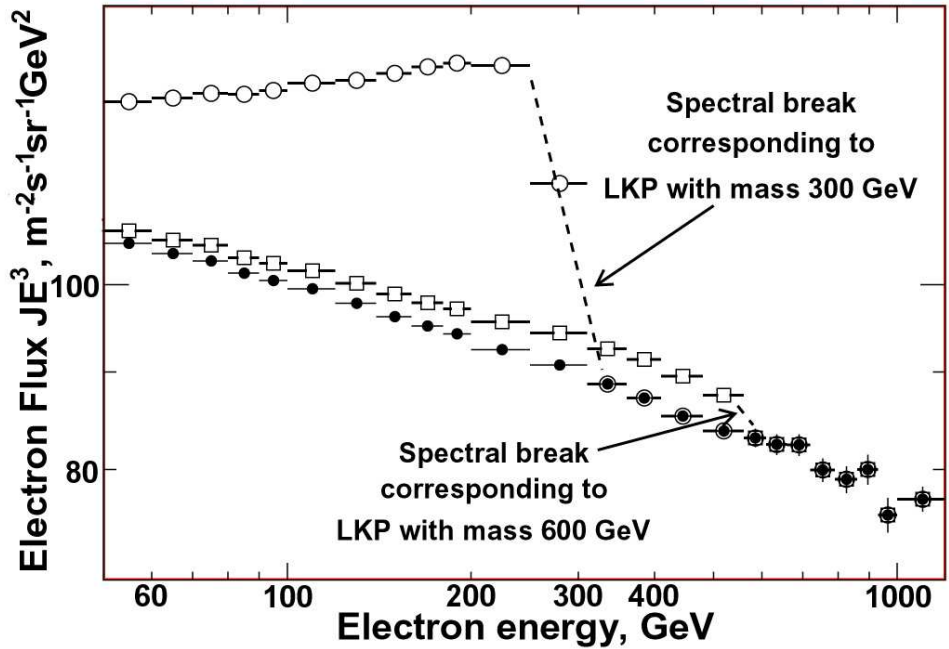}
\caption{Simulated detection of LKPs with masses of 300 GeV and 600 GeV in the LAT electron spectrum to be collected in {\bf 5 years of operation}. Filled circles --- ``conventional'' electron flux; open circles --- the same but with added signal from 300 GeV LKP, and open squares --- the same with added signal from 600 GeV LKP. For this signal, only the nearest clump at a distance of 100 pc is considered. The breaks are also shown by dashed lines to guide the eye.}
\label{fig:moiseev-fig3}
\end{center}
\end{figure}


\section{Discussion \& Conclusions}
Using the current state-of-the-art Monte Carlo and event reconstruction software developed within the LAT collaboration, we present preliminary  sensitivity calculations for several astrophysical searches of DM annihilation. In particular, we present sensitivities for detecting DM annihilation from the GC, Galactic and extragalactic diffuse emission, Galactic known and unknown satellites, point sources and a dedicated search for the 2$\gamma$ line signal.  We exemplify the possibility to constrain specific particle physics scenarios (especially within in mSUGRA and UED frameworks) on the search for DM annihilation \gray \ emission from the GC and by using the LAT not as a \gray \ detector, but employing its very good electron/positron detection capabilities at very high energy.\\

\noindent
The sensitivities presented here are based on analyses which are idealized
in the sense that systematic uncertainties in the instrument performance
estimates are neglected and that uncertainties in the background estimate are
quantitatively taken into account only  in a preliminary manner. For a
given particle physics model, the dominant uncertainty in the
sensitivities presented here is due to the lack of knowledge on the DM
density distribution. Depending on which DM halo profile and/or
substructure is assumed sensitivities can easily vary by one or two orders
of magnitude. For most examples we consider ``standard'' DM halo
structure, i.e. we are conservative in the sense that we do not include 
density enhancements that might boost the expected annihilation signals.\\ 

\noindent
Using simplifying assumptions, the uncertainty in the background 
prediction (most crucial in the search for DM signal from the halo) 
leads to a decrease in sensitivity by between 25 \% and 45 \% depending on the mass of the WIMP. 
For the GC analysis, the  dominant background will be from
sources in the vicinity of the GC, removal of which will lead to
systematic uncertainties. Assessing those without the GLAST data does
not make much sense, given uncertainties in extrapolating into the GLAST
energy region. For the diffuse extragalactic and high-latitude source
searches the charged particle background and uncertainties therein are
potentially important. Prior to launch, the levels of the
charged particle background and its uncertainty are very difficult to assess. 
Estimates have to rely on poorly constrained Monte Carlo simulations. 
The charged particle background included at the level of
roughly 10 \% of the extragalactic background, which is in compliance
with specifications, leads to about  a 10 \% decrease in
sensitivity for a signal in the EGRB and  in the number of detectable
satellites. Uncertainties in the charged particle background are
negligible for the sensitivity for both signals (assuming they are at
the level of $\sim$ 20 \%).\\

\noindent
For the Galactic and extragalactic diffuse background the range of possible backgrounds is illustrated by assuming several representative models, which are discussed in detail. The sensitivities are preliminary in the sense that estimates of the instrument performance, analysis methods and estimates of the expected backgrounds are being continuously improved. In context of specific particle physics scenarios, also the estimates of the signals are continuously updated: For example, calculations of the \gray \ flux for Supersymmetric Dark Matter annihilation incorporating QED corrections, indicate that for part of the parameter space \gray \ yields might be boosted by three or four orders of magnitude and and lead to distinct spectral signatures \cite{Bringmann:2007nk}.\\

 \noindent
We conclude that the LAT has good potential to discover DM
 annihilation for a significant fraction of interesting parameter space,
 i.e. for values of annihilation cross-section of between $<\sigma v>
 \simeq 10^{-26}$ cm$^3$s$^{-1}$  and $<\sigma v> \simeq 10^{-24}$
 cm$^3$s$^{-1}$ depending on WIMP masses in the range between 40 and 500 GeV.
For less conservative assumptions on the Dark Matter density (for example
additional substructure or a Moore profile) the sensitivity improves by
one to two order of magnitudes. Our conclusions are consistent
with previous work that employed cruder representations of the GLAST
response  and/or less thorough treatments of the backgrounds (see e.g
\cite{reviews},\cite{Bergstrom:1997fj}, \cite{Dodelson:2007gd} and
references therein). If indeed a significant DM signal is present, GLAST
will be able to image the DM structure in our Galaxy.\\

\ack{ We thank the members of the LAT collaboration for providing many interesting discussions and elements of the software used in this analysis. Sergio Colafrancesco, Steve Ritz and Julie McEnery are thanked for careful reading of the manuscript. Numerous discussions with Jeff Scargle are gratefully acknowledged. This work was supported by the U.S. Department of Energy contract number DE-AC02-76SF00515, the U.S. Department of Energy contract number, DE-FG02-91ER40690,  NASA, Vetenskapsr{\aa}det and the Swedish Space Board. IM acknowledges partial support by the NASA APRA program. The LAT is being developed by an international collaboration with primary hardware 
and software responsibilities at (in alphabetical order) The Agenzia Spaziale Italiana, 
Centre National de la Recherche Scientifique / Institut National de Physique Nucl\'eaire et de Physique des Particules,
 Commissariat \`a l'Energie Atomique, Goddard Space Flight Center, Hiroshima University, 
Istituto Nazionale di Astrofisica, Istituto Nazionale di Fisica Nucleare, Naval Research Laboratory, 
Ohio State University, Kalmar University, Royal Institute of Technology - Stockholm, Stanford Linear Accelerator Center, 
Stanford University, University of California at Santa Cruz, and University of Washington. The LAT project is managed by the Stanford Linear Accelerator Center, which is also the location of the Instrument Science Operations Center, and the LAT Principal Investigator
is Peter Michelson at Stanford University.Other institutions that have made significant contributions to  the instrument development include the Institute of Space and Astronautical Science, Stockholm University, University of Tokyo, and Tokyo Institute of Science and Technology.\\

}

\section*{References}


\begin{thebibliography}{00}

\bibitem{Atwood:1993zn}
  W.~B.~Atwood  [GLAST Collaboration],
  Nucl.\ Instrum.\ Meth.\  A {\bf 342}, 302 (1994).

\bibitem{Michelson:1999}
 P.~F.~Michelson {\it et al.}, GLAST Proposal to NASA, Response to AO 99-OSS-03, Stanford University, (1999).

\bibitem{Michelson:2007zz}
  P.~F.~Michelson  [GLAST-LAT Collaboration],
  AIP Conf.\ Proc.\  {\bf 921} (2007) 8.

\bibitem{Meegan:2007zz}
  C.~Meegan {\it et al.},
  AIP Conf.\ Proc.\  {\bf 921} (2007) 13.



\bibitem{glast_website} {\tt http://www-glast.staford.edu}



\bibitem{reviews}

  G.~Jungman, M.~Kamionkowski and K.~Griest,
  Phys.\ Rept.\  {\bf 267}, 195 (1996)
  [arXiv:hep-ph/9506380];
  L.~Bergstr\"om,
  Rept.\ Prog.\ Phys.\  {\bf 63}, 793 (2000)
  [arXiv:hep-ph/0002126];
  G.~Bertone, D.~Hooper and J.~Silk,
  Phys.\ Rept.\  {\bf 405}, 279 (2005)
  [arXiv:hep-ph/0404175].
\bibitem{KK}
  D.~Hooper and S.~Profumo,
  Phys.\ Rept.\  {\bf 453}, 29 (2007)
  [arXiv:hep-ph/0701197].


\bibitem{EGRET} http://cossc.gsfc.nasa.gov/docs/cgro/cgro/egret.html


\bibitem{baltz}

 E.~A.~Baltz, M.~Battaglia, M.~E.~Peskin and T.~Wizansky,
  Phys.\ Rev.\  D {\bf 74}, 103521 (2006)
  [arXiv:hep-ph/0602187].

\bibitem{Spergel:2006hy}
  D.~N.~Spergel {\it et al.}  [WMAP Collaboration],
  Astrophys.\ J.\ Suppl.\  {\bf 170}, 377 (2007)
  [arXiv:astro-ph/0603449].

\bibitem{sn}
  S.~Perlmutter {\it et al.}  [Supernova Cosmology Project Collaboration],
  Astrophys.\ J.\  {\bf 517}, 565 (1999)
  [arXiv:astro-ph/9812133];
  A.~G.~Riess {\it et al.}  [Supernova Search Team Collaboration],
  Astron.\ J.\  {\bf 116}, 1009 (1998)
  [arXiv:astro-ph/9805201].

\bibitem{Tegmark:2006az}
  M.~Tegmark {\it et al.},
  Phys.\ Rev.\  D {\bf 74}, 123507 (2006)
  [arXiv:astro-ph/0608632].

\bibitem{Sanchez:2005pi}
  A.~G.~Sanchez {\it et al.},
  Mon.\ Not.\ Roy.\ Astron.\ Soc.\  {\bf 366}, 189 (2006)
  [arXiv:astro-ph/0507583].

\bibitem{Bradac:2006er}
  M.~Bradac {\it et al.},
  Astrophys.\ J.\  {\bf 652}, 937 (2006)
  [arXiv:astro-ph/0608408].


\bibitem{Hannestad:2003xv}
  S.~Hannestad,
  JCAP {\bf 0305}, 004 (2003)
  [arXiv:astro-ph/0303076].


\bibitem{Bergstrom:1997fj}
  L.~Bergstr\"om, P.~Ullio and J.~H.~Buckley,
  Astropart.\ Phys.\  {\bf 9}, 137 (1998)
  [arXiv:astro-ph/9712318].


\bibitem{Gustafsson:2007pc}
  M.~Gustafsson, E.~Lundstr\"om, L.~Bergstr\"om and J.~Edsj\"o,
  Phys.\ Rev.\ Lett.\  {\bf 99} 041301 (2007)
  [arXiv:astro-ph/0703512].



\bibitem{Moiseev:2007hk}
  A.~A.~Moiseev \etal,
  Astropart.\ Phys.\  {\bf 27} (2007) 339
  [arXiv:astro-ph/0702581].


\bibitem{st}
http://glast.gsfc.nasa.gov/ssc/data/
\bibitem{perf}
http://www-glast.slac.stanford.edu/software/IS/glast\_lat\_performance.htm




\bibitem{Allanach:2003jw}
  B.~C.~Allanach, S.~Kraml and W.~Porod,
  JHEP {\bf 0303} (2003) 016
  [arXiv:hep-ph/0302102].


\bibitem{Belanger:2005jk}
  G.~Belanger, S.~Kraml and A.~Pukhov,
  Phys.\ Rev.\  D {\bf 72}, 015003 (2005)
  [arXiv:hep-ph/0502079].


\bibitem{Gondolo:2004sc}
  P.~Gondolo, J.~Edsj\"o, P.~Ullio, L.~Bergstr\"om, M.~Schelke and E.~A.~Baltz,
  JCAP {\bf 0407} (2004) 008
  [arXiv:astro-ph/0406204].

\bibitem{Cesarini:2003nr}
  A.~Cesarini, F.~Fucito, A.~Lionetto, A.~Morselli and P.~Ullio,
  Astropart.\ Phys.\  {\bf 21} (2004) 267


\bibitem{pythia}
  Pythia program package, 
  see T. Sj\"ostrand, Comp.  Phys. Comm. {\bf 82} (1994) 74. 



\bibitem{Navarro:1995iw}
J.~F.~Navarro, C.~S.~Frenk and S.~D.~White,
Astrophys.\ J.\  {\bf 462} (1996) 563


\bibitem{Moore:1999gc}
B.~Moore, T.~Quinn, F.~Governato, J.~Stadel and G.~Lake,
Mon.\ Not.\ Roy.\ Astron.\ Soc.\  {\bf 310} (1999) 1147


\bibitem{Yao:2006px}
  W.~M.~Yao {\it et al.}  [Particle Data Group],
  J.\ Phys.\ G {\bf 33} (2006) 1.





\bibitem{Strong:2006hf}
  A.~W.~Strong,
  Astrophys.\ Space Sci.\  {\bf 309} (2007) 35
  [arXiv:astro-ph/0609359].



\bibitem{Strong:1998fr}
  A.~W.~Strong, I.~V.~Moskalenko and O.~Reimer,
  Astrophys.\ J.\  {\bf 537} (2000) 763


\bibitem{Strong:2004de}
  A.~W.~Strong, I.~V.~Moskalenko and O.~Reimer,
  Astrophys.\ J.\  {\bf 613}, 962 (2004)
  [arXiv:astro-ph/0406254].


\bibitem{Galprop}
Galprop,2008, http://galprop.stanford.edu

\bibitem{Strong:2007nh}
  A.~W.~Strong, I.~V.~Moskalenko and V.~S.~Ptuskin,
  Ann.\ Rev.\ Nucl.\ Part.\ Sci.\  {\bf 57}, 285 (2007)
  [arXiv:astro-ph/0701517].


\bibitem{M04rev}
\pubproc{I.~V.~Moskalenko, A.~W.~Strong, and O.~Reimer}
{ed K S Cheng and G E Romero}
{Astrophysics and Space Science Library \rm vol 304}
{Dordrecht: Kluwer}
{279--310}
{2004}
{Diffuse Gamma Rays: Galactic and Extragalactic Diffuse Emission}


\bibitem{Hunter:1997we}
  S.~D.~Hunter {\it et al.},
  Astrophys.\ J.\  {\bf 481}, 205 (1997).

\bibitem{Moskalenko:2006zy}
  I.~V.~Moskalenko, S.~W.~Digel, T.~A.~Porter, O.~Reimer and A.~W.~Strong,
  Nucl.\ Phys.\ Proc.\ Suppl.\  {\bf 173} (2007) 44
  [arXiv:astro-ph/0609768].


\bibitem{Stecker:2007xp}
  F.~W.~Stecker, S.~D.~Hunter and D.~A.~Kniffen,
  arXiv:0705.4311 [astro-ph].
 


\bibitem{Baughman:2007ck}
  B.~M.~Baughman, W.~B.~Atwood, R.~P.~Johnson, T.~A.~Porter and M.~Ziegler,
  arXiv:0706.0503 [astro-ph].
, Contributed to 30th International Cosmic Ray Conference (ICRC 2007), Merida, Yucatan, Mexico, 3-11 Jul 2007.




\bibitem{yiou97}
\pubjournal{F.~Yiou \etal}
{\jgr--Oceans}
{102}
{26783}
{1997}

\bibitem{knie04}
\pubjournal{K.~Knie \etal}
{\prl}
{93}
{171103}
{2004}
{Fe-60 anomaly in a deep-sea manganese crust and implications for a nearby 
supernova source}

\bibitem{case96}
\pubjournal{G.~Case and D.~Bhattacharya}
{\aap\ Suppl.}
{120C}
{437}
{1996}



\bibitem{Moskalenko:2005xu}
  I.~V.~Moskalenko and A.~W.~Strong,
  AIP Conf.\ Proc.\  {\bf 801} (2005) 57
  [arXiv:astro-ph/0509414].


\bibitem{Strong:1998pw}
  A.~W.~Strong and I.~V.~Moskalenko,
  Astrophys.\ J.\  {\bf 509}, 212 (1998)
  [arXiv:astro-ph/9807150].

\bibitem{SeoPtuskin1994}
\pubjournal{E.~S.~Seo and V.~S.~Ptuskin}
{\apj}
{431}
{705}
{1994}
{}


\bibitem{Orito:1999re}
  S.~Orito {\it et al.}  [BESS Collaboration],
  Phys.\ Rev.\ Lett.\  {\bf 84}, 1078 (2000)
  [arXiv:astro-ph/9906426].



\bibitem{Moskalenko:2001ya}
  I.~V.~Moskalenko, A.~W.~Strong, J.~F.~Ormes and M.~S.~Potgieter,
  Astrophys.\ J.\  {\bf 565}, 280 (2002)
  [arXiv:astro-ph/0106567].


\bibitem{Hams:2007}
  T.~Hams {\it et al.}  [BESS Collaboration],
   {Proc.\ 30th \icrc\ (Merida)}


\bibitem{Moskalenko:2002yx}
  I.~V.~Moskalenko, A.~W.~Strong, S.~G.~Mashnik and J.~F.~Ormes,
  Astrophys.\ J.\  {\bf 586}, 1050 (2003)
  [arXiv:astro-ph/0210480].


\bibitem{Moskalenko:2003kq}
  I.~V.~Moskalenko, A.~W.~Strong, S.~G.~Mashnik and J.~F.~Ormes,
  {Proc.\ 28th \icrc\ (Tsukuba)}
{4}
{1921}
{2003}
{}
  arXiv:astro-ph/0306368.


\bibitem{Thompson92}
\pubjournal{D.~J.~Thompson and C.~E.~Fichtel}
{\aap}
{109}
{352}
{1982}

\bibitem{Sreekumar:1997un}
   P.~Sreekumar {\it \etal}  [EGRET Collaboration],
   Astrophys.\ J.\  {\bf 494}, 523 (1998)
   [arXiv:astro-ph/9709257].


\bibitem{Strong:2004ry}
  A.~W.~Strong, I.~V.~Moskalenko and O.~Reimer,
  Astrophys.\ J.\  {\bf 613}, 956 (2004)
  [arXiv:astro-ph/0405441].

\bibitem{Moskalenko:1998gw}
  I.~V.~Moskalenko and A.~W.~Strong,
  Astrophys.\ J.\  {\bf 528}, 357 (2000)
  [arXiv:astro-ph/9811284].


\bibitem{Moskalenko:2006ta}
  I.~V.~Moskalenko, T.~A.~Porter and S.~W.~Digel,
  Astrophys.\ J.\  {\bf 652}, L65 (2006)
  [Erratum-ibid.\  {\bf 664}, L143 (2007)]
  [arXiv:astro-ph/0607521].


\bibitem{Orlando:2006zs}
  E.~Orlando and A.~Strong,
  Astrophys.\ Space Sci.\  {\bf 309}, 359 (2007)
  [arXiv:astro-ph/0607563].

\bibitem{Orlando:2007qx}
  E.~Orlando, D.~Petry and A.~Strong,
  AIP Conf.\ Proc.\  {\bf 921}, 502 (2007)
  [arXiv:0704.0462 [astro-ph]].




\bibitem{Moskalenko:2007tk}
  I.~V.~Moskalenko {\it et al.},
  arXiv:0712.2015 [astro-ph].


\bibitem{Stecker:1996ma}
  F.~W.~Stecker and M.~H.~Salamon,
  Astrophys.\ J.\  {\bf 464}, 600 (1996)
  [arXiv:astro-ph/9601120].

\bibitem{Elsaesser:2004ap}
  D.~Elsaesser and K.~Mannheim,
  Phys.\ Rev.\ Lett.\  {\bf 94}, 171302 (2005)
  [arXiv:astro-ph/0405235].


\bibitem{deBoer:2004es}
  W.~de Boer,
  New Astron.\ Rev.\  {\bf 49}, 213 (2005)
  [arXiv:hep-ph/0408166].



\bibitem{HessGC}
\pubjournal{F.~Aharonian \etal\ (HESS Collaboration)}
{\aap}
{425}
{L13}
{2004}
{H.E.S.S. observations of the Galactic Center region
and their possible dark matter interpretation}


\bibitem{Albert06} J.~Albert \etal\ (MAGIC Collaboration) 
2006, \apjl, 638, L101 


\bibitem{VeritasGC}  
\pubjournal{K.~Kosack, \etal\ (VERITAS Collaboration)}
{\apj}
{608}
{L97}
{2004} 
{}

\bibitem{CangarooGC}  
\pubjournal{T.~Tsuchiya, \etal\ (CANGAROO Collaboration)}
{\apj}
{606}
{L115}
{2004} 
{}

\bibitem{Profumo:2005xd}
  S.~Profumo,
  Phys.\ Rev.\  D {\bf 72} (2005) 103521
  [arXiv:astro-ph/0508628].

\bibitem{Bergstrom:2004cy}
  L.~Bergstr\"om, T.~Bringmann, M.~Eriksson and M.~Gustafsson,
  Phys.\ Rev.\ Lett.\  {\bf 94}, 131301 (2005)
  [arXiv:astro-ph/0410359].


\bibitem{vanEldik:2007yi}
  C.~van Eldik, O.~Bolz, I.~Braun, G.~Hermann, J.~Hinton and W.~Hofmann
                  (H.E.S.S. Collaboration),
  arXiv:0709.3729 [astro-ph], Contributed to 30th International Cosmic Ray Conference (ICRC 2007), Merida, Yucatan, Mexico, 3-11 Jul 2007. 

\bibitem{Hooper:2002ru}
  D.~Hooper and B.~L.~Dingus,
  Phys.\ Rev.\  D {\bf 70} (2004) 113007
  [arXiv:astro-ph/0210617].


\bibitem{Dodelson:2007gd}
  S.~Dodelson, D.~Hooper and P.~D.~Serpico,
  arXiv:0711.4621 [astro-ph].


\bibitem{Mayer}  
  H.~Mayer-Hasselwander \etal,  Astron. Astrophys. {\bf 335} (1998) 161.

\bibitem{DarkSusy}
P.~Gondolo \etal ,
JCAP, {\bf 0407} (2004) 008

\bibitem{Hooper:2007gi}
  D.~Hooper, G.~Zaharijas, D.~P.~Finkbeiner and G.~Dobler,
  arXiv:0709.3114 [astro-ph].


\bibitem{Finkbeiner:2003im}
  D.~P.~Finkbeiner,
  Astrophys.\ J.\  {\bf 614} (2004) 186
  [arXiv:astro-ph/0311547].


\bibitem{Regis:2008ij}
  M.~Regis and P.~Ullio,
  arXiv:0802.0234 [hep-ph].


\bibitem{Blumenthal:1984bp}
  G.~R.~Blumenthal, S.~M.~Faber, J.~R.~Primack and M.~J.~Rees,
  Nature {\bf 311}, 517 (1984).

\bibitem{Peebles:1984ge}
  P.~J.~E.~Peebles,
  Astrophys.\ J.\  {\bf 284}, 439 (1984).


\bibitem{Profumo:2006bv}
  S.~Profumo, K.~Sigurdson and M.~Kamionkowski,
  Phys.\ Rev.\ Lett.\  {\bf 97} (2006) 031301
  [arXiv:astro-ph/0603373].

\bibitem{Hayashi:2002qv}
  E.~Hayashi, J.~F.~Navarro, J.~E.~Taylor, J.~Stadel and T.~Quinn,
  Astrophys.\ J.\  {\bf 584} (2003) 541
  [arXiv:astro-ph/0203004].


\bibitem{Taylor:2004gq}
  J.~E.~Taylor and A.~Babul,
  Mon.\ Not.\ Roy.\ Astron.\ Soc.\  {\bf 364}, 535 (2005)
  [arXiv:astro-ph/0410049].
  J.~E.~Taylor and A.~Babul,
  Mon.\ Not.\ Roy.\ Astron.\ Soc.\  {\bf 364}, 515 (2005)
  [arXiv:astro-ph/0410048].



\bibitem{Eke:2000av}
  V.~R.~Eke, J.~F.~Navarro and M.~Steinmetz,
  Astrophys.\ J.\  {\bf 554}, 114 (2001)
  [arXiv:astro-ph/0012337].






\bibitem{Diemand:2005vz}
  J.~Diemand, B.~Moore and J.~Stadel,
  Nature {\bf 433}, 389 (2005)
  [arXiv:astro-ph/0501589].



\bibitem{Jungman:1995df}
  G.~Jungman, M.~Kamionkowski and K.~Griest,
  Phys.\ Rept.\  {\bf 267}, 195 (1996)
  [arXiv:hep-ph/9506380].

\bibitem{Pieri:2007ir}
  L.~Pieri, G.~Bertone and E.~Branchini,
  arXiv:0706.2101 [astro-ph].

\bibitem{Diemand:2006ik}
  J.~Diemand, M.~Kuhlen and P.~Madau,
  Astrophys.\ J.\  {\bf 657} (2007) 262
  [arXiv:astro-ph/0611370].


\bibitem{Taylor:2002zd}
  J.~E.~Taylor and J.~Silk,
  Mon.\ Not.\ Roy.\ Astron.\ Soc.\  {\bf 339} (2003) 505
  [arXiv:astro-ph/0207299].

\bibitem{bertin96} 
E.~Bertin and  S.~Arnouts, S., 1996, 

\aaps, 117, 393. See also http://terapix.iap.fr/rubrique.php?id\_rubrique=91/.

\bibitem{Baltz:2006sv}
  E.~A.~Baltz, J.~E.~Taylor and L.~L.~Wai,
  arXiv:astro-ph/0610731.


\bibitem{Evans:2003sc}
  N.~W.~Evans, F.~Ferrer and S.~Sarkar,
  Phys.\ Rev.\  D {\bf 69}, 123501 (2004)
  [arXiv:astro-ph/0311145].


\bibitem{Hartman:1999fc}
  R.~C.~Hartman {\it et al.}  [EGRET Collaboration],
  Astrophys.\ J.\ Suppl.\  {\bf 123} (1999) 79.


\bibitem{Kleyna}
J.~T.~Kleyna, M.~I.~Wilkinson, N.~W.~Evans and G.~Gilmore
Astrophys.\ J.\ {\bf 563}, L115 (2001)
[arXiv:astro-ph/0111329].

\bibitem{Ibata}
R.~A.~Ibata, R.~F.~G. Wyse, G.~Gilmore, M.~J.~Irwin, N.~B.~Suntzeff, 
Astron.\ J.\ {\bf 113}, 634 (1997)
[arXiv:astro-ph/9612025].

\bibitem{IMBH2}
  G.~Bertone, A.~R.~Zentner and J.~Silk,
  Phys.\ Rev.\ D {\bf 72}, 103517 (2005)
  [arXiv:astro-ph/0509565].

\bibitem{Stoehr:2003hf}
  F.~Stoehr, S.~D.~M.~White, V.~Springel, G.~Tormen and N.~Yoshida,
  Mon.\ Not.\ Roy.\ Astron.\ Soc.\  {\bf 345} (2003) 1313
  [arXiv:astro-ph/0307026].


\bibitem{Ullio:2002pj}
  P.~Ullio, L.~Bergstr\"om, J.~Edsj\"o and C.~G.~Lacey,
  Phys.\ Rev.\  D {\bf 66}, 123502 (2002)
  [arXiv:astro-ph/0207125].

\bibitem{Primack:2000xp}
   J.~R.~Primack, R.~S.~Somerville, J.~S.~Bullock and  
J.~E.~G.~Devriendt,
   AIP Conf.\ Proc.\  {\bf 558}, 463 (2001)
   [arXiv:astro-ph/0011475].


\bibitem{Stecker:2007nv}
   F.~W.~Stecker,
   AIP Conf.\ Proc.\  {\bf 921}, 237 (2007)
   [arXiv:astro-ph/0703505].



\bibitem{Bullock:1999he}
   J.~S.~Bullock {\it \etal},
   Mon.\ Not.\ Roy.\ Astron.\ Soc.\  {\bf 321} (2001) 559
   [arXiv:astro-ph/9908159].


\bibitem{Diemand:2007qr}
   J.~Diemand, M.~Kuhlen and P.~Madau,
   Astrophys.\ J.\  {\bf 667}, 859 (2007)
   [arXiv:astro-ph/0703337].

\bibitem{Hall:1983iz}
  L.~J.~Hall, J.~D.~Lykken and S.~Weinberg,
  Phys.\ Rev.\  D {\bf 27}, 2359 (1983).


\bibitem{Ohta:1982wn}
  N.~Ohta,
  Prog.\ Theor.\ Phys.\  {\bf 70}, 542 (1983).

\bibitem{Appelquist:2000nn}
  T.~Appelquist, H.~C.~Cheng and B.~A.~Dobrescu,
  Phys.\ Rev.\  D {\bf 64} (2001) 035002
  [arXiv:hep-ph/0012100].


\bibitem{Baltz:2004ie}
  E.~A.~Baltz and D.~Hooper,
  JCAP {\bf 0507}, 001 (2005)
  [arXiv:hep-ph/0411053].


\bibitem{Torii:2001aw}
  S.~Torii {\it et al.},
  Astrophys.\ J.\  {\bf 559} (2001) 973.


\bibitem{Moiseev:2007js}
  A.~A.~Moiseev, J.~F.~Ormes and I.~V.~Moskalenko,
   {Proc.\ 30th \icrc\ (Merida)}
  arXiv:0706.0882 [astro-ph].

\bibitem{moiseev-frascati} A.~A.~Moiseev \etal,  Frascati Phys.Ser.45:1-328,2007. Edited by A. Lionetto, A. Morselli (
ISBN 978-88-86409-54-0)

\bibitem{ATIC} J.~Chang \etal, 2005, {Proc.\ 30th \icrc\ (Pune)}
\bibitem{BETS} K.~Yoshida \etal, 2005, {Proc.\ 30th \icrc\ (Pune)}

\bibitem{Bringmann:2007nk}
  T.~Bringmann, L.~Bergstr\"om and J.~Edsj\"o,
  arXiv:0710.3169 [hep-ph].


\bibitem{Cheng:2002ej}
  H.~C.~Cheng, J.~L.~Feng and K.~T.~Matchev,
  Phys.\ Rev.\ Lett.\  {\bf 89} (2002) 211301
  [arXiv:hep-ph/0207125].


\end{thebibliography}
\end{document}